\documentclass[11pt]{article}
\usepackage[T1]{fontenc}
\usepackage[utf8]{inputenc}
\usepackage{lmodern,microtype}
\usepackage[margin=1in]{geometry}
\usepackage{amsmath,amssymb,amsthm,mathtools}
\usepackage{booktabs,array,tabularx,enumitem}
\usepackage{float}
\floatstyle{ruled}
\newfloat{algorithm}{tbp}{loa}
\floatname{algorithm}{Algorithm}
\newcounter{algline}
\newlength{\algindent}
\newlength{\algorithmicindent}
\newenvironment{algorithmic}[1][]{%
 \small\setcounter{algline}{0}\setlength{\algindent}{0pt}%
 \begin{list}{}{\setlength{\leftmargin}{1.8em}\setlength{\labelwidth}{1.4em}%
 \setlength{\labelsep}{.4em}\setlength{\itemsep}{2pt}\setlength{\parsep}{0pt}}%
}{\end{list}}
\newcommand{\State}{\stepcounter{algline}\item[\scriptsize\thealgline:]\hspace*{\algindent}}
\newcommand{\Statex}{\item[]\hspace*{\algindent}}
\newcommand{\Require}[1]{\item[]\textbf{Input:} #1}
\newcommand{\Return}{\textbf{return}\ }
\newcommand{\For}[1]{\State\textbf{for} #1 \textbf{do}\addtolength{\algindent}{1em}}
\newcommand{\EndFor}{\addtolength{\algindent}{-1em}\State\textbf{end for}}
\newcommand{\If}[1]{\State\textbf{if} #1 \textbf{then}\addtolength{\algindent}{1em}}
\newcommand{\Else}{\addtolength{\algindent}{-1em}\State\textbf{else}\addtolength{\algindent}{1em}}

\newcommand{\EndIf}{\addtolength{\algindent}{-1em}\State\textbf{end if}}

\usepackage{tikz}
\usetikzlibrary{arrows.meta,positioning,calc,patterns}
\usepackage[dvipsnames]{xcolor}
\usepackage[colorlinks=true,linkcolor=MidnightBlue,citecolor=MidnightBlue,urlcolor=MidnightBlue]{hyperref}
\usepackage{url}
\setlist{topsep=4pt,itemsep=2pt,parsep=0pt}
\newtheorem{theorem}{Theorem}[section]
\newtheorem{lemma}[theorem]{Lemma}
\newtheorem{corollary}[theorem]{Corollary}
\newtheorem{proposition}[theorem]{Proposition}
\theoremstyle{definition}

\theoremstyle{remark}

\newcommand{\diam}{\operatorname{diam}}
\newcommand{\poly}{\operatorname{poly}}

\newcommand{\R}{\mathbb R}
\newcommand{\Q}{\mathbb Q}

\newcommand{\Work}{\mathsf{Work}}

\usepackage{placeins}

\hypersetup{pdftitle={Finding the Farthest in High Dimensions: Dynamic Diameter and Farthest-Neighbor Search},pdfauthor={Mohammad Hajiaghayi and Morteza Monemizadeh}}

\title{Finding the Farthest in High Dimensions:\\
Dynamic Diameter and Farthest-Neighbor Search
}
\author{
Mohammad Hajiaghayi\\
University of Maryland, College Park\\
\texttt{hajiagha@umd.edu}
\and
Morteza Monemizadeh\\
TU Eindhoven, the Netherlands\\
\texttt{m.monemizadeh@tue.nl}
}
\date{November 15, 2026}
\begin{document}
\maketitle

\begin{abstract}
We give algorithms for maintaining diameter under point insertions
and deletions, and for answering farthest-neighbor queries on a fixed
point set, in high-dimensional Euclidean space. For every fixed
$1<c<\sqrt2$, capacity $n$, update horizon $t$, and error probability
$\delta$, we maintain an actual factor-$c$ diameter pair with
worst-case update time
$n^{2/c^2-1}\poly_c(d,\log(2nt/\delta))$ and space
$n^{2/c^2}\poly_c(d,\log(2nt/\delta))$ in the arithmetic-RAM model.
Here $n$ bounds the number of active point identities. The update
exponent improves Indyk's SODA~2003 bound~\cite{indyk} from $1/c^2$
to $2/c^2-1$, while the space exponent improves from $1+1/c^2$
to $2/c^2$. At $c=\sqrt{3/2}$, the update exponent decreases
from $2/3$ to $1/3$. A separate construction achieves approximation
exactly $\sqrt2$ with update time polynomial in
$d$ and $\log(2nt/\delta)$. With probability at least $1-\delta$,
every maintained answer during the first $t$ updates is correct,
even when updates depend on all previous answers and random choices.

A second construction improves the update exponent when the space
exponent is close to one. On rational inputs, it has an always-correct
variant with expected amortized update bounds over every fixed finite
prefix of an adaptive execution, without an advance update horizon. The dynamic structures store a current answer, so reporting
its two point identifiers takes constant time.

The geometric method also gives static diameter algorithms and a
farthest-neighbor structure for a fixed set that answers arbitrary
adaptive queries. At approximation $\sqrt2-\varepsilon$, the
diameter time is $n^{1+O(\varepsilon)}$, and the query time is
$n^{O(\varepsilon)+o(1)}$ with near-linear preprocessing and space,
up to polynomial input-parameter factors. 
On binary points, this
improves the $n^{O(\sqrt\theta)}$ overhead of
Andoni, Jiang, and Zharkov~\cite{ajz2026} (STOC~2026) for Hamming approximation
$2-\theta$ to $n^{O(\theta)+o(1)}$.

The main geometric step augments a centerpoint with a second-moment
coordinate. The resulting variance identity supplies many distance
witnesses that survive deletions, and reduces the remaining searches
to a common angular separation. Implicit representations of the
filter lists give the improved space tradeoff.
\end{abstract}

\section{Introduction}
\label{sec:introduction}

High-dimensional data are often represented by vectors, and distances
between these vectors measure how different two observations are.
Finding a farthest point is a basic operation in diversity maximization,
outlier detection, and iterative learning procedures~\cite{adaptiveafn}.
The diameter summarizes the largest separation within the data and is
closely related to enclosing balls and clustering~\cite{agarwal,dynamic}.
For example, consider a collection of feature vectors that changes as
observations arrive or expire. A pair of well-separated current vectors
describes the collection's extent; a farthest-neighbor query asks for
an observation most different from a specified vector. Recomputing
either answer by scanning the entire collection can be expensive when
such operations are repeated.

For a finite indexed family $P$ of points in $\mathbb R^d$, its
Euclidean diameter is $\diam(P)=\max_{x,y\in P}\|x-y\|$, with value
zero when $|P|\le1$. For $|P|\ge2$, a factor-$c$ diameter pair
consists of distinct active identities $x,y\in P$ with
$\|x-y\|\ge\diam(P)/c$; distinct identities may have equal coordinates. For a query $q\in\mathbb R^d$,
a factor-$c$ farthest neighbor is a point $p\in P$ with
$\|q-p\|\ge\max_{x\in P}\|q-x\|/c$; the query need not belong to $P$.
We study fully dynamic diameter, where individual points may be inserted
or deleted, and also obtain a farthest-neighbor structure for a fixed
point set. The dimension is an input parameter, and our bounds have
polynomial dependence on $d$.

\paragraph{Why deletions and adaptivity matter.}
An insertion can create a new farthest pair, while a deletion can remove
an endpoint of the pair currently stored by the algorithm. After such a
deletion, the structure must find a sufficiently long pair among the
remaining points, even if it has never returned that pair before.
Moreover, an application may choose its next update or query after
seeing earlier answers. An \emph{oblivious} adversary fixes the entire
operation sequence in advance; an \emph{adaptive} adversary may use the
history of the computation. We allow the latter to see all previous
random choices as well as all answers. Only the fresh randomness used
after the next operation is fixed remains unknown. This is the robust
dynamic model studied in~\cite{dynamic}. It is also useful when a
geometric data structure is called by another algorithm whose later
requests depend on its earlier answers.

\paragraph{Previous dynamic and streaming bounds.}
Indyk's asymmetric embedding~\cite{indyk}\footnote{The asymmetry is
in the expansion and contraction guarantees of the random-projection
embedding. The diameter application maintains ordered lists along
many Gaussian projection directions and tests extreme projected
candidates.}
gives a fully dynamic factor-$c$ diameter approximation with
$\widetilde O(dn^{1/c^2})$ amortized update time. Its projection lists
use $\widetilde O(dn+n^{1+1/c^2})$ space. Here $n$ bounds the number
of active points, and the tilde suppresses logarithmic factors.
The guarantee does not provide the exposed-randomness robustness
considered here.

A different line achieves an approximation slightly larger than
$\sqrt2$ with much faster updates. Agarwal and Sharathkumar's
blurred-ball method~\cite{agarwal} is an insertion-only streaming
algorithm: it gives a $(\sqrt2+\varepsilon)$-approximate diameter
using $O(d\varepsilon^{-3}\log(1/\varepsilon))$ space and amortized
update time. Chan and Pathak~\cite{chanpathak} extend the underlying
enclosing-ball approach to fully dynamic updates. For fixed
$\varepsilon>0$, the resulting diameter structure has
$O_\varepsilon(d\log n)$ expected amortized update time against an
oblivious adversary; the diameter consequence is stated explicitly
in~\cite[Appendix~A]{dynamic}. Thus insertion-only streaming and
fully dynamic maintenance have different guarantees and should be
distinguished.

Banihashem et al.~\cite{dynamic} obtain a factor-$2$ approximation with
worst-case $\poly(d,\log t)$ update time over $t$ updates against an
adaptive adversary that sees previous random choices. Their central idea is to use a
centerpoint as a representative that remains useful after deletions.\footnote{
A point has halfspace depth at least $\alpha m$ in an $m$-point set
if every closed halfspace containing it contains at least $\alpha m$
input points. We call such a point an $\alpha$-centerpoint; see
Section~\ref{sec:preliminaries}.}
Their representative is taken in the original space. Our work is
motivated by this connection between centerpoints and robust dynamic
diameter. We add a squared-norm coordinate before choosing the
representative, thereby retaining a variance term that is absent
from the original-space argument.

These results leave a clear quantitative question. Above $\sqrt2$,
known structures have update time almost linear in the dimension and
logarithmic in the set size. Below $\sqrt2$, Indyk's update bound has
a positive power of $n$ that approaches $1/2$ as $c$ approaches
$\sqrt2$. We ask:
\begin{center}
\fbox{\begin{minipage}{.91\linewidth}
Can one maintain a $c$-approximate diameter pair below $\sqrt2$ with
an update exponent that tends to zero at $\sqrt2$, while supporting
arbitrary deletions and fully adaptive updates?
\end{minipage}}
\end{center}
Our answer is affirmative. The same geometry also gives faster
farthest-neighbor queries on a fixed data set, with one success event
valid for every query vector.

\subsection{Main results}
\label{main:thm}

We first state the dynamic diameter bounds. The capacity $n\ge2$ is an
upper bound on $|P|$ throughout the execution. For a Monte Carlo
structure, $t\ge1$ is an upper bound on the number of updates and
$\delta\in(0,1/2)$ is the allowed probability that any answer is
incorrect. All guarantees below permit the adaptive adversary just
defined. A diameter query accesses a cached header containing two compact
point handles and a reference to their previously computed squared
distance in $O(1)$ RAM-cell accesses. With fewer than two active
points it reports \textsc{None} and squared diameter zero.
Serializing handles, coordinates, or the exact squared distance
is charged to the output length, in particular in the bit model.

The notation $\poly_c(d,L)$ denotes a polynomial in $d$ and $L$ whose
degree and coefficients may depend on the fixed approximation factor
$c$. In particular, it contains no additional power of $n$. All statements use the computation and parameter conventions of
Section~\ref{sec:model}. Fixed approximation and space parameters
are rational or admit polynomial-time certified rational
approximations. The arithmetic model counts exact field operations,
comparisons, word operations, and fair random bits. For rational
inputs, bounds that display $B$ count bit operations and bits of
space, including intermediate numbers. The bounds include empty
initialization; an initial nonempty set is constructed by insertions.

The following is our main update bound.
\begin{theorem}[Fully dynamic diameter below $\sqrt2$]
\label{thm:paper-balanced}
For every fixed $1<c<\sqrt2$, a fully dynamic structure maintains a
factor-$c$ diameter pair with worst-case update time
$n^{2/c^2-1}\poly_c(d,\log(2nt/\delta))$ and space
$n^{2/c^2}\poly_c(d,\log(2nt/\delta))$.
With probability at least $1-\delta$, all answers during the first
$t$ adaptive updates are correct.
\end{theorem}

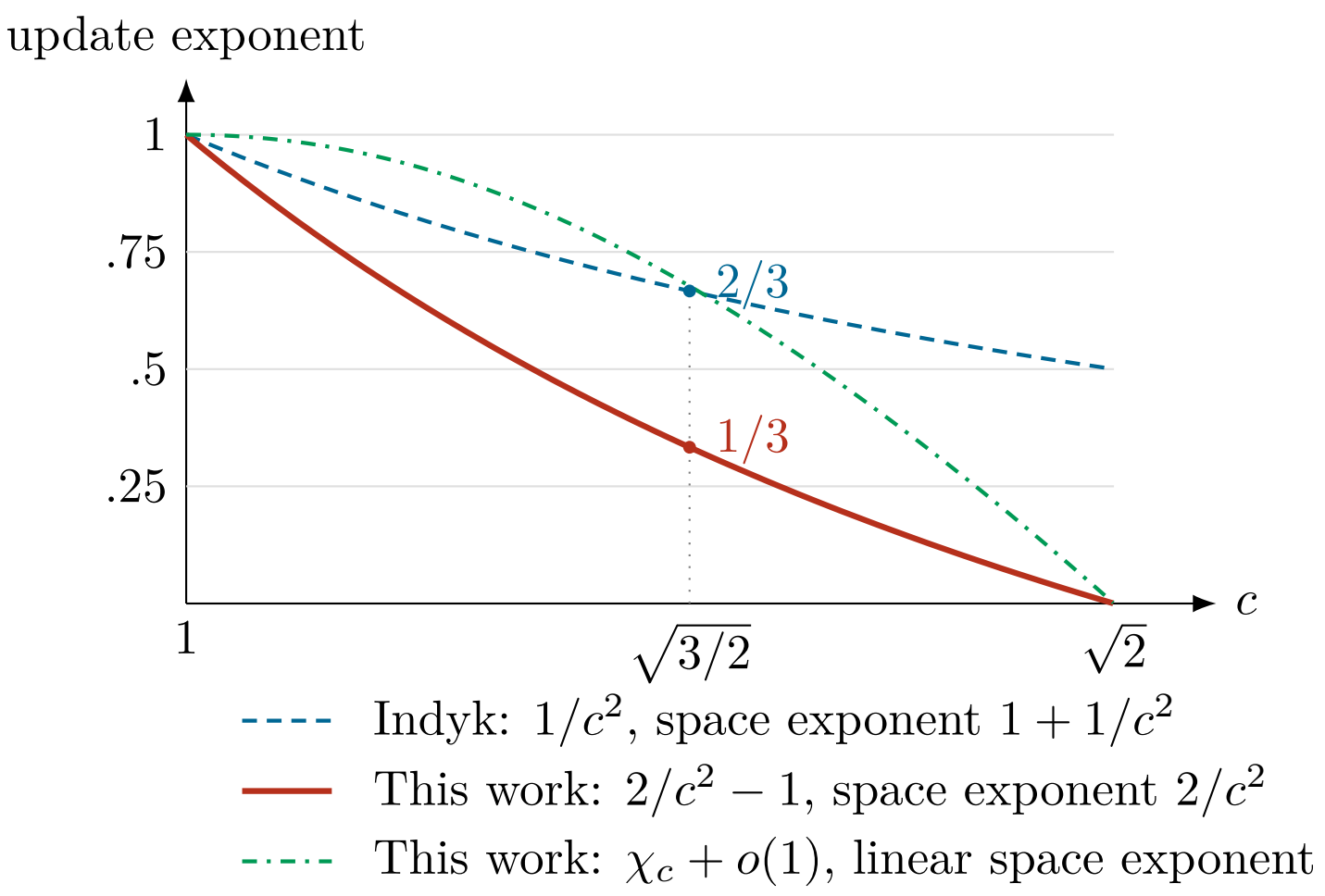
\begin{figure}[tbp]
\centering
\begin{tikzpicture}[x=16.0cm,y=3.35cm,>=Latex,font=\small]
 \draw[->] (0,0)--(.46,0) node[right] {$c$};
 \draw[->] (0,0)--(0,1.12) node[above] {update exponent};
 \foreach \y in {.25,.5,.75,1} {
   \draw[gray!25] (0,\y)--(.41421356,\y);
   \node[left] at (0,\y) {$\y$};
 }
 \node[below] at (0,0) {$1$};
 \node[below] at (.22474487,0) {$\sqrt{3/2}$};
 \node[below] at (.41421356,0) {$\sqrt2$};
 \draw[densely dashed,thick,MidnightBlue,domain=1:1.41421356,samples=100]
   plot ({\x-1},{1/(\x*\x)});
 \draw[very thick,BrickRed,domain=1:1.41421356,samples=100]
   plot ({\x-1},{2/(\x*\x)-1});
 \draw[thick,ForestGreen,dash dot,domain=1:1.41421356,samples=100]
   plot ({\x-1},{2-sqrt(1+3*(\x*\x-1)^2)});
 \draw[gray,dotted] (.22474487,0)--(.22474487,.6666667);
 \fill[MidnightBlue] (.22474487,.6666667) circle (1.3pt);
 \fill[BrickRed] (.22474487,.3333333) circle (1.3pt);
 \node[right,MidnightBlue] at (.228,.68) {$2/3$};
 \node[right,BrickRed] at (.228,.35) {$1/3$};
 \draw[densely dashed,thick,MidnightBlue] (.025,-.25)--(.065,-.25);
 \node[right] at (.075,-.25) {Indyk: $1/c^2$, space exponent $1+1/c^2$};
 \draw[very thick,BrickRed] (.025,-.40)--(.065,-.40);
 \node[right] at (.075,-.40) {This work: $2/c^2-1$, space exponent $2/c^2$};
 \draw[thick,ForestGreen,dash dot] (.025,-.55)--(.065,-.55);
 \node[right] at (.075,-.55) {This work: $\chi_c+o(1)$, linear space exponent};
\end{tikzpicture}
\caption{Update exponents as a function of the approximation factor.
The red curve improves both exponents of Indyk's dynamic diameter
bound. The green curve uses less space and is the limiting exponent
of our always-correct construction; its time has an additional
vanishing $o(1)$ term. The three curves have the distinct space
bounds shown in the legend.}
\label{fig:paper-exponents}
\end{figure}

Compared with Indyk's exponents, both the update and space exponents
decrease by $1-1/c^2$ throughout $1<c<\sqrt2$. At
$c=\sqrt{3/2}$, the update exponent decreases from $2/3$ to $1/3$,
and the space exponent from $5/3$ to $4/3$. As $c$ increases to
$\sqrt2$, our update exponent tends to zero and the space exponent
tends to one. Figure~\ref{fig:paper-exponents} shows this comparison.
The theorem also gives worst-case updates and robustness to exposed
previous randomness.

The endpoint requires a separate argument: the angular search used
below $\sqrt2$ is unnecessary there. To state an explicit polynomial
bound, put $D=d+3$ and $H=\lceil\log_2(2Dnt/\delta)\rceil$.
\begin{theorem}[The exact $\sqrt2$ endpoint]
\label{thm:paper-endpoint}
A fully dynamic structure maintains a factor-$\sqrt2$ diameter pair
with worst-case update time $O(D^{19}H^{18})$ and space
$O(nD^{19}H^{18})$. With probability at least $1-\delta$, every answer
during the first $t$ adaptive updates is correct.
For rational coordinates of at most $B$ bits, an always-correct
variant uses $n\poly(d,B,\log(2n))$ space and
$\poly(d,B,\log(2n))$ expected amortized update time, without a
prescribed bound on the number of updates.
\end{theorem}

The displayed polynomial is one explicit bound for the construction;
its degrees are not optimized. Lemma~\ref{lem:explicit-endpoint-cost}
derives it by substituting the centerpoint construction time and the
exact-leaf cutoff into the replacement schedule. The significance
of the endpoint is its polylogarithmic dependence on $n$ at
approximation \emph{exactly} $\sqrt2$.

\paragraph{Time--space tradeoffs.}
The main theorem balances the update exponent and the extra exponent
above linear space. More generally, let $\sigma$ specify the space
bound $n^{1+\sigma}$, apart from polynomial factors. Write
$g=c^2-1$ and $\rho=2/c^2-1$. For $0\le\sigma\le\rho$, storing
the filter lists explicitly gives update exponent
\[
Q_c(\sigma)=\bigl(\sqrt{1-g^2}-g\sqrt\sigma\bigr)^2.
\]
Thus Theorem~\ref{thm:paper-balanced} extends to the bounds
\[
 \begin{aligned}
 \text{update time}&\quad n^{Q_c(\sigma)}
          \poly_{c,\sigma}(d,\log(2nt/\delta)),\\
 \text{space}&\quad n^{1+\sigma}
          \poly_{c,\sigma}(d,\log(2nt/\delta)).
 \end{aligned}
\]
At $\sigma=\rho$ both exponents beyond the leading linear space
factor equal $\rho$; at $\sigma=0$ the update exponent is
$Q_c(0)=c^2(2-c^2)$. These bounds have no $n^{o(1)}$ overhead.

Our second construction improves the update exponent when the
space is close to linear. It represents point--filter incidences
implicitly and recovers only those needed by a search. The quantities
below describe its curve: $\chi_c$ is the exponent at linear space,
and $u_c$ is the point where the improved linear branch meets $Q_c$.
Set
\[
\chi_c=2-\sqrt{1+3g^2},\qquad
 u_c=\frac{g^2(1-g^2)}{(1+g^2+\sqrt{1+3g^2})^2},
\]
and
\[
\Lambda_c(\sigma)=
 \begin{cases}
 \chi_c-(3-\chi_c)\sigma,&0\le\sigma\le u_c,\\
 Q_c(\sigma),&u_c\le\sigma\le\rho.
 \end{cases}
\]
The first branch is tangent to $Q_c$ at $u_c$, and is strictly below
it for $0\le\sigma<u_c$. This improvement comes with a vanishing
additive term in the time exponent. It also admits an always-correct
(Las Vegas) implementation: randomness affects the work, but never
the validity of an answer.

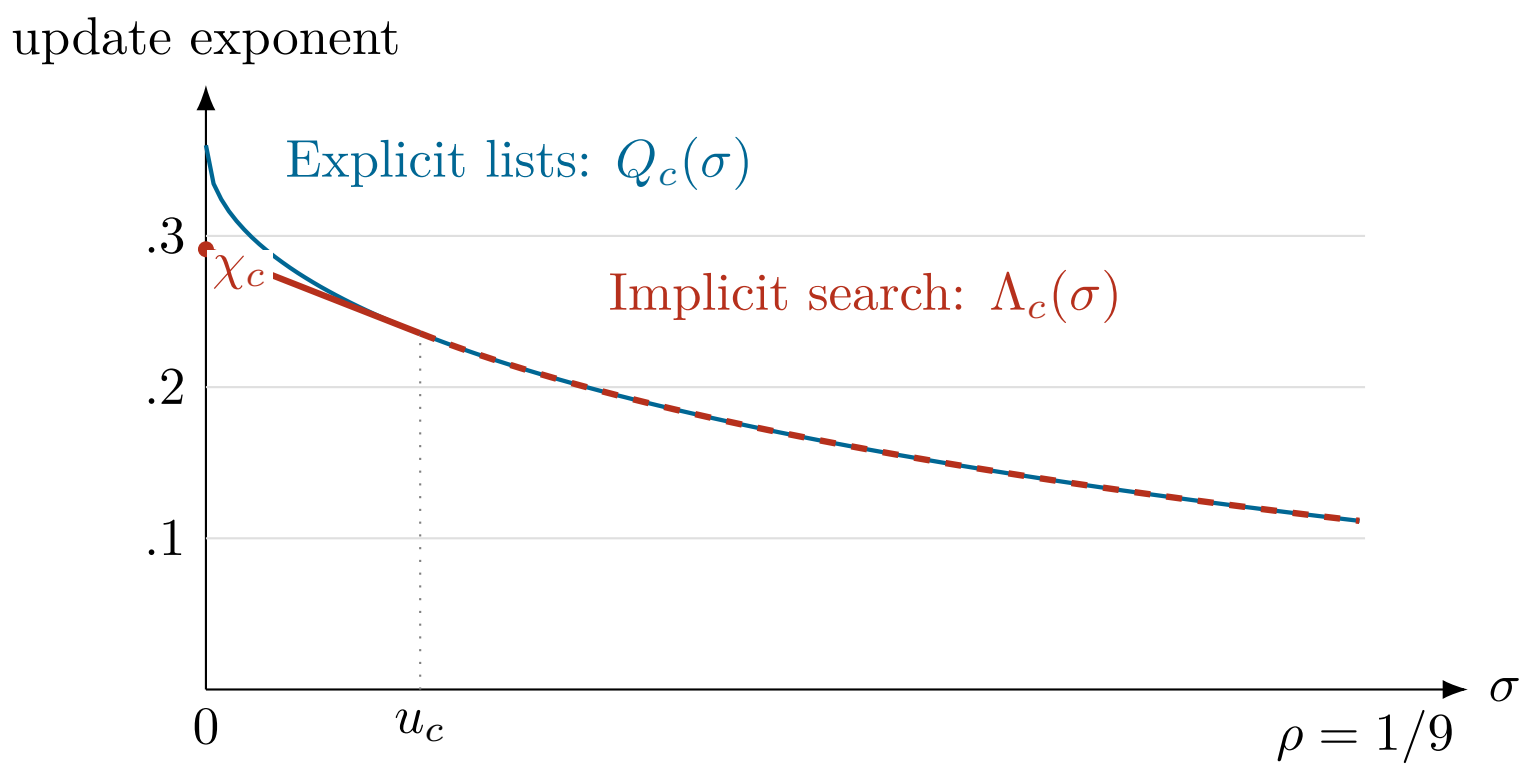
\begin{figure}[tbp]
\centering
\begin{tikzpicture}[x=69cm,y=10cm,>=Latex,font=\small]
 \draw[->] (0,0)--(.121,0) node[right] {$\sigma$};
 \draw[->] (0,0)--(0,.40) node[above] {update exponent};
 \foreach \y in {.1,.2,.3} {
   \draw[gray!25] (0,\y)--(.1111111,\y);
   \node[left] at (0,\y) {$\y$};
 }
 \node[below] at (0,0) {$0$};
 \node[below] at (.020545,0) {$u_c$};
 \node[below] at (.1111111,0) {$\rho=1/9$};
 \draw[gray,dotted] (.020545,0)--(.020545,.2355);
 \draw[thick,MidnightBlue,domain=0:.1111111,samples=150]
   plot (\x,{(.6-.8*sqrt(\x))^2});
 \draw[very thick,BrickRed,domain=0:.020545,samples=25]
   plot (\x,{(2-sqrt(2.92))-(1+sqrt(2.92))*\x});
 \draw[very thick,BrickRed,dashed,domain=.020545:.1111111,samples=100]
   plot (\x,{(.6-.8*sqrt(\x))^2});
 \fill[BrickRed] (0,{2-sqrt(2.92)}) circle (1.5pt);
 \node[below right,BrickRed,fill=white,inner sep=1pt] at (0,{2-sqrt(2.92)}) {$\chi_c$};
 \node[anchor=west,MidnightBlue,fill=white,inner sep=1pt] at (.007,.348)
   {Explicit lists: $Q_c(\sigma)$};
 \node[anchor=west,BrickRed,fill=white,inner sep=1pt] at (.038,.26)
   {Implicit search: $\Lambda_c(\sigma)$};
\end{tikzpicture}
\caption{The time--space tradeoff at $c=\sqrt{9/5}$.
Space is $n^{1+\sigma}$ up to polynomial factors. The implicit
construction replaces the initial part of the explicit curve by
its tangent at $u_c\approx0.0205$, improving the update exponent
near linear space. Its running time has an additional vanishing
$o(1)$ term; the space exponent stays $1+\sigma$.}
\label{fig:paper-space-tradeoff}
\end{figure}

See also Figure~\ref{fig:paper-space-tradeoff} for the two curves
at a fixed approximation factor.

\begin{theorem}[Improved tradeoff with zero error]
\label{thm:paper-certified}
For fixed $1<c<\sqrt2$ and $0\le\sigma\le\rho$, rational points
with coordinate bit length at most $B$ admit an always-correct fully
dynamic factor-$c$ diameter structure with space
$n^{1+\sigma}\poly_{c,\sigma}(d,B,\log(2n))$ and expected amortized
update time
$n^{\Lambda_c(\sigma)+o(1)}\poly_{c,\sigma}(d,B,\log(2n))$.
It supports adaptive update sequences of arbitrary length.
For a prescribed $t$ and $\delta$, a Monte Carlo variant has the
same exponents, worst-case update time, and polynomial logarithmic
dependence on $t/\delta$.
\end{theorem}

Expected amortized time bounds the expected total work over every
fixed finite prefix of an adaptive execution starting from the empty
set. It includes construction, unsuccessful attempts, and rebuilding.
The space bound includes all simultaneously live structures and their
workspace. The following quantitative version makes the $o(1)$ term
in Theorem~\ref{thm:paper-certified} explicit.

\begin{corollary}[Explicit vanishing overhead]
\label{cor:paper-uniform-overhead}
The $o(1)$ in Theorem~\ref{thm:paper-certified} can be replaced by
$C_{c,\sigma}(\log(2n))^{-1/100}$ for one algorithm and a constant
$C_{c,\sigma}$ depending only on $c,\sigma$. The polynomial degrees
in $d,B,\log(2n)$ are fixed, and the space exponent remains
$1+\sigma$.
\end{corollary}

The proof is Theorem~\ref{thm:uniform-subpolynomial-density} in
Appendix~\ref{sec:uniform-subpolynomial-density}. In particular,
increasing the dimension does not change the displayed exponent of
$n$. At linear space, that exponent tends to $\chi_c$; for
$c=1+\varepsilon$ it equals
$1-6\varepsilon^2+O(\varepsilon^3)$. The accuracy is fixed in these
statements: the constants need not be polynomial in $1/\varepsilon$.

\paragraph{Static diameter and arbitrary farthest-neighbor queries.}
Inserting all input points immediately gives static diameter
algorithms. The geometric reduction also supports arbitrary queries
to a fixed data set. These are separate consequences: the fully
dynamic theorems maintain one pair, while the static query structure
returns a farthest neighbor of any supplied $q\in\mathbb R^d$.

\begin{corollary}[Static Euclidean diameter]
\label{cor:paper-static-diameter}
For fixed $1<c<\sqrt2$, a factor-$c$ diameter pair of $n$ points
can be found with high probability in
$n^{1+c^2(2-c^2)}\poly_c(d,\log n)$ time and
$n\poly_c(d,\log n)$ space. For rational inputs, an always-correct
algorithm has expected time
$n^{1+\chi_c+o(1)}\poly_c(d,B,\log n)$ and
$n\poly_c(d,B,\log n)$ space.
\end{corollary}
\begin{proof}
Insert the points into the explicit structure with $\sigma=0$, or
into Theorem~\ref{thm:paper-certified} with $\sigma=0$, and return
the stored pair.
\end{proof}

For a fixed set, Corollary~\ref{cor:paper-static-fns} gives
$n^{1+o(1)}\poly_c(d,\log(n/\delta))$ preprocessing and space, and
$n^{c^2(2-c^2)+o(1)}\poly_c(d,\log(n/\delta))$ query time.
With probability at least $1-\delta$, \emph{every} real query is
answered correctly, including queries chosen after the entire
structure and its random choices have been inspected. No upper
bound on the number of queries is needed. A zero-error counterpart
for rational inputs is stated in
Corollary~\ref{cor:paper-static-fns-las-vegas}.

At $c=\sqrt2-\varepsilon$, these results give static diameter time
$n^{1+O(\varepsilon)}$ and query time
$n^{O(\varepsilon)+o(1)}$, with near-linear preprocessing and space
for the query structure and polynomial factors in the input
parameters. On binary inputs, these Euclidean guarantees also improve
the Hamming exponents of Andoni, Jiang, and Zharkov~\cite{ajz2026}.
Squared Euclidean distance equals Hamming distance on binary vectors. Taking
$c=\sqrt{2-\theta}$ therefore yields the direct comparison in
Table~\ref{tab:static-comparison}.

\begin{table}[tbp]
\centering\small
\begin{tabular}{@{}lll@{}}
\toprule
Hamming approximation $2-\theta$ & Diameter time & Farthest-neighbor query\\
\midrule
Andoni--Jiang--Zharkov~\cite{ajz2026}
 & $n^{1+O(\sqrt\theta)+o(1)}$ & $n^{O(\sqrt\theta)+o(1)}$\\
This work, explicit filters
 & $n^{1+2\theta-\theta^2}$ & $n^{2\theta-\theta^2+o(1)}$\\
This work, zero error
 & $n^{1+\frac32\theta+O(\theta^2)+o(1)}$
 & $n^{2\theta-\theta^2+o(1)}$\\
\bottomrule
\end{tabular}
\caption{Static bounds for fixed $0<\theta<1$, suppressing polynomial
dimension and logarithmic factors. Our diameter algorithms have
linear space exponent; our query structures have near-linear
preprocessing and space. Zero-error bounds are expected total-work
bounds. The exact zero-error diameter exponent is
$3-\sqrt{4-6\theta+3\theta^2}+o(1)$.}
\label{tab:static-comparison}
\end{table}

\paragraph{Limits on faster updates.}
We also explain why the dependence on the approximation factor
cannot simply be removed. The randomized Orthogonal Vectors
Conjecture (OVC) states that, for every constant $\eta>0$, orthogonal
pair detection between two $n$-point binary sets in a suitable
dimension $O(\log n)$ has no bounded-error
$O(n^{2-\eta})$-time algorithm~\cite{chenwilliams}.
Under this hypothesis, a single $(1+\varepsilon)$-approximation
scheme cannot have update time
$n^{1-\gamma}\poly(d,1/\varepsilon,\log n)$ for a fixed
$\gamma>0$, a constant-time answer, and fixed polynomial
initialization. Section~\ref{sec:paper-lower-bounds} proves this by
an explicit reduction that pays for initialization only once.
This does not conflict with our bounds, whose saving tends to zero
as $c$ tends to one.

Memory lower bounds concern a different resource. Adapting the
insertion-only INDEX reduction of Agarwal--Sharathkumar~\cite{agarwal}
gives a lower bound of
$\Omega_c(n)$ bits below $\sqrt2$, already on sign vectors in
$O_c(\log n)$ dimensions (Proposition~\ref{prop:paper-index-space}).
Thus our linear-space exponent is optimal
on those bounded-coordinate inputs, up to logarithmic factors.
The bound counts the stored data as part of memory; it gives no
corresponding lower bound on update time. Our structures explicitly
store the active points.

\subsection{Related work}

\paragraph{Geometric reductions and robust representatives.}
Goel, Indyk, and Varadarajan~\cite{goel2001} reduce static
farthest-neighbor search to nearest-neighbor search using enclosing
balls, radial shells, and reflection. Chan and Pathak's
hierarchy~\cite{chanpathak} supplies a dynamic organization, while
Banihashem et al.~\cite{dynamic} show how centerpoint depth protects
a representative against adaptive deletions. The additional
squared-norm coordinate in our construction preserves a variance
identity as well as depth. This gives a stronger distance witness
than the original-space representative and makes the remaining
searches instances of one common angular gap.

\paragraph{Filtering and implicit search.}
Our angular indices use locality-sensitive filtering
\cite{christiani,alrw2017}. The time--space curve for explicitly
stored incidences is consistent with the filtering tradeoffs
of~\cite{alrw2017}; their lower bounds concern the specified search
model, rather than arbitrary dynamic diameter structures.
Our zero-error construction uses finite norm-splitting families
from Wei~\cite{wei2019lasvegas}. To reduce space, we combine
implicit-set search~\cite{aronov2024implicit} and McCauley's
function-inversion approach~\cite{mccauley2024} with complete-preimage
recovery~\cite{bbm2026} and dynamic phase techniques
\cite{kuszmaul2026}. We require two additional properties: construction
workspace must fit the final space bound, and the cost of recovering
a bucket must be charged to its \emph{current} members after
deletions. Recent static preimage-reporting bounds~\cite{vannimmen2026}
do not by themselves supply this dynamic guarantee.

\paragraph{Static and adaptive query algorithms.}
Polynomial methods and fast matrix multiplication give offline
approximate pair algorithms~\cite{acw16,acw20}.
Andoni--Jiang--Zharkov's regularity method~\cite{ajz2026} gives
Hamming $(2-\theta)$-diameter in
$n^{1+O(\sqrt\theta)+o(1)}d^{O(1)}$ time and the corresponding
farthest-neighbor structure; Table~\ref{tab:static-comparison}
records our improvement. Their approximate Orthogonal Vectors
algorithm has the different bound $n^{1+O(\theta)+o(1)}$.
Adversarially robust farthest-neighbor search on a fixed data set is
studied in~\cite{adaptiveafn}. Our fixed-set result similarly allows
arbitrary adaptive queries, while our dynamic results concern
maintenance of the diameter under point insertions and deletions.

\paragraph{Streaming and other update models.}
Dynamic streaming $\varepsilon$-kernels give near-exact extent
approximations in fixed dimension~\cite{chan2016kernels}, with
dimension dependence outside the bounds sought here. Small
insertion-only diameter summaries are studied in~\cite{hmv2025}.
The expiration-streaming model~\cite{expiration2026} supplies deletion
times when points arrive; our model does not. The merge-only
partition structure of Bathie--Lagarde~\cite{bathielagarde2025} has
another update model. Polynomial streaming-space lower bounds for
general metrics~\cite{kpsw2025} and Euclidean strict-turnstile
streams~\cite{psw2026} further describe the memory needed when the
input is not retained. The latter allows arbitrary stream length
and multiplicities. Polynomial-length sketch simulations
\cite{jly2026} have separate assumptions, which we keep distinct in
Section~\ref{sec:paper-lower-bounds}.

\FloatBarrier

\section{Our approach and techniques}
\label{sec:techniques}

The proof has three main ingredients. First, we strengthen the
centerpoint method for adaptive dynamic diameter
of Banihashem et al.~\cite{dynamic}: a center chosen after adding one
coordinate also records the spread of the points. This gives an exact
$\sqrt2$-approximation that survives deletions. Second, any pair that
this approximation misses at a factor $c<\sqrt2$ satisfies a useful
angular separation. We search for such pairs by filtering, and maintain
a small set of candidate pairs by a matching. Third, we improve the
space bound by recovering filter lists when they are needed, instead
of storing every list explicitly. The first two ingredients yield the
balanced bound; the third yields the improved tradeoff near linear
space. We explain the geometry first and then the dynamic implementation.

\subsection{A center that records the spread of the points}

For a center $o$ in the convex hull of a point set, every query $q$ has
some point at distance at least $\|q-o\|$. This is the lower bound used
by the original-space centerpoint approach~\cite{dynamic}; combined
with the triangle inequality, it gives factor two. To improve that
argument, we retain a second quantity. Suppose nonnegative weights
$\lambda_p$ sum to one and satisfy $\sum_p\lambda_pp=o$. Their
\emph{variance about $o$} is $V=\sum_p\lambda_p\|p-o\|^2$. Then
\[
 \sum_p\lambda_p\|p-q\|^2=\|q-o\|^2+V.
\]
Consequently some point is at distance at least
$\sqrt{\|q-o\|^2+V}$ from $q$. The extra term $V$ is the improvement
we need. The difficulty is to preserve it when the adversary deletes
points: one deletion can invalidate a particular convex combination.

We obtain many witnesses to the same inequality by taking a
centerpoint in one higher dimension. At the construction of a local
structure, let $P_0$ be its fixed set of $m$ points, called its
\emph{birth set}. Replace each $p$ by its paraboloid lift
$p^+=(p,\|p\|^2)$. Choose a point $z=(o,s)\in\mathbb R^{d+1}$
of halfspace depth at least $\alpha m$, where $s$ is a real number and
$\alpha^{-1}$ is polynomial in the dimension and the logarithmic
accuracy parameters.\footnote{The halfspace depth of $z$ is the minimum
number of input points in a closed halfspace containing $z$. Thus an
$\alpha$-centerpoint lies in no closed halfspace containing fewer than
an $\alpha$ fraction of the input. We use the polynomial-time
approximate centerpoint constructions discussed in
Section~\ref{sec:paper-geometry}; see also~\cite{harpeledjones2019}.
Computing this point is a geometric optimization problem; it is not
in general the centroid.}
Write $V=s-\|o\|^2$. Since positive depth puts $z$ in the convex hull
of the lifted points, $V$ is the variance of every convex combination
representing $z$, and in particular $V\ge0$.

The benefit of the lift is that squared-distance inequalities become
linear inequalities in $(p,\|p\|^2)$. Two halfspaces whose boundaries
pass through $z$ therefore give
\[
 \begin{aligned}
 |\{p\in P_0:\|p-o\|^2\le V\}|&\ge\alpha m,\\
 |\{p\in P_0:\|p-q\|^2\ge\|q-o\|^2+V\}|&\ge\alpha m
       &&\text{for every }q\in\mathbb R^d.
 \end{aligned}
\]
The first set is the part of the data inside the \emph{core ball}
$B(o,\sqrt V)$. The second consists of points that witness the improved
distance lower bound for $q$. The witnesses need not belong to the
core. After $a$ birth points are deleted, both counts decrease by at
most $a$, simultaneously for every $q$. This deterministic statement
is what allows the query and the deletions to depend on earlier answers
and random choices.

\begin{figure}[tbp]
\centering
\begin{tikzpicture}[>=Latex,font=\small]
 \begin{scope}[x=.92cm,y=.92cm]
  \fill[ForestGreen!7] (0,0) circle (1);
  \draw[ForestGreen,thick] (0,0) circle (1);
  \draw[->,gray] (-2.45,0)--(2.5,0);
  \draw[->,gray] (0,-2.35)--(0,2.5);
  \foreach \p in {(0,0),(-1,0),(1,0),(0,-1),(0,1)}
    \fill[ForestGreen] \p circle (2.2pt);
  \foreach \p in {(-2,0),(2,0),(0,-2),(0,2)}
    \fill[BrickRed] \p circle (2.2pt);
  \node[below left] at (0,0) {$o$};
  \node[anchor=west,fill=white,inner sep=1pt] at (.75,.8) {$\sqrt V=1$};
  \node[align=center,text width=5.4cm] at (0,-2.95)
    {(a) Core points lie in the ball.\\The residual consists of the four outer points.};
 \end{scope}
 \begin{scope}[xshift=7.3cm,x=.92cm,y=.68cm]
  \draw[->,gray] (-2.45,0)--(2.5,0) node[right] {$p$};
  \draw[->,gray] (0,-.4)--(0,5.0) node[above] {$h$};
  \draw[gray,domain=-2.15:2.15,samples=50] plot (\x,{\x*\x});
  \draw[dashed,ForestGreen,thick] (-2.4,1)--(2.4,1);
  \draw[BrickRed,thick] (-.45,-.35)--(1.3,4.9);
  \foreach \x in {-2,-1,0,1,2}
    \fill[MidnightBlue] (\x,{\x*\x}) circle (2.2pt);
  \fill (0,1) circle (2.5pt);
  \node[above left] at (0,1) {$z=(0,1)$};
  \node[anchor=west,fill=white,inner sep=1pt] at (.8,3.1) {$h=3p+1$};
  \node[align=center,text width=5.9cm] at (0,-1.95)
    {(b) A planar picture of the lift.\\Each halfspace through $z$ contains at least two points.};
 \end{scope}
\end{tikzpicture}
\caption{Two examples of the certificate. In (a), the nine points
$0,\pm e_1,\pm e_2,\pm2e_1,\pm2e_2$ admit $o=0,V=1$; the core
and residual are shown in the original plane. In (b), the one-dimensional
points $-2,-1,0,1,2$ lift to $(p,p^2)$. The point $z=(0,1)$ lies in
the convex hulls of two disjoint blocks, $\{-1,1\}$ and
$\{-2,0,2\}$, so its depth is at least two. The halfspace $h\le1$
selects the core. For $q=3/2$, the halfspace $h\ge3p+1$ selects exactly
the two leftmost distance witnesses.}
\label{fig:paper-lift}
\end{figure}

\paragraph{A chain of cores.}
A \emph{node} is one local structure with a fixed center $o$, variance
$V$, and a changing input set $S$. Its current core is
$C=S\cap B(o,\sqrt V)$; its child receives the \emph{residual}
$S\setminus C$. Repeating this construction gives a chain, ending when
the residual is small enough to inspect all pairs. Each node is replaced
after at most $\alpha m/16$ changes to its input. During this lifetime,
its core contains a constant fraction of $\alpha m$ surviving birth
points. Hence each successive input loses an $\Omega(\alpha)$ fraction,
and the chain has $O(\alpha^{-1}\log n)$ nodes.

At a node, a balanced search tree maintains the point $y\in S$
maximizing the radius $R=\|y-o\|$. Apply the witness inequality with
$q=y$, and store an actual pair $\{y,p\}$ of length
$L\ge\sqrt{R^2+V}$. Fresh uniform samples from $S$ find such a point
$p$ with high probability: many birth witnesses still survive, even
when the adversary chose $y$ after seeing all earlier randomness.
The update uses new random choices after $y$ has been fixed.

To see why these local pairs suffice, consider a pair $x,y$ attaining
the global diameter. Follow the chain until the first core containing
either endpoint, say $x$. Both endpoints are still in that node's
input, so $\|x-o\|\le\sqrt V$ and $\|y-o\|\le R$. Therefore
\[
 \|x-y\|\le\sqrt V+R\le\sqrt2\sqrt{V+R^2}\le\sqrt2 L.
\]
If both endpoints reach the terminal set, its exact diameter suffices.
Thus the longest stored local pair and the exact terminal answer give
factor $\sqrt2$. The approximation is applied only at the first core
containing an endpoint; it is not multiplied along the chain.

\begin{figure}[tbp]
\centering
\begin{tikzpicture}[>=Latex,font=\small,
 every node/.style={align=center}]
 \node[draw,rounded corners,text width=2.6cm,minimum height=1.15cm] (a)
   at (0,0) {$x,y\in S_1$\\core $C_1$};
 \node[draw,rounded corners,text width=2.6cm,minimum height=1.15cm] (b)
   at (3.7,0) {$S_2=S_1\setminus C_1$\\core $C_2$};
 \node[draw,rounded corners,fill=ForestGreen!7,text width=2.6cm,
 minimum height=1.15cm] (c)
   at (7.4,0) {$x,y\in S_i$\\$x\in C_i$};
 \node[draw,rounded corners,text width=2.6cm,minimum height=1.15cm] (d)
   at (10.8,0) {terminal set\\exact diameter};
 \draw[->,thick] (a)--(b);
 \draw[->,thick,dashed] (b)--(c);
 \draw[->,thick,dashed] (c)--(d);
 \node[text width=5.7cm,below=4mm of c] {At this first core, compare the
 pair $x,y$ with the local stored pair once.};
\end{tikzpicture}
\caption{The cores partition the active set together with the terminal
set. A diametral pair is analyzed at the first core containing one of
its endpoints. The other endpoint is still available in that node's
input.}
\label{fig:paper-core-chain}
\end{figure}
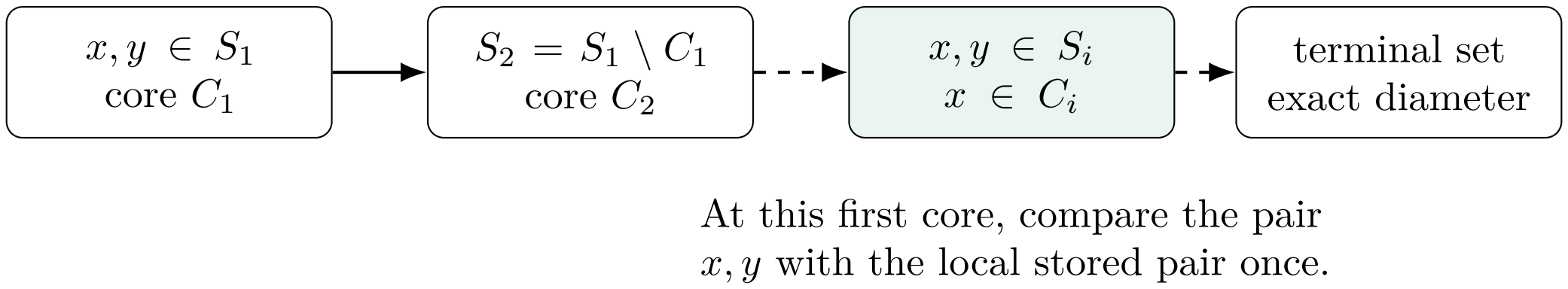

\paragraph{Making the witnesses always correct.}
For the zero-error structure, the center comes with many disjoint
small blocks of birth points. For each block we store nonnegative
coefficients that sum to one and express $z$ as the weighted average
of its lifted points. These equalities can be checked exactly. An
\emph{intact block} is one from which no point has been deleted; its
coefficients prove the variance identity for every query. Scanning
one intact block therefore finds a valid witness without a probability
of error. A deletion invalidates at most one block because the blocks
are disjoint. We accept a randomly constructed center only after
checking the block certificates and the core size, retrying otherwise.
This changes randomness from a possible source of incorrect answers
into a source of variation in construction time.

\subsection{Improving the approximation by searching directions}

The $\sqrt2$ construction already answers many of the searches needed
for a better factor. Suppose a core point $p$ is at distance at least
$\tau$ from a query $q$, but the stored baseline gives no distance as
large as $\tau/c$. Put $r=\|p-o\|$ and $Q=\|q-o\|$, and group points
with similar radii into shells. For the moment, consider an exact
shell of radius $r>0$ and assume $Q>0$; zero radii are handled by the
baseline. Let $u=(p-o)/r$ and $z=-(q-o)/Q$. Negating the query direction
turns a large distance into a large inner product:
\[
             \|p-q\|^2=r^2+Q^2+2rQ\langle u,z\rangle.
\]
A point at distance at least $\tau$ has correlation at least $a$,
whereas a point at distance at most $\tau/c$ has correlation at most
$b$, for
\[
 a=\frac{\tau^2-r^2-Q^2}{2rQ},\qquad
 b=\frac{\tau^2/c^2-r^2-Q^2}{2rQ}.
\]
Here the core condition gives $r^2\le V$, and failure of the baseline
gives $\tau^2/c^2>V+Q^2$. Thus $b\ge0$. The same facts imply
$a\ge c^2b+(c^2-1)$. This is the crucial separation: it is stronger
than an arbitrary pair of angular thresholds.

Append one coordinate with opposite signs to the two unit vectors.
This changes their correlation $w$ to $(w-b)/(1+b)$. The points we
must find now have correlation at least $g=c^2-1$, while all
unacceptable answers have correlation at most zero. Every shell pair
therefore reduces to the same search problem, independent of $r$ and
$Q$. Section~\ref{sec:paper-geometry} implements this idealized
calculation with inward radial rounding, rational coordinates, and
small margins at the thresholds.

For a concrete illustration, take $r=Q=\sqrt V=1$,
$c=\sqrt{3/2}$, and $\tau=2$. Then $a=1$, $b=1/3$, and the
baseline guarantee $\sqrt2$ is smaller than the required distance
$\tau/c=\sqrt{8/3}$. Padding maps the target correlation to
$(1-1/3)/(1+1/3)=1/2=c^2-1$, while every unacceptable correlation
becomes nonpositive. Thus the padded target lies at angle $60^\circ$
from the padded query, whereas every unacceptable point lies at
angle at least $90^\circ$.

A filter is a subset of directions selected by a random Gaussian
projection. A point and a query with sufficiently large correlation
are likely to belong to a common filter. We use different projection
thresholds for stored points and queries, trading the number of copies
of each point against query work. With threshold ratio $x$, the data
replication and query exponents are
\[
 u=\frac{(x-g)^2}{1-g^2},\qquad
 v=\frac{(1-gx)^2}{1-g^2}.
\]
Thus a birth set of size $m$ uses about $m^{1+u}$ space and
$m^v$ work per search, apart from polynomial and logarithmic factors.
Choosing $x=g+\sqrt{\sigma}\sqrt{1-g^2}$ sets $u=\sigma$ and
$v=Q_c(\sigma)$. The symmetric choice $x=1$ gives
$u=v=(1-g)/(1+g)=2/c^2-1$, explaining the balanced exponent.
The formal proof chooses margins of order $1/\log n$; their effect on
an exponent is $O_c(1/\log n)$ and therefore changes the running time
by only a constant factor. This is why the explicit construction has
the stated powers of $n$ without an additional $n^{o(1)}$ factor.

\subsection{Maintaining a pair, rather than answering one search}

A farthest-neighbor search starts with a specified query point.
Maintaining the diameter is different: after the reported pair loses
an endpoint, neither endpoint of its replacement is known. At each
node and distance scale, we use a bipartite graph whose left vertices
are core points and whose right vertices are eligible input points.
A \emph{strong edge} is a pair that must be detected; a \emph{weak edge}
is a pair whose distance is already good enough to report at that
scale. Every strong edge is weak. The angular search finds an actual
weak neighbor whenever a strong neighbor exists.

Maintain a matching of weak edges, with unmatched sets $U_L$ and
$U_R$, subject to the invariant
\[
 \boxed{\text{No strong edge has one endpoint in $U_L$ and one in $U_R$.}}
\]
This does not require a maximum matching. It requires only one search
when a vertex first becomes unmatched. If a strong edge exists, the
matching cannot be empty, so any matching edge supplies a candidate
pair.

\begin{algorithm}[tbp]
\caption{The local matching rule (overview)}
\label{alg:paper-matching-overview}
\begin{algorithmic}[1]
\State When a vertex $v$ becomes unmatched, search the opposite unmatched set.
\If{a weak neighbor $w$ is found}
 \State Match $v$ with $w$ and remove $w$ from its unmatched set.
\Else
 \State Keep $v$ in its unmatched set.
\EndIf
\State On deletion, remove the incident matching edge, if any, and apply the rule to its surviving endpoint.
\State Report any matching edge, if one exists.
\end{algorithmic}
\end{algorithm}

A point has at most one copy on each side. Deleting it frees at most
two surviving partners, so one point update causes only constantly
many searches per configuration. The detailed algorithm removes both
copies before repairing either side, and verifies every candidate's
original distance.

The distance scales are measured relative to $\sqrt V$, not to the
largest input coordinate. If the maximum radius $R$ is much larger
than $\sqrt V$, the baseline already gives factor $c$. Otherwise,
all relevant radii lie in a bounded interval around $\sqrt V$ and
can be rounded inward. A rounded pair is justified either by its
original distance or by the current baseline. Consequently deleting
a faraway point does not force every shell to be rescaled, and the
bounds do not depend on the spread of the coordinates.

\paragraph{Why exposed randomness does not invalidate the searches.}
An adversary that sees a filter family's random choices could choose
a newly inserted point specifically for that family. We avoid this
dependence by choosing each family's randomness only after fixing its
birth set. An old family subsequently handles deletions from that set.
We divide the chronological sequence of insertion occurrences into
intervals of lengths $1,2,4,\ldots$ and build a family for a completed
interval. Reinserting a deleted point creates a new occurrence. Smaller
completed intervals continue to answer searches while a larger family
is being built. The success event for each family holds for every
query direction and every subset of its birth set, so revealing its
random choices does not weaken that event.

The center chain also needs replacements. Building a new chain from
top to bottom without a schedule could leave a small residual's center
too old by the time the entire chain is ready. Instead, all nodes in
a replacement chain share an activation deadline. A residual of size
$m$ is opened only when $O(\alpha m)$ updates remain before that
deadline, and its construction finishes in $O(\alpha^2m)$ updates.
The input shrinks by an $\Omega(\alpha)$ fraction at each level, so
these opening times leave room for the next node. Each activated node
retains its actual age. A chain under construction does not start
further replacement chains. Section~\ref{sec:paper-dynamization}
proves that these rules finish the work on time with only polynomial
and logarithmic overhead.

\subsection{Using less space by recovering filter lists}

Explicit filters store a separate record whenever a point belongs to
a filter. The improved tradeoff keeps the same geometric search idea,
but represents these point--filter incidences implicitly. Two issues
must be resolved: a family must cover all valid queries, and its lists
must remain efficient to recover after arbitrary deletions.

\paragraph{Coverage that can be checked.}
We divide the coordinates into mutually orthogonal blocks. A pair
with large correlation has large normalized correlation in at least
one block. Within a block, the finite splitting families of
Wei~\cite{wei2019lasvegas} reduce coverage to spheres of much smaller
dimension. On each small sphere, a finite, sufficiently fine set of
directions allows us to check that a sampled family covers every
relevant pair. A positive margin extends the checked property from
the finite set to the whole sphere. We retry until the check succeeds.
Every accepted family is therefore correct. Random rotations are used
to bound the number of irrelevant candidates examined during a search;
a bad rotation can increase work, but it cannot remove the guarantee
that a good candidate is covered.

The resulting family is a product of small families. Its labels can
be counted and selected without writing out the product: two lists
of lengths two and three, for instance, define six pairs, and a rank
identifies one pair by quotient and remainder. Applying the same
mixed-radix calculation to all blocks gives a subpolynomial-time
oracle for the $j$th label of a point. We assign each point to an
accepted family in which it has at most $R$ labels. All its labels
in that family are retained; none is discarded to obtain the bound.

\paragraph{One function represents every incidence.}
For a birth set of $m$ points, define a virtual function on
$\{1,\ldots,m\}\times\{1,\ldots,R\}$: at $(p,j)$ it returns the
$j$th filter label of $p$, or a null value if that occurrence is absent.
The function is evaluated by the rank oracle, not stored as an array.
For a query label, recovering its complete inverse image reveals the
corresponding point occurrences. Making the rank part of the function's
domain lets one inverse query recover the entire list.

Let $M=mR$ and let $\tau$ be the inversion parameter. The inverse data
structure uses roughly $M/\tau$ space and recovers a bounded list in
roughly $\tau^3$ function evaluations. Both its final representation
and the workspace used to construct it must fit the space bound.
A deletion changes every occurrence of that point to null, which
creates the next difficulty: a list that was large at construction
may later have very few surviving members.

We split lists into buckets with certified birth sizes between
$\tau/16$ and $\tau$. A bucket remains implicit while at least
$\tau/32$ members survive. If a query rejects all of these members,
its recovery cost can be charged at $\widetilde O(\tau^2)$ per
current rejected occurrence. When a bucket becomes smaller, we recover
and explicitly store its remaining members. At least $\tau/32$
deletions have then occurred in that bucket, paying for the small
explicit remainder. Rebuilding after $O(M/\tau)$ deletions keeps all
such remainders within the space bound. If a search generates $Q$
labels and encounters $F$ rejected current occurrences, its cost is
\[
             \widetilde O\bigl((Q+1)\tau^3+F\tau^2\bigr).
\]
Occurrences are counted with multiplicity: a point lying in several
queried filters contributes once for each such incidence. The displayed
bound is deterministic and remains valid after adaptive deletions.

Reducing filter density decreases the space for incidences but changes
both the number of query labels and the number of rejected candidates.
Balancing these two terms with the $M/\tau$ space bound gives the
linear branch of $\Lambda_c(\sigma)$, tangent to the explicit-filter
curve. At linear space its exponent is
$\chi_c=2-\sqrt{1+3(c^2-1)^2}$. Section~\ref{sec:paper-parameters}
performs the optimization and verifies that construction, updates,
and rebuilding fit the same bound.

Finally, correct certificates alone do not bound running time against
an adversary that sees every random choice. We race independent
certified copies for a prescribed amount of work and fall back to an
exact scan if all copies are slow. Amplifying over a finite set of
representative directions makes this fallback rare simultaneously for
all queries. A bad construction can still be selected by the adversary,
so the proof charges all work it can cause during its lifetime to its
construction. This gives the expected total-work bound for every fixed
finite prefix of an arbitrarily long adaptive sequence. For the
worst-case Monte Carlo version, constructors and searches are stopped
at fixed budgets; the probability of exhausting a budget is included
in the global failure probability.

\section{Model and notation}
\label{sec:model}

This section specifies the updates, the information available to the
adversary, and the computational costs used in our statements. The
geometric terminology needed for the proofs follows in
Section~\ref{sec:preliminaries}.

\paragraph{Points and updates.}
The active data form a finite indexed family $P$ in $\mathbb R^d$,
with $|P|\le n$ at all times. An insertion supplies $d$ coordinates
and returns a compact handle; a deletion supplies a currently active
handle. Each insertion is a new logical identity, called its
\emph{insertion lifetime}, even if its coordinates equal those of
another insertion. Coordinates are immutable throughout that lifetime.
No future deletion time is supplied. If an application uses additional
external names, their dictionary and input/output costs are charged
separately to their encoding lengths; the stated geometric bounds use
the handle interface.

A handle is reused only after its lifetime is inactive and no live
snapshot, certificate, pending construction, or queued retirement
record refers to it. Consequently historical coordinates cannot be
changed by a later insertion. Logical lifetime labels need not be
stored as an ever-increasing global counter. A free list of the
bounded set of retained records, together with counters local to a
node generation, implements them. Local counters are reset when that
generation is discarded. This distinction is necessary for space
bounds independent of the length of an unbounded stream.

The diameter is $\diam(P)=\max_{x,y\in P}\|x-y\|$, with value zero
when $|P|\le1$. For $|P|\ge2$, the structure returns two distinct
active identities, including when all coordinates coincide. A
factor-$c$ answer has distance between $\diam(P)/c$ and $\diam(P)$.
The algorithm stores the answer's identities and exact squared
distance during updates. The header is accessible in a constant number of RAM-cell accesses;
serializing its contents is charged to their bit length. If $|P|\le1$,
the header stores \textsc{None} and squared distance zero. If
$|P|\ge2$ and the maximum candidate length is zero, two distinct
active handles are supplied from a maintained fallback pair.

\paragraph{Adaptive adversaries and probability of error.}
Before an external operation begins, the adversary specifies that
operation using the complete preceding history, including every
previous random choice and answer. The algorithm then uses fresh
randomness to process it. Internal searches and matching repairs may
also depend on earlier random choices. An oblivious adversary is the
special case in which the operation sequence is fixed in advance.

A Monte Carlo statement is supplied with a capacity $n$, an upper
bound $t$ on the number of updates, and a total error probability
$\delta\in(0,1/2)$. With probability at least $1-\delta$, every answer
in that execution satisfies its approximation guarantee. An
always-correct, or Las Vegas, structure is supplied with no update
horizon: all its answers are valid, and each operation terminates
with probability one. For every fixed integer $T\ge1$, its expected amortized bound $U$
means
\[
 \mathbb E\!\left[\Work_{\rm init}+\sum_{i=1}^{T}\Work_i\right]
 \le TU
\]
for each valid adaptive adversary, starting from the empty set.
The expectation includes all rejected construction attempts and
rebuilds. The integer $T$ is fixed before the random execution;
no assertion about arbitrary random stopping times is needed. It does not
assert a conditional expected-time bound for each operation after
particular random choices have been exposed. Unless stated otherwise,
executions start empty; constructing an initial set by insertion is
charged to those insertions.

\paragraph{Arithmetic and finite representations.}
The arithmetic RAM supports exact field operations
$+,-,\times,/$ (division only by a nonzero value), exact comparisons,
integer word operations, and independent fair random bits at unit
cost. Square roots, exponentials, and logarithms are not primitive
instructions. They describe parameters or geometric quantities;
certified rational approximations and sign-guarded squared
comparisons implement their uses. Empty initialization allocates
only the header and parameter records. Tables and workspaces are
allocated lazily, so no array of length $n$ is eagerly cleared.

For rational inputs, $B\ge1$ bounds the binary length of each reduced
numerator magnitude and each positive denominator; a signed
coordinate thus has at most $2B+1$ bits. Bounds displaying $B$ charge
bit operations and total stored bits, including addresses, handles,
and intermediate numerators and denominators. Where a subsidiary
word-operation count is stated, a word is large enough to address
the live memory, and multiword arithmetic and fair-bit generation
are charged explicitly. In particular, drawing a uniform rank in a
population of size $m$ uses $O(\log(2m))$ expected random bits.
Polynomial logarithmic costs of these operations are included in
the main bounds. Linear-programming costs are included; no strongly
polynomial linear-programming algorithm is assumed. Linear programs
use original lifted rational coordinates, so their bit lengths are
polynomial in $d$ and $B$.

The approximation and space parameters are fixed independently of
$n,d,B$. They are either rational, or can be approximated in
polynomial time by rational intervals of length $2^{-k}$ that are
guaranteed to contain the parameter. This is the meaning of a
certified $k$-bit approximation. We round these parameters in the
direction required by the inequalities in the proof.

\paragraph{Suppressed factors.}
We use $\mathcal P$ for a polynomial in $d$, the coordinate bit length
when relevant, and the logarithmic parameters of the statement. Its
coefficients and degree may depend on the fixed approximation and
space parameters, but not on the input. The polynomial may be enlarged
between inequalities. Any dependence on an additional exponent
allowance is stated explicitly. The notation $\widetilde O$ suppresses
logarithmic factors only; powers such as $n^\eta$ and $n^{o(1)}$ are
shown explicitly, including in construction workspace. Space counts
the active input as well as all auxiliary structures, simultaneous
rebuilds, temporary memory, and records awaiting reclamation.
It is a bound on peak live memory on every outcome, including failed
attempts. Temporary memory is released or reused before a retry.
The geometric constructions below are described for $d\ge1$;
for $d=0$ all distances vanish and the handle dictionary with the
distinct-pair convention suffices, within the stated bounds.

\section{Geometric preliminaries}
\label{sec:preliminaries}

The basic certificate has two purposes: to place many points in a
bounded region and to supply many actual points witnessing a distance
lower bound. We define its objects here so that the statements in the
next section can be concise. Figures~\ref{fig:paper-lift}
and~\ref{fig:paper-core-chain} illustrate the geometry and the way these
certificates are organized.

\paragraph{Halfspace depth and centerpoints.}
Let $Q$ be a nonempty finite indexed point set in $\mathbb R^k$,
where $k\ge1$. A closed halfspace has the form
$\{x:\langle v,x\rangle\le b\}$ with $v\ne0$. The
\emph{halfspace depth}, also called \emph{Tukey depth}, of $z$ in $Q$
is
\[
 \operatorname{depth}_Q(z)
   =\min\{|Q\cap H|:H\text{ is a closed halfspace containing }z\}.
\]
Point identities are counted with multiplicity if coordinates coincide.
A point of depth at least $\alpha|Q|$ is an
\emph{$\alpha$-centerpoint}. This definition asks for a depth guarantee,
not for maximum possible depth. We use approximate centerpoint
constructions whose running time is polynomial in the dimension;
the precise guarantees and finite representations are given in
Appendix~\ref{sec:paper-center-constructions}.

For a concrete planar example, the five lifted points in
Figure~\ref{fig:paper-lift}(b) have two disjoint subsets whose convex
hulls contain $z$. Every closed halfspace containing $z$ must meet
each subset, so the depth is at least two. More generally, such
representations provide a certificate of depth. Positive depth also
implies that $z$ lies in the convex hull of $Q$: otherwise a separating
halfspace containing $z$ would contain no point of $Q$.

\paragraph{Birth sets, lifting, and variance.}
When a local structure is constructed, its initial fixed set $P_0$
is called its \emph{birth set} or \emph{birth pool}; write $m=|P_0|$.
The word ``birth'' distinguishes this fixed set from the points that
remain after later updates. The \emph{paraboloid lift} maps a point
$p\in\mathbb R^d$ to $p^+=(p,\|p\|^2)\in\mathbb R^{d+1}$; let
$P_0^+=\{p^+:p\in P_0\}$. Write a positive-depth point of this lifted
set as $z=(o,s)$, with $o\in\mathbb R^d$ and $s\in\mathbb R$, and set
$V=s-\|o\|^2$. If $\lambda_p\ge0$ sum to one and
$z=\sum_p\lambda_pp^+$, then
\[
 \sum_p\lambda_pp=o,\qquad
 V=\sum_p\lambda_p\|p-o\|^2\ge0.
\]
Thus $V$ is the variance associated with the representation. It need
not be the variance of the uniform distribution on $P_0$.

\paragraph{Core and witness halfspaces.}
The core ball is $B(o,\sqrt V)$. A \emph{variance witness} for a query
$q$ is a point $p$ satisfying
$\|p-q\|^2\ge V+\|q-o\|^2$. These two geometric conditions are
linear in the lifted coordinates. For a general lifted-space point
$(x,h)\in\mathbb R^{d+1}$, define
\[
 \begin{aligned}
 H_{\rm core}
   &=\{(x,h):h-2\langle o,x\rangle\le s-2\|o\|^2\},\\
 H_q
   &=\{(x,h):h-2\langle q,x\rangle\ge s-2\langle q,o\rangle\}.
 \end{aligned}
\]
Both boundaries pass through $z$, and both normals are nonzero:
their last coordinates are respectively $1$ and $-1$ when the
inequalities are written in the same orientation. Expanding squared distances shows
that $p^+\in H_{\rm core}$ exactly when $p\in B(o,\sqrt V)$, and
$p^+\in H_q$ exactly when $p$ is a variance witness for $q$. The next
section applies halfspace depth to these two sets, then counts how
many witnesses remain after deletions.

\paragraph{Nodes, cores, and residuals.}
A \emph{node} stores one center $o$, one variance $V$, and auxiliary
structures for a changing input set $S$. Its current \emph{core} is
$C=S\cap B(o,\sqrt V)$, and its \emph{residual} is $S\setminus C$.
The residual becomes the input to the next node in a chain; a small
terminal set is handled exactly. The birth set determines the node's
center, while its current input determines which points and candidate
pairs are still present. Its \emph{age} is the number of changes to
its input since the snapshot used to construct it. Rebuilding replaces
the node before too many of its birth witnesses can disappear.

A \emph{block certificate} for $z$ is a subset of birth identities
with explicitly stored nonnegative convex coefficients expressing
$z$ in that block's lifted convex hull. It is \emph{intact} when none
of those identities has been deleted. Disjoint block certificates
are useful because deleting one identity invalidates at most one
block. Their role in obtaining always-correct distance witnesses is
proved in the next section.

\section{Variance certificates and local witness matchings}
\label{sec:paper-geometry}

This section proves the geometric guarantee independently of the
search implementation. At each node of a chain, we remove a large
core from the node's input and maintain a pair that approximates
every distance from that core to the input. Every pair of points is
assigned to the first core containing one of its endpoints, so these
local guarantees imply the global diameter guarantee without
multiplying approximation factors along the chain.

We proceed in four steps. First, the lifted centerpoint supplies a
distance witness that remains available after deletions; this alone
gives factor $\sqrt2$. Second, for smaller $c$ we restrict attention
to a bounded radial range and round the radii. Third, a matching
maintains one witness at each relevant distance scale. Finally, we
reduce every resulting search to the same angular problem. The
definitions of birth pools, cores, and witnesses are given in
Section~\ref{sec:preliminaries}; Figures~\ref{fig:paper-lift}
and~\ref{fig:paper-core-chain} illustrate their roles.

\subsection{A centerpoint that also certifies variance}

Fix a nonempty node birth pool $P_0$ of $m\ge1$ point identities
and a depth parameter $0<\alpha\le1$. Its coordinates
are fixed before the node is constructed. Lift each point to
$p^+=(p,\|p\|^2)\in\mathbb R^{d+1}$, and let $z=(o,s)$ have
halfspace depth at least $\alpha m$ in $P_0^+$. Here
$o\in\mathbb R^d$ and $s\in\mathbb R$. Since positive depth places
$z$ in the convex hull of the lifted points, some convex coefficients
$\lambda_p$ satisfy $o=\sum_p\lambda_pp$ and
$s=\sum_p\lambda_p\|p\|^2$. Consequently
$V=s-\|o\|^2=\sum_p\lambda_p\|p-o\|^2\ge0$.
This explains the term \emph{variance certificate}: the point $z$
records both a mean and the second moment of a distribution on the
birth points.

The usefulness of depth is that we need not preserve this particular
convex combination. The next lemma gives many witnesses for every
query at once, so deleting a few points cannot destroy the guarantee.

\begin{lemma}[Core points and distance witnesses]
\label{lem:paper-lifted-certificate}
At least $\alpha m$ birth points lie in $B(o,\sqrt V)$.
For every $q\in\mathbb R^d$, at least $\alpha m$ birth points satisfy
\begin{equation}\label{eq:paper-variance-witness}
                    \|p-q\|^2\ge V+\|q-o\|^2.
\end{equation}
After $a$ deletions, each count is at least $\alpha m-a$,
simultaneously for all $q$. Insertions preserve these lower bounds.
\end{lemma}
\begin{proof}
Write $(x,r)$ for a point of the lifted space. Consider the closed
halfspace
$H_{\rm core}=\{(x,r):r-2\langle o,x\rangle\le s-2\|o\|^2\}$.
Its boundary contains $z=(o,s)$. For a lifted input point, membership
is equivalent to
$\|p-o\|^2\le s-\|o\|^2=V$. Depth therefore gives the core count.

For a fixed query $q$, use the other closed halfspace
$H_q=\{(x,r):r-2\langle q,x\rangle\ge s-2\langle q,o\rangle\}$.
Its boundary also contains $z$. Adding $\|q\|^2$ to the inequality
for $p^+$ gives
$\|p-q\|^2\ge s-2\langle q,o\rangle+\|q\|^2
=V+\|q-o\|^2$. Depth gives at least $\alpha m$ such points.
This holds for every halfspace $H_q$, not just for a fixed list of
queries. A deletion removes at most one point from each counted
set; an insertion removes none. The simultaneous survival statement
follows deterministically.
\end{proof}

\paragraph{Computing the certificate.}
Finding a point of high halfspace depth is a separate algorithmic
problem; taking the ordinary centroid does not give the required
depth. In the arithmetic model, the Har-Peled--Jones
constructor~\cite[Theorem~18]{harpeledjones2019}, applied in
dimension $d+1$, gives $\alpha=1/[4(d+3)^2]$ with failure
probability $\varphi$ in
$O((d+1)^7\log^3(d+2)\log^3(2/\varphi))$ arithmetic operations,
given uniform sampling access to the birth pool.
For rational coordinates, Appendix~\ref{sec:paper-center-constructions}
gives an exact sampled linear program with a depth parameter satisfying
$\alpha^{-1}=\poly(d,\log(1/\varphi))$, using
$\poly(d,\log(1/\varphi))$ uniform-sampling calls and
$\poly(d,B,\log(1/\varphi))$ additional bit operations.
Implementing sampling access contributes the logarithmic population
and rejection-sampling costs specified in Section~\ref{sec:model}.
This avoids an assumption of strongly polynomial linear programming.

For zero error, the same appendix constructs at least $\beta m$
pairwise disjoint blocks, each containing at most
$s_0=\poly(d)$ birth points, with $\beta^{-1}=\poly(d)$.
It supplies exact convex coefficients expressing the same $z$ in
the lifted convex hull of every block. All coefficients and the
core size are checked before a construction is accepted. The expected construction time is
$m\poly(d,B,\log(2m))$ in the bit model, including random-bit
generation and handle manipulation.

\paragraph{The current node and its residual.}
During a node's lifetime, let $S$ denote its current input. Its core
is $C=S\cap B(o,\sqrt V)$, and its child receives exactly
$S\setminus C$. The center and $V$ stay fixed until replacement.
We permit at most $\alpha m/16$ changes to the node's input.
Lemma~\ref{lem:paper-lifted-certificate} then leaves at least
$15\alpha m/16$ birth points in the current core, while
$|S|\le(1+\alpha/16)m$. In particular,
\[
 |S\setminus C|\le\left(1-\frac{15\alpha}{17}\right)|S|,
\]
because $|S|\le17m/16$. Thus the child is smaller than the input
by a factor $1-\Omega(\alpha)$. Repeating this operation gives
$O(\alpha^{-1}\log(n+1))$ nodes and ends in a leaf of polynomial
size, where all distances can be checked exactly. In the certified
version the lifetime is set by $\beta$ instead. The replacement
schedule in Section~\ref{sec:paper-dynamization} enforces these
lifetime bounds, including while new nodes are being constructed.

\paragraph{Maintaining the baseline pair.}
A balanced search tree stores the current squared radii
$\|p-o\|^2$ for $p\in S$. If $S=\varnothing$, the node has no
baseline pair and contributes length zero. Otherwise let $y\in S$
have maximum radius $R_S=\max_{p\in S}\|p-o\|$. We maintain a
current pair whose length $L$ satisfies
\begin{equation}\label{eq:local-variance-baseline}
                            L^2\ge V+R_S^2.
\end{equation}
A repeated identity is permitted in this internal baseline only
when its distance is zero; the final output uses the distinct-handle
convention. If $S$ is a singleton on a valid node, the witness
inequality forces $V=R_S=0$.

In the Monte Carlo construction, draw
$\lceil2\alpha^{-1}\log(1/\psi)\rceil$ fresh uniform samples from $S$ and
take the farthest sampled point from $y$. Conditional on the entire
history and a valid center, the witness fraction is at least
$(15\alpha m/16)/(17m/16)=15\alpha/17\ge\alpha/2$.
The probability of missing all witnesses is therefore at most
$\exp(-\alpha k/2)\le\psi$ for the stated sample count $k$. The point $y$ may
depend on every earlier random choice: the sampling is performed
after $y$ is fixed. Subtree sizes in the membership tree support
uniform rank sampling.

In the zero-error construction, choose any certificate block whose
points all remain active. Its exact coefficients give, for every $q$,
\begin{equation}\label{eq:paper-certified-average}
 \sum_p\lambda_p\|p-q\|^2=\|q-o\|^2+V.
\end{equation}
Scanning that block finds a point whose squared distance is at least
this weighted average. A deletion invalidates at most one block because the blocks
are disjoint, so an intact block remains throughout the permitted
lifetime. A linked list of intact blocks and one block pointer per
birth identity suffice for the bookkeeping.

The next lemma completes the geometric part of the endpoint result.
Its proof assigns a diametral pair to one node, where the triangle
inequality and~\eqref{eq:local-variance-baseline} give the entire
approximation factor.

\begin{lemma}[A $\sqrt2$-approximate diameter pair]
\label{lem:paper-sqrt-two-cache}
If every nonempty current node satisfies~\eqref{eq:local-variance-baseline},
with the empty-node convention above,
the longest of its baseline pairs and the exact leaf pair has length
in $[\diam(P)/\sqrt2,\diam(P)]$.
\end{lemma}
\begin{proof}
For $|P|\le1$, use the zero-diameter convention. Otherwise choose
distinct identities $x,y\in P$ with $\|x-y\|=\diam(P)$; this is
possible also when all coordinates coincide. If both reach the leaf,
the leaf's exact pair has this length. Otherwise let node $i$ be
the first node whose core contains either endpoint, say $x$.
Both endpoints still belong to its input $S_i$: neither was removed
by an earlier core. Hence $\|x-o_i\|\le\sqrt{V_i}$ and
$\|y-o_i\|\le R_{S_i}$. Therefore
\[
 \diam(P)\le\sqrt{V_i}+R_{S_i}
 \le\sqrt2\sqrt{V_i+R_{S_i}^2}\le\sqrt2 L_i.
\]
The middle inequality follows from $(a+b)^2\le2(a^2+b^2)$.
Every candidate is a pair of active points, so its length is at most
$\diam(P)$. Before exposing the answer after an update, refresh
every affected node; in particular no deleted endpoint remains in
a stored pair.
\end{proof}

We keep two distinct active identifiers as a fallback when the
maximum stored distance is zero. With fewer than two active points
we use the zero-diameter convention. Thus coincident coordinates
do not force an output to repeat one identifier.

\subsection{Restricting the radii and distance scales}

To obtain a factor $c<\sqrt2$, we must improve the estimate of
$D_C=\max_{x\in C,y\in S}\|x-y\|$ at each node. Set $D_C=0$
if $C$ is empty. The baseline already handles points very far from
the center. This permits a fixed local radial grid, independent
of the coordinate spread and of later deletions.

Fix $1<c<\sqrt2$ and choose rational $e>0$, $\gamma>1$, and
$\lambda>1$ such that
\[
\kappa=\gamma+2e<c,\qquad \kappa\lambda<c,
 \qquad K\in\mathbb Z,\quad K\ge2/(c-1).
\]
The parameter $e$ controls radial rounding, $\gamma$ separates the
strong and weak distance thresholds, and $\lambda$ is the ratio
between consecutive distance scales. Section~\ref{sec:paper-parameters}
gives concrete choices.

\begin{lemma}[Distances outside the local range]
\label{lem:bounded-radial-range}
If $V=0$ or $R_S>K\sqrt V$, then $D_C<cL$ unless $D_C=L=0$.
\end{lemma}
\begin{proof}
For $V=0$, every core point equals $o$, so $D_C\le R_S\le L$.
For $R_S>K\sqrt V$, the triangle inequality gives
\[
D_C\le\sqrt V+R_S<(1+1/K)R_S<cL,
\]
because $1+1/K<c$ and $R_S\le L$.
\end{proof}

Accordingly, the local search structures include on their right
side only points of radius at most $K\sqrt V$. If an outlier is
present, the baseline suffices. When the last outlier is deleted,
every remaining point is already in the local structures.

Assume now $V>0$, and put $m_0=\lceil1/e\rceil$. The fixed shell
radii are $r_j=(j/m_0)\sqrt V$ for $0\le j\le Km_0$. Round each
eligible point inward to the largest radius no greater than its
own, and write the rounded point as $\widetilde p$.
The displacement is at most $e\sqrt V$. We store only the original
point and its shell number: comparisons of
$m_0^2\|p-o\|^2$ with $j^2V$ determine the shell exactly.

Use distance scales $R_\ell=\lambda^\ell\sqrt V/2$ for
$0\le\ell\le J$, where $J$ is the first index with
$\lambda^J/2\ge K+1$. A \emph{configuration} is a triple
$(j,k,\ell)$ of a positive core shell, a positive input shell,
and a distance scale. Retain it only if
\begin{equation}\label{eq:local-eligible-configuration}
                  r_j^2+r_k^2\le R_\ell^2,
                  \qquad \gamma R_\ell\le r_j+r_k.
\end{equation}
Its left vertices are copies of the core points in shell $j$; its
right vertices are copies of the eligible input points in shell $k$.
A \emph{weak edge} has rounded distance greater than $R_\ell$.
A \emph{strong edge} has rounded distance greater than
$\gamma R_\ell$ and is therefore also weak.

For nonzero directions $u_p=(p-o)/\|p-o\|$, the identity
$\|\widetilde x-\widetilde y\|^2
=r_j^2+r_k^2+2r_jr_k\langle-u_x,u_y\rangle$
turns distance into correlation. Define
\[
a=\frac{\gamma^2R_\ell^2-r_j^2-r_k^2}{2r_jr_k},\qquad
 b=\frac{R_\ell^2-r_j^2-r_k^2}{2r_jr_k}.
\]
Then strong edges have correlation greater than $a$, and the exact
weak-edge test is $\langle-u_x,u_y\rangle>b$. All factors of $V$
cancel, so $a,b$ are rational numbers fixed when the configuration
is created. With $g=\gamma^2-1$, eligibility gives
\begin{equation}\label{eq:paper-angular-premise}
 0\le b<a\le1,\qquad
 a-(1+g)b=g\frac{r_j^2+r_k^2}{2r_jr_k}\ge g.
\end{equation}

\subsection{Maintaining one witness at each scale}

An angular query returns a verified weak neighbor whenever a strong
neighbor exists. It may also return a weak neighbor in the absence
of a strong one. Such a search operation does not, by itself,
maintain a diameter pair after deletions. A matching supplies the
missing invariant with only a constant number of searches per
membership change.

For a configuration, let $V_L,V_R$ be its active vertex sets and
$E_{\rm weak},E_{\rm strong}$ its two edge sets. Vertices are
side-tagged copies of identities, so the two copies of a point are
distinct vertices. The thresholds are the actual $\gamma$ and
$R_\ell$ of that configuration. Maintain a matching
$M$, and let $U_L,U_R$ be the unmatched vertices on the two sides.
The invariant is
\[
\boxed{\quad M\subseteq E_{\rm weak}\text{ is a matching},\qquad
 E_{\rm strong}\cap(U_L\times U_R)=\varnothing.\quad}
\]
Each unmatched set has an angular index. The birth-pool guarantees
of the next section hold for all later subsets and queries, and
Section~\ref{sec:paper-dynamization} permits fully dynamic membership
changes. Thus these searches remain valid even when matching repairs
depend on exposed earlier randomness.

\begin{algorithm}[tbp]
\caption{Maintaining one local matching}
\label{alg:local-matching-operations}
\begin{algorithmic}[1]
\Statex \textbf{Attach$(v)$}: $v$ is active, unmatched, and absent from its unmatched index.
\State Search the opposite unmatched index in direction $-u_v$, accepting exactly correlations $>b$.
\If{a returned $w$ is currently unmatched on the opposite side, has a different identity, and passes the exact weak-edge test}
 \State Remove $w$ from its unmatched index; insert edge $vw$ in $M$ and record its exact squared distance.
\Else
 \State Insert $v$ into its unmatched index with a fresh local search-set insertion lifetime.
\EndIf
\Statex \textbf{InsertOriginal$(p)$}:
\State Create its applicable left and right copies, in that order, and call \textsc{Attach} on each.
\Statex \textbf{DeleteOriginal$(p)$}:
\State Remove both copies of $p$ from their unmatched indices or matching edges.
\State Collect the surviving former partners in $F$; call \textsc{Attach} on each still-unmatched member of $F$.
\Statex \textbf{StoredEdge$()$}: return the first edge in the matching list, or empty.
\end{algorithmic}
\end{algorithm}

The weak-edge and matching conditions hold on every execution
because positive answers are validated. The no-unmatched-strong-edge
condition is asserted on the uniform success event of all invoked
search indices, with the configuration's actual strong threshold.
To verify this latter invariant, consider \textsc{Attach}. If it finds a weak
edge, removing the two endpoints from the unmatched sets cannot create
a strong edge between those sets. If it finds none, the query promise
ensures that the newly unmatched vertex has no strong neighbor in
the opposite unmatched set. Deleting an unmatched vertex preserves
the invariant. Deleting a matched vertex makes only its former
partner newly unmatched, which \textsc{Attach} handles. An original
point has at most two copies, so $|F|\le2$. We remove both before
repairing either partner. A self-edge between two copies of one
identity has rounded distance zero and cannot pass the weak test.

An empty matching certifies that the configuration has no strong
edge. A nonempty matching supplies an actual current pair from any
one of its weak edges. For the node, let $W$ be the maximum of the
baseline length $L$ and the original lengths of one stored edge
from each nonempty matching. The next lemma explains why inward
rounding does not invalidate an old edge when the baseline decreases.

\begin{lemma}[Radial rounding preserves a distance witness]
\label{lem:two-sided-radial-domination}
For any current identities $x,y\in S$ to which inward radial
rounding is applied,
\[
 \|\widetilde x-\widetilde y\|
 \le\max\{\|x-y\|,\|x-o\|,\|y-o\|\}
 \le\max\{\|x-y\|,L\}.
\]
In particular, a current weak edge at scale $R_\ell$ implies
$W>R_\ell$.
\end{lemma}
\begin{proof}
Inward rounding means
$\widetilde x=\xi x+(1-\xi)o$ and
$\widetilde y=\upsilon y+(1-\upsilon)o$ for
$\xi,\upsilon\in[0,1]$. Subtracting gives
\[
 \widetilde x-\widetilde y
 =\xi\upsilon(x-y)+\xi(1-\upsilon)(x-o)
  +(1-\xi)\upsilon(o-y).
\]
The three coefficients are nonnegative and sum to at most one.
The triangle inequality proves the first bound. Both radii are at
most the current $R_S\le L$, proving the second. A weak edge has
rounded length greater than $R_\ell$, so either its original
length or the current baseline is greater than $R_\ell$.
\end{proof}

We can now show that some maintained scale detects every core--input
distance that is too long for the baseline. The gaps between $c$,
$\kappa$, and $\gamma$ pay, respectively, for scale rounding and
radial rounding.

\begin{lemma}[The local diameter guarantee]
\label{lem:local-matching-diameter}
If the baseline and angular searches satisfy their promises, then
$D_C/c\le W\le\diam(S)$.
\end{lemma}
\begin{proof}
If $V=0$, an outlier is present, or $D_C\le cL$, the baseline
suffices by Lemma~\ref{lem:bounded-radial-range}. Otherwise
$V>0$, $R_S\le K\sqrt V$, and $D_C>cL$.

Choose the grid scale immediately below $D_C/\kappa$, taking the
predecessor when this value is itself on the grid. It satisfies
\[
 R_\ell<D_C/\kappa\le\lambda R_\ell,
 \qquad R_\ell\ge D_C/(\kappa\lambda)>D_C/c>L.
\]
Such a scale exists because the grid starts at $\sqrt V/2$, while
$D_C>c\sqrt V$, and ends at or above $(K+1)\sqrt V\ge D_C$.

Let $x\in C,y\in S$ realize $D_C$. Rounding both radii loses at
most $2e\sqrt V<2eR_\ell$. The rounded distance is therefore
greater than $(\kappa-2e)R_\ell=\gamma R_\ell$.
Neither rounded radius is zero, since then the rounded distance
would be at most $R_S\le L<R_\ell$. If their shells are $j,k$,
then $r_j^2+r_k^2\le V+R_S^2\le L^2<R_\ell^2$; their rounded
distance is at most $r_j+r_k$. These two facts establish eligibility
in~\eqref{eq:local-eligible-configuration}.

The configuration contains the strong edge $xy$, so its matching
is nonempty. Lemma~\ref{lem:two-sided-radial-domination} yields
$W>R_\ell>D_C/c$. Every stored candidate is an actual current
pair, proving the upper bound.
\end{proof}

The first-core argument from Lemma~\ref{lem:paper-sqrt-two-cache}
now proves the global factor-$c$ guarantee. Each point update causes
only constantly many angular operations per applicable configuration.
When a pending node scans a point, it completes both matching copies,
the point's child membership, and its processed flag within one
external update. Section~\ref{sec:paper-dynamization} specifies how
these scans include insertions and deletions arriving during construction.

\subsection{One angular gap for all configurations}

Recall the inequality
$a-(1+g)b\ge g$ in~\eqref{eq:paper-angular-premise}, where
$g=\gamma^2-1$. It permits the same search exponent for every
shell pair: data directions $u$ and query directions $z$ can be
padded so that correlations above $a$ become at least $g$, whereas
correlations at most $b$ become nonpositive. Formally, padding by
$(u,\sqrt b)/\sqrt{1+b}$ and $(z,-\sqrt b)/\sqrt{1+b}$ changes
correlation $r$ into $(r-b)/(1+b)$. We use rational padding with a
small margin to accommodate rounding of queries.

\begin{lemma}[A common angular gap]
\label{lem:common-gap-lift}
Assume~\eqref{eq:paper-angular-premise} and let $0<e\le g/16$
be rational. One can compute rational $A>0,B\ge0$ with
$A^2+B^2=1$ such that, for unit vectors $u,z,\widehat z$ with
$\|\widehat z-z\|\le e/8$, the padded correlation
$\langle(A\widehat z,-B),(Au,B)\rangle$ is at least
$h=g-11e$ when $\langle z,u\rangle\ge a$, and is negative
when $\langle z,u\rangle\le b$.
More quantitatively, a rejected original pair has padded correlation
at most $-e$. The computation is polynomial in the bit length of $b$
and $\log(1/e)$.
\end{lemma}
\begin{proof}
Use the rational parametrization
$A=(1-v^2)/(1+v^2)$, $B=2v/(1+v^2)$.
Choose $v\in[0,1/2]\cap\mathbb Q$ so that
$B^2/A^2\in[b+4e,b+5e]$. The function
$4v^2/(1-v^2)^2$ is increasing on this interval, has derivative
at most $12$, and exceeds $1$ at $v=1/2$. Moreover $b\le1-g$
and $5e\le5g/16$, so the target interval lies below $1$.
Binary search on a rational mesh of spacing at most $e/16$
therefore finds a suitable $v$.

Write $\Delta=B^2/A^2-b\in[4e,5e]$. Before rounding $z$, a
correlation $r$ becomes $(r-b-\Delta)/(1+b+\Delta)$.
For $r\le b$ this is at most $-2e$, because
$1+b+\Delta<2$. For $r\ge a$, the premise gives a lower bound
$g-(1+g)\Delta/(1+b+\Delta)\ge g-10e$.
Replacing $z$ by $\widehat z$ changes the padded correlation
by at most $e/8$. Both asserted inequalities follow.
\end{proof}

\paragraph{Queries and exact comparisons.}
Use a normalized lattice net of radius $e/8$ in the original
direction space. Its logarithmic cardinality is $O(d\log(d/e))$;
it is used for the analysis and never enumerated. Round each query
before padding, but continue to accept candidates by the original
test $\langle z,u\rangle>b$. Thus every rejected candidate has
negative padded correlation, exactly the separation required by the
two search structures in the next section.

For rational coordinates, normalization does not require storing
irrational coordinates. For a raw vector $w$, a cap comparison has
the form
$A\langle G_{1:d},w\rangle\ge(a_0-BG_{d+1})\|w\|$.
Writing $L_0=A\langle G_{1:d},w\rangle$,
$R_0=a_0-BG_{d+1}$, and $S_0=\|w\|^2>0$, this is exactly
\[
 (R_0\ge0\ \wedge\ L_0\ge0\ \wedge\ R_0^2S_0\le L_0^2)
 \ \vee\
 (R_0<0\ \wedge\ (L_0\ge0\ \vee\ L_0^2\le R_0^2S_0)).
\]
The signs are checked before squaring. Zero raw vectors are handled
by the baseline and never normalized.
Section~\ref{sec:paper-parameters} chooses $\gamma=c-O(e)$,
$\lambda=1+e/(4\kappa)$, a fixed integer $K\ge2/(c-1)$, and
$e=\Theta_c(1/\log(2n))$. There are then $O_c(\log n)$ shells
and scales and $O_c(\log^3n)$ configurations per node. Their common
good correlation is $h=c^2-1-O_c(1/\log n)$, which preserves
the exact powers of $n$ in the explicit-filter bounds.

\section{Angular search: explicit lists and implicit incidences}
\label{sec:paper-search}

The preceding section reduces each local distance search to a single
problem on unit vectors: find a point whose inner product with the
query is positive whenever some point has inner product at least a
fixed $h>0$. This section gives two implementations. Explicit lists
prove the balanced bound with exact powers of $n$. A second
implementation certifies that no target can be missed and stores
point--filter memberships implicitly; it yields the improved bound
near linear space. We first establish coverage and query costs, then
account for construction and deletions.

\paragraph{The search interface.}
A \emph{generation} is an index constructed on a fixed indexed
family $P_0$ of $m\ge1$ unit vectors, with $0<h<1$. Empty pools
return \textsc{None} without constructing an index. The set is fixed before this generation's random
choices, and later operations may delete its members. A query supplies
a unit vector $z$ and an exact acceptance test in the original
geometry. A \emph{target} is a current point $p$ with
$\langle p,z\rangle\ge h$. The test is \emph{compatible} if every
rejected point satisfies $\langle p,z\rangle\le0$; the certified
construction uses the stronger bound $\langle p,z\rangle\le-\delta$
for a specified $\delta>0$. Thus every target must be accepted.
The required answer is an accepted current identity, or \textsc{None}
when the search finds none. On its correctness event, \textsc{None}
implies that no target exists. Query directions have already been
rounded before the common-gap lift, using its reserved error margin.

\paragraph{The two search exponents.}
The parameter $x\in[h,1]$ controls the relative thresholds for data
and queries. Define
\[
I_h(x)=\frac{1+x^2-2hx}{1-h^2},\qquad
 u=I_h(x)-1=\frac{(x-h)^2}{1-h^2},\qquad
 q=I_h(x)-x^2=\frac{(1-hx)^2}{1-h^2}.
\]
The meaning of these quantities is as follows: a point belongs to
roughly $m^u$ filters, and a query examines roughly $m^q$ filter
labels and rejected occurrences. The next lemma records the resulting
curve and an inequality needed to pay for updates.

\begin{lemma}[The angular tradeoff]
\label{lem:angular-tradeoff}
As $x$ ranges from $h$ to $1$, $u$ ranges from $0$ to
$\rho_h=(1-h)/(1+h)$, and
\[
q=q_h(u)=\bigl(\sqrt{1-h^2}-h\sqrt u\bigr)^2.
\]
Moreover,
\[
q-u=1-x^2\ge0.
\]
\end{lemma}
\begin{proof}
From the formula for $u$, $x=h+\sqrt{u(1-h^2)}$.
Substitution into the expression for $q$ gives the curve.
The endpoint $x=1$ gives $u=(1-h)/(1+h)$.
Finally, subtracting $u=I_h(x)-1$ from $q=I_h(x)-x^2$
gives the last identity.
\end{proof}

Throughout this section, $\mathcal P$ denotes a polynomial factor of fixed
degree in dimension, input precision, and the stated logarithmic
parameters. Fixed accuracy parameters may affect its coefficients.
All powers of the pool size $m$ are displayed explicitly.

\subsection{Explicit filter lists}
\label{sec:filtering}

A filter is a bucket label defined by a random direction. A point is
stored under that label when its projection exceeds a threshold; a
query visits the label when its own projection exceeds a possibly
different threshold. The two thresholds let us reduce point replication
at the cost of more query work. This is the asymmetric Gaussian
filtering approach of Christiani~\cite{christiani}.

Let $G$ be a standard Gaussian vector and let $t\ge1$.
The data test is $\langle G,p\rangle\ge t$, and the query test is
$\langle G,z\rangle\ge xt$. Write $p_u,p_q$ for their marginal
probabilities, $p_1$ for a lower bound on their joint probability
when $\langle p,z\rangle\ge h$, and $p_2$ for an upper bound when
$\langle p,z\rangle\le0$.

\begin{lemma}[Gaussian filter probabilities]
\label{lem:gaussian-filter-probabilities}
For $h\le x\le1$ and $t\ge1$,
\[
p_u\le e^{-t^2/2},\qquad p_q\le e^{-x^2t^2/2},\qquad
 p_1\ge c_h t^{-2}e^{-I_h(x)t^2/2},\qquad
 p_2\le e^{-(1+x^2)t^2/2},
\]
where $c_h>0$ can be chosen uniformly when $h$ stays in a compact
subinterval of $(0,1)$.
\end{lemma}
\begin{proof}
The Chernoff bound $\Pr[Z\ge s]\le e^{-s^2/2}$ for a standard
normal $Z$ holds for every $s\ge0$, including $s<1$; it gives the
two marginal bounds.

For the joint bounds, put $r=\langle p,z\rangle$. When $|r|<1$,
the two projections have the same law as
$(X,rX+\sqrt{1-r^2}Y)$ for independent standard normals $X,Y$.
If $r\le0$, the event $X\ge t$ and
$rX+\sqrt{1-r^2}Y\ge xt>0$ implies $Y\ge xt$.
Its probability is therefore at most the product of the marginal
tail bounds.

If $h\le r<1$, set
$a_r=\max\{0,(x-r)/\sqrt{1-r^2}\}$.
Then $0\le a_r\le(x-h)/\sqrt{1-h^2}\le1$: on $r<x$ the
derivative of $(x-r)/\sqrt{1-r^2}$ is nonpositive, and on
$r\ge x$ the numerator is nonpositive. The rectangle
\[
 [t,t+1/t]\times[a_rt,a_rt+1/t]
\]
in the $(X,Y)$ plane lies inside the joint event. On this rectangle,
$(X^2+Y^2)/2\le(1+a_r^2)t^2/2+3\le I_h(x)t^2/2+3$.
Integrating its product density proves the lower bound with the
universal constant $c_h=e^{-3}/(2\pi)$. If $r=1$, a direct integral
over $X\in[t,t+1/t]$ gives the same bound; if $r=-1$, the joint
event is empty. This also covers collinear vectors in dimension one.
Conservative rational approximations to these bounds suffice for
the implementation.
\end{proof}

We combine a fixed number $\ell$ of independent base families. A
full filter is a tuple containing one cap from each family. Evaluating
the base lists avoids examining all tuples. Explicit caps on list
length and search work give deterministic resource bounds; independent
copies will supply the desired success probability.

\begin{lemma}[A capped explicit index]
\label{lem:fixed-depth-exact-space}
Fix an integer $\ell\ge1$. A pool of size $m$ admits an explicit
index with data work $(m^u+m^{I_h(x)/\ell})\mathcal P$, query work
$(m^q+m^{I_h(x)/\ell})\mathcal P$, and space
$(m^{1+u}+m^{I_h(x)/\ell})\mathcal P$. These bounds hold on every outcome.
For a fixed query and target, a copy succeeds with constant probability;
the same event works for every later subset of the birth pool and
every compatible exact acceptance predicate.
\end{lemma}
\begin{proof}
Choose rational $t\ge1$ with
$\ell t^2/2=\log(2m)+O_\ell(1)$. Certified rational
approximations to logarithms and exponentials determine this parameter
and conservative probability and cap bounds in
$\poly_\ell(\log(2m))$ work. Multiplicative constant slack in those
bounds is included in the caps below; no transcendental operation
is assumed to have unit cost. Each of $\ell$ independent base families
contains $M_0=\lceil\widehat L/p_{\rm rat}\rceil$ caps, where
$\widehat L\ge\log(100\ell)$ is a rational upper bound and
$p_{\rm rat}$ is a positive rational lower bound on $p_1$, each
within a constant factor of its analytic value. A full filter is
a tuple of one cap from each family. Put $A=M_0^\ell$ and
\[
 E_u=Ae^{-\ell t^2/2}\le m^u\log^{O(\ell)}(2m),\qquad
 E_q=Ae^{-\ell x^2t^2/2}\le m^q\log^{O(\ell)}(2m).
\]
Compute rational upper bounds $\widehat E_u\ge E_u$ and
$\widehat E_q\ge E_q$, within a constant factor. Compute a point's
eligible base lists first. If their product has
more than $\lceil100(\widehat E_u+1)\rceil$ tuples, omit this point from this copy;
otherwise store its tuples in buckets with exact deletion handles.
Only nonempty buckets are allocated. Put
$C_q=\lceil100(\widehat E_q+1)\rceil$. A query enumerates at most
$C_q$ tuple labels and examines at most $C_q+1$ current point--tuple
occurrences, testing each candidate before applying the occurrence
cap. Thus a target immediately following exactly $C_q$ rejected
occurrences is still tested. A cap exhaustion without an accepted
candidate returns \textsc{None} in this copy. The base-list computation
costs $\ell M_0\le m^{I_h(x)/\ell}\mathcal P$; enumerating a Cartesian
product is output sensitive.

The expected point and query tuple counts are at most $E_u,E_q$.
For a rejected birth point the shared-tuple probability is at most
$p_2^\ell$, whence
$mAp_2^\ell\le E_q$. A target shares a base cap in every family
with probability at least $0.99$. Markov's inequality for the three
caps leaves constant success probability. Here the three bad events
are an overlong target list, an overlong query list, and too many
shared occurrences with nonpositive-correlation birth points.
On their complement, the query has at most $C_q$ labels and at
most $C_q$ potentially rejected birth occurrences. The stored
target is therefore reached within $C_q+1$ examined occurrences,
unless an acceptable point has already been returned. This argument
counts potential rejected occurrences against all of $P_0$.
Every later subset and every compatible acceptance test rejects only
occurrences in this same set. Thus one event works for all those
choices; no monotonicity of the acceptance test is needed.
\end{proof}

\paragraph{One event for all adaptive queries.}
For a raw nonzero direction record, the lattice rounding described
below can be performed by sign-guarded squared comparisons to the
finitely many coordinate breakpoints. It does not require an exact
square-root or a real floor primitive.

A fixed-query success probability is insufficient when the random
directions are exposed: an adversary could choose a later query that
is unusually bad for those directions. We instead prove that the
index succeeds simultaneously for a finite cover of all directions.
To obtain this event, round the original unit
direction to a normalized rational lattice of mesh
$\eta/(8\lceil\sqrt d\rceil)$. Its error is at most $\eta/8$ and
its logarithmic cardinality is $O(d\log(d/\eta))$. Only the chosen
rounded point is computed. The common-gap lift reserves this error
before specifying $h$ and the rejected correlation. Amplify with
$O(\log(m/\varepsilon_0)+d\log(d/\eta))$ independent copies and
union bound over the net and the $m$ birth targets. The resulting
event holds for every future query, every subset, and every
compatible predicate, even after the seeds are exposed. This
query-cover argument also appears in adversarially robust farthest
neighbor search~\cite{adaptiveafn}; here polynomial dimension
dependence permits checking all copies.

\paragraph{Finite precision and exact exponents.}
The preceding argument uses ideal Gaussian variables only to analyze
success probability. Finite Gaussian approximations preserve the event
with polynomial
precision. There are finitely many birth-vector and net comparisons;
one-dimensional Gaussian anti-concentration bounds the chance that
any ideal comparison lies in its rounding interval. A union bound
only adds the logarithm of their number to the precision. Actual
cap tests use signs and squared rational comparisons, and resource
caps hold regardless of the coupling. A bounded implementation of
the finite sampler is given in
Appendix~\ref{sec:bounded-gaussian-sampling}.
Section~\ref{sec:paper-parameters}
chooses fixed $\ell\ge16/\rho_c$, where $\rho_c=2/c^2-1$,
and $h=g_c-O_c(1/\log n)$ with $g_c=c^2-1$, and
$x$ with $u\le\sigma$ and $q\le Q_c(\sigma)+O_c(1/\log n)$.
Thus the explicit update and space powers are exactly $Q_c(\sigma)$
and $1+\sigma$, with polynomial logarithmic factors.

\subsection{Certified finite families}
\label{sec:finite-word-las-vegas-filters}

The explicit index may omit a target on an unlucky outcome. An
always-correct search needs a finite test that certifies coverage of
\emph{every} target direction. We obtain such a test by decomposing
the sphere into small-dimensional blocks, where an explicit finite
net can be checked. The resulting product family can have many labels,
but its description remains small. Section~\ref{sec:implicit-certified-tensor-lists}
will store and search these labels without expanding their products.

A \emph{label} identifies one filter; data and query labels use the
same naming convention. Sharing a label makes a point a candidate,
which must still pass the exact original acceptance test. Fix a
negative rejection margin $\delta>0$ and a density parameter
$0<\kappa\le1$. To control the number of labels and the number of
rejected candidates separately, replace the threshold by
\[
t^2=2\kappa\log m+O(1).
\]
The pool still contains $m$ points: in particular, an expected
collision count must be summed over all $m$ birth points, regardless
of $\kappa$. The exponent allowance $\eta>0$ below pays for finite
nets, splitting families, and exact representations. It may be made
arbitrarily small for fixed accuracy parameters.

\begin{proposition}[Certified angular labels]
\label{prop:finite-word-lv-angular}
\label{lem:implicit-certified-labels}
For every fixed $\eta>0$, a finite-word randomized constructor
terminates almost surely and returns labels with the following guarantees. Every target pair shares a label on every
accepted outcome. A fixed data point has expected label count at
most $m^{\kappa u+\eta}\mathcal P$; a fixed query has expected routing work
and label count at most $m^{\kappa q+\eta}\mathcal P$; and its expected
shared-label occurrences with all birth points of correlation at
most $-\delta$ are at most

\begin{equation}\label{eq:paper-certified-bad-moment}
                         m^{1+\kappa(q-1)+\eta}\mathcal P.
\end{equation}
The description size and workspace are at most $m^\eta \mathcal P$, and
expected certification time is at most $m^\eta \mathcal P$. The full raw label universe has a deterministic
size bound $m^{\kappa(1+u)+\eta}\mathcal P$. Its size and each label's integer
code can be computed without enumerating that universe.
\end{proposition}

\begin{proof}
The construction has three steps. First, an orthogonal decomposition
puts every target into at least one moderate-dimensional block.
Second, a finite splitting family reduces that block to small spheres.
Third, checked cap families cover those spheres. We prove deterministic
coverage first and then bound expected work. Exact finite certificates
and arithmetic details appear in Appendix~\ref{sec:paper-search-certificates}. Bounded $m$ is handled
by a scan. If $d\ge m$, scanning also costs only $O(d^2)$, so assume
$d<m$. These scan cases are trivial single-label indices; their
routing and rejected-occurrence counts are absorbed by the polynomial
dimension factor or the fixed small-pool constant.

\paragraph{Preserving a target in one moderate block.}
Apply an exact rational orthogonal rotation and divide the
padded space into blocks of dimension
$B_0=\Theta_\delta(\log m)$. Every positive-correlation pair survives
in some normalized block, for every rotation: if all its nonzero
block correlations were less than $h$, then
\[
 \langle p,z\rangle
 <h\sum_i\|p_i\|\|z_i\|\le h.
\]
For a fixed query, the negative margin $\langle p,z\rangle\le-\delta$
leaves room for projection error. Concentration of the projected norms
of $p,z,p-z$, followed by a union bound over rejected birth points,
shows that their correlations remain nonpositive in every normalized
block with probability at least $1-m^{-30}$. A poor rotation can
increase the number of rejected candidates, but cannot remove every
block containing a target: that fact follows from orthogonality alone.
Thus rotations affect running time, while coverage is deterministic.

\paragraph{Reducing to small spheres.}
Within a moderate block use the finite splitting family of
Wei~\cite[Proposition~11]{wei2019lasvegas}. Take
$b=\Theta(\log^{2/3}m)$ and $s=B_0/b$, with dimensions rounded to
powers of four. There are $\exp(O(s\log B_0))=m^{o(1)}$ splitters.
For every three vectors, one splitter preserves all squared block
norms to relative error $\epsilon=O(b^{-1/2})$. Admit a vector only
when every block norm squared lies in $[(1-\epsilon)/s,
(1+\epsilon)/s]$. A splitter preserving $p,z,p-z$ admits both and
leaves every normalized block correlation at least $h-2\epsilon$.
The final splitter matrices are rational.

\paragraph{Checking coverage on each small sphere.}
Set $t_0=t/\sqrt s$. Independently in each small block, sample a finite
family of rational Gaussian caps with thresholds $t_0,xt_0$. Its
certificate checks all pairs in a $b^{-2}$-net of the small sphere,
using strengthened thresholds and a checked upper bound on every
cap vector's norm. The net has $\exp(O(b\log b))=m^{o(1)}$ points.
A family is accepted only after this finite test proves coverage
of every pair of correlation at least $h-2\epsilon$. The certificate
accepts with probability at least $1/2$, and its family size and
verification cost are $m^{o(1)}$. A full label specifies an outer
block, a splitter, and one cap from each of its $s$ small blocks.

\paragraph{Bounding the expected label counts.}
A raw label is any tuple of base caps, before testing whether a point
or query belongs to it. The raw number of labels in a moderate block is
$\exp(I_h(x)t^2/2+o(\log m))=m^{\kappa I_h(x)+o(1)}$.
For one fixed label under the unconditioned law, concatenate its
independent Gaussian vectors. The result is a standard Gaussian in
the moderate block. The norm gates and positive thresholds imply
a total data cap at least $(1-\epsilon)t$ and a total query cap at
least $(1-\epsilon)xt$. Therefore the data, query, and nonpositive-pair
probabilities are at most
\[
 e^{-t^2/2+o(\log m)},\qquad
 e^{-x^2t^2/2+o(\log m)},\qquad
 e^{-(1+x^2)t^2/2+o(\log m)}.
\]
Conditioning the $s$ independent base families on their certificates
inflates these bounds by at most $2^s=m^{o(1)}$. Multiplication by
the raw-label count gives exponents $\kappa u$, $\kappa q$, and
$\kappa(q-1)$, respectively. The last expression is summed over
the actual $m$ birth points, yielding
\eqref{eq:paper-certified-bad-moment}. Splitters, outer blocks,
and finite precision contribute only the stated overhead.
The exceptional rotation contributes negligibly: its probability
is $m^{-30}$, whereas full routing and occurrence multiplicities
have a deterministic polynomial bound. No subset is involved in
this estimate. Increasing the fixed cutoff bounds all subpolynomial
terms by $m^\eta \mathcal P$.
\end{proof}

\paragraph{Accessing one label without listing all labels.}
The product representation also supports counting and random access.
In a chunk
(an outer block and splitter), let $L_1(p),\ldots,L_s(p)$ be the
eligible base-cap lists. Its label count is
$\prod_i|L_i(p)|$. Compute these counts without expanding products;
locate a requested rank by subtracting chunk counts, then decode it
in mixed radix and rescan the base families. The complete operation
costs $E=m^\eta \mathcal P$ time and workspace. No point-specific base lists
need be stored. For full-family sizes $M_{a,i}$, a raw label in
chunk $a$ has the injective code
\[
1+\operatorname{offset}(a)
       +\sum_i(j_i-1)\prod_{r>i}M_{a,r}.
\]
Offsets are sums of full products. Notice the distinction between
the eligible-list radices used for a point's rank and the full-family
radices used for a label's code.

\subsection{One function for all point--label incidences}
\label{sec:implicit-certified-tensor-lists}

The preceding proposition proves coverage, but explicitly storing
all point--label memberships may use too much space. We represent
those memberships by a function that can be evaluated on demand.
To answer a query label, we then recover its entire preimage.
There are two requirements: every point must retain enough labels
for coverage, and its retained count must satisfy a deterministic
bound. We meet both by assigning each point to one of several
independent certified families.

\paragraph{A bounded number of labels per point.}
An expected count alone permits an unusually long point list. Construct
$L_f=\lceil\log_2(4m)\rceil$ independent complete certified families,
tagging labels by their family. Choose a computable bound
$B_m=m^{\kappa u+\eta}\mathcal P$ on one point's expected count, with $\mathcal P\ge1$;
in particular, choose $B_m\ge m^{\kappa u}$. Let $U$ be the tagged
raw universe size. Set
\[
R=\max\{\lceil2B_m\rceil,\lceil U/m\rceil,1\}.
\]
Assign each point to its first family with at most $R$ labels; if
some point has no such family, resample the whole group. Markov's
inequality and independence bound a point's failure by $2^{-L_f}$,
so group acceptance has probability at least $3/4$. Store only its
chosen family, exact count, and generation-membership bit. Each point
retains every label of its chosen family, so coverage is deterministic.
Queries use all families. Assignment only removes incidences, and
conditioning on group acceptance costs one factor at most $4/3$
for the entire group.

Write $a=\kappa u$, $b_*=\kappa q$, and
$f_*=1+\kappa(q-1)$. The construction gives
$m^a\le R\le m^{a+\eta}\mathcal P$, because
$\kappa(1+u)-1\le\kappa u$. All metadata and expected assignment
work fit $m\mathcal P$ and $m^{1+\eta}\mathcal P$, respectively. In particular,
neither assignment nor a later inverse construction materializes
the point--label incidence relation.

\paragraph{The occurrence function.}
An occurrence records a point together with the rank of one of its
labels. The rank belongs in the \emph{domain} of the function: otherwise
a query could need a separate inversion for each rank. Put $M=mR+1$
and introduce virtual identities
$\iota(p,k)=(p-1)R+k$, for $p\in[m]$, $k\in[R]$. They are integer
addresses, not an allocated array. If $\lambda_k(p)$ is the $k$th
retained label, define
\begin{equation}\label{eq:implicit-product-function}
 F(\iota(p,k))=
 \begin{cases}
  \operatorname{Code}(\lambda_k(p)),&p\text{ is a current member and rank }k\text{ exists},\\
  \bot,&\text{otherwise}.
 \end{cases}
\end{equation}
The dummy identity also maps to $\bot$. For example, if $R=2$ and
two points have label lists $(a,c)$ and $(b,c)$, then
$F^{-1}(c)=\{\iota(1,2),\iota(2,2)\}$. One inverse query recovers
both occurrences, without storing the two lists explicitly.
Since $U\le mR$, this is
exactly a function $F:[M]\to[M-1]\cup\{\bot\}$. Evaluation costs
$E=m^\eta \mathcal P$ by the rank oracle. No fingerprint, alias, or range
padding to a larger exponent is required.
During a point deletion, a pending point and rank cursor additionally
make its already processed ranks null; this override is part of the
current-function oracle, before the point's membership bit is cleared.

\paragraph{Queries and deletions.}
A query enumerates its raw labels and inverts each label code once.
Every inverse output is an exact current point--label incidence;
test the original geometric predicate and stop at its first success.
An exhausted search is correct because every target shares a retained
label. Repeated reports of one point are counted as occurrences.
Membership is local to the frozen generation: removing a point from
an unmatched set can leave it globally active. A later restoration
creates a fresh local search-set insertion lifetime and never revives a deleted value
of this function.

\subsection{Certified inversion with a density bound}
\label{sec:lv-dynamic-inversion}

The occurrence function can be much larger than the space available
for the index. Function inversion permits a smaller representation:
with parameter $\tau$, we retain roughly $M/\tau$ words and spend
roughly $\tau^3$ forward evaluations to recover a small inverse fiber.
The main issue under deletions is that a once-large fiber can become
sparse. We first certify complete recovery during a phase, then show
how to charge recovery costs to surviving rejected occurrences.

Here a \emph{fiber} is the preimage of one nonnull function value,
and a \emph{value deletion} changes one domain address to $\bot$.
A phase freezes the function at its start and records subsequent
value deletions. We use the bounded-fiber Las Vegas inverse of Bibbens, Borevitz, and
McCauley~\cite[Theorem~5 and Section~4.3.2]{bbm2026}, together with
the phase organization of Kuszmaul and
Kuszmaul~\cite[Section~6]{kuszmaul2026}. The precise interface matters:
the null fiber is unrestricted, recovery returns the complete
preimage, and constructor workspace can be limited by batching.
In particular, Theorem~5 of~\cite{bbm2026} imposes its fiber-size
condition only on nonnull outputs of a function
$[M]\to[M-1]\cup\{\bot\}$; absent ranks require no sanitization.

\begin{lemma}[Verified inverse phases]
\label{lem:lv-dynamic-inversion}
\label{cor:verified-lv-dynamic-inversion}
Let $F:[M]\to[O(M)]\cup\{\bot\}$ have evaluation time $E$.
For $1\le\tau\le M^{1/3}$, a phase allowing $O(M/\tau)$ value
deletions has all-time space $\widetilde O(M/\tau)$, expected
certified construction time $\widetilde O(M\tau^3E)$, and
worst-case deletion work $\widetilde O(\tau^3E)$ excluding
reconstruction. Every accepted phase reports exact current preimages
with worst-case initialization, next-item, and termination delay
$\widetilde O(\tau^3E)$. Sequential phases have expected amortized
deletion time $\widetilde O(\tau^4E)$.
\end{lemma}
\begin{proof}
The proof separates large fibers, certifies the resulting bounded
fibers, and records deletions without modifying the certified snapshot.

\paragraph{Splitting large fibers within the space budget.}
Freeze the phase-birth function $F_0$. For a heavy nonnull fiber of
size $k>\tau$, choose the least power of two at least $2k/\tau$
split buckets. A fresh logarithmic-wise independent hash has expected
bucket size in $(\tau/4,\tau/2]$. Count all buckets and retry until
every heavy bucket has size in $[\tau/16,\tau]$. Bounded-independence
Chernoff bounds give constant acceptance probability for
$\tau\ge C\log^2(2M)$; below this cutoff explicit lists fit the
claimed space. There are $O(M/\tau)$ heavy buckets. Light outputs
keep their names; heavy outputs use split names. Constant-factor
domain padding gives a derived function with all nonnull fibers at
most $\tau$ and an unrestricted null fiber.
To identify heavy outputs without a dense count table, process the
$O(M)$ possible names in batches of $\Theta(M/\tau)$, scanning the
virtual domain for each batch and retaining only heavy names and
their exact counts. This costs $\widetilde O(M\tau E)$ time and
$\widetilde O(M/\tau)$ workspace. A further scan counts the
$O(M/\tau)$ allocated heavy buckets.

\paragraph{Certifying the immutable inverse.}
The static primitive stores $\widetilde O(M/\tau)$ words and
recovers a whole derived fiber in $\widetilde O(\tau^3E)$ time.
Its low-workspace constructor takes expected
$\widetilde O(M\tau E)$ time. Its repair dictionary contains
precisely unrecovered identities; cap its size at a sufficiently
large constant times $M/\tau$ and retry if necessary. The expected
repair count is $O(M/\tau)$, so this preserves constant acceptance.

We additionally certify every possible derived output, including
empty outputs. Process output names in batches of size $M/\tau$.
Scan the virtual domain to obtain exact counts for a batch; run each
output's recovery under its $\widetilde O(\tau^3E)$ cap; check all
returned identities by evaluating the oracle, check distinctness in
an $O(\tau)$ temporary set, and compare the count. These conditions
certify equality with the complete fiber. The scans cost
$\widetilde O(M\tau E)$ and all recovery checks cost
$\widetilde O(M\tau^3E)$. Workspace is
$\widetilde O(M/\tau+\tau)=\widetilde O(M/\tau)$, with no forward
array of length $M$. Accepted recoveries are deterministic thereafter.
All attempts obey the same space cap.
Appendix~\ref{sec:paper-search-certificates} also gives a direct
implementation that compares actual raw-chain recoveries with exact
batched fibers and stores precisely their missing identities.

\paragraph{Answering after deletions.}
During a phase, keep the old value of each deleted virtual identity
in a dictionary $A$; consult $A$ before the current oracle to
reconstruct $F_0$ exactly. Its size is $O(M/\tau)$. For a heavy
output, maintain a list of split buckets having current survivors.
A deletion can recover its one affected bucket to update this list
in $\widetilde O(\tau^3E)$ work. An iterator recovers one nonempty
bucket at a time, discards identities in $A$, and buffers at most
$\tau$ survivors. Every heavy-bucket recovery yields an output;
a light output, possibly empty, needs only one bounded recovery.
This gives the claimed prefix delay. Rebuild after $\Theta(M/\tau)$ deletions;
its conditional expected cost charges $\widetilde O(\tau^4E)$
per deletion. All statements survive exposed-seed adaptive deletions.
\end{proof}

A complete inverse fiber may contain many points rejected by the
geometric acceptance test. Charging $\tau^3$ to every reported point
would lose the space improvement. The next bound charges that cost
once to a group of $\Omega(\tau)$ current points. When a group becomes
smaller, we store its survivors explicitly and charge their space to
the deletions that made it small.

\begin{corollary}[Density-sensitive predicate search]
\label{cor:density-sensitive-lv-inversion}
For $Q$ queried nonnull names, let $F$ count current-preimage
occurrences rejected before the search stops. The phase finds an
accepted identity or correctly reports none in
\[
\widetilde O\bigl(((Q+1)\tau^3+F\tau^2)E\bigr),
\]
where $F$ counts rejected current-preimage occurrences. This bound
is deterministic for every accepted phase and every deletion history.
\end{corollary}

\begin{proof}
Call a heavy bucket \emph{thick} while it has at least $\tau/32$
current members, and \emph{thin} afterward. Maintain its exact current size. When it first drops
below $\tau/32$, recover its survivors and store them explicitly.
It had at least $\tau/16$ birth identities, so the fewer than
$\tau/32$ new records can be charged to distinct preceding deletions
in that bucket. Buckets are disjoint and become thin once per phase;
all thin lists and deletion pointers therefore use $O(M/\tau)$
space. The transition costs one recovery, within the deletion bound.

For a query, a completely rejected thick bucket contains
$\Omega(\tau)$ current rejected identities, paying for its
$\widetilde O(\tau^3E)$ recovery at cost
$\widetilde O(\tau^2E)$ per occurrence. Thin lists are explicit.
Charge light-fiber setup once per queried name. At most one final
thick bucket contains an accepted identity; the additive
$\tau^3E$ term pays for it. Deleted identities never pay this
charge: thickness is defined by current survivors.
\end{proof}

\subsection{The pool interface and its optimal parameters}
\label{sec:scaled-density-las-vegas}

We now combine the certified labels with the dynamic inverse and
choose their parameters. Let $a=\kappa u$ be the point-label exponent,
$b_*=\kappa q$ the query-label exponent, and
$f_*=1+\kappa(q-1)$ the rejected-occurrence exponent. To obtain
space exponent $1+\sigma$, the inverse must save a factor of about
$m^{(a-\sigma)_+}$ from the occurrence domain. Use

\[
v=(a-\sigma)_+,\qquad \tau=\lceil m^{v+2\eta}\rceil,
 \qquad 3v+6\eta<1+a.
\]
The extra $2\eta$ in $\tau$ pays for the description and rank-oracle
overhead without increasing the requested space exponent. There is
no dimension factor in $\tau$. The final inequality ensures
$\tau\le M^{1/3}$ above a fixed cutoff, as required by the inverse.
The following proposition records all costs needed by the dynamic
construction, including construction work and deletions.

\begin{proposition}[Implicit birth-pool interface]
\label{prop:scaled-density-product-index}
The resulting generation is always correct and uses at most
$m^{1+\sigma}\mathcal P$ space. Its expected construction work is $m^{1+a+3v+8\eta}\mathcal P$, and its expected amortized deletion work is
$m^{a+4v+10\eta}\mathcal P$ per original point. For each fixed rounded query $z$, a random variable $Z_z$, determined
at generation construction, bounds every later query with that direction,
for all surviving subsets, compatible acceptance tests, and accepted
inverse phases, and
\[
\mathbb E Z_z
                 \le m^{\max\{b_*+3v,\ f_*+2v\}+8\eta}\mathcal P.
\]
The bounds include all rejected construction attempts and low-workspace
certification. Independent copies use independent complete
constructions, including outer rotations and point assignments.
\end{proposition}
\begin{proof}
The domain has $m^{1+a}\le M\le m^{1+a+\eta}\mathcal P$ addresses and the forward oracle
costs $E=m^\eta \mathcal P$. Hence $M/\tau\le m^{1+a-v-\eta}\mathcal P
\le m^{1+\sigma}\mathcal P$. Family descriptions, point records, and temporary
rank work fit as well. Construction costs
$\widetilde O(M\tau^3E)$; deleting a point changes at most $R$
values, at cost $\widetilde O(R\tau^4E)$ including reconstruction.

To obtain a bound that remains valid after adaptive deletions,
fix $z$ and count against the entire birth set. Let $Q_z$ count
full query routing work and labels, and let
$J_z$ count shared labels with all birth points satisfying
$\langle p,z\rangle\le-\delta$. Thus $J_z$ is defined before any
later subset or acceptance predicate is chosen. Their
expectations are at most $m^{b_*+\eta}\mathcal P$ and $m^{f_*+\eta}\mathcal P$.
Set $Z_z=\widetilde O(((Q_z+1)\tau^3+J_z\tau^2)E)$.
The actual rejected current occurrences are a subset of these birth
occurrences. The density lemma bounds work for every accepted phase
by this same $Z_z$, regardless of later phase seeds. Taking
expectations proves the claim. This is stronger than a bound for
one subset fixed before the random choices.
\end{proof}

\begin{algorithm}[tbp]
\caption{A certified generation on an implicit occurrence domain}
\label{alg:implicit-product-inverse}
\begin{algorithmic}[1]
\Statex \textbf{Build$(P_0,h,\delta,x,\kappa,\sigma,\eta)$:}
\State Construct $L_f$ independent certified families; compute $U$ and $R$ by full-product counts.
\State Assign every point to its first family of count at most $R$; retry the group if an assignment fails.
\State Store only point records, assignments, counts, and generation-membership bits.
\State Define the rank oracle $F$ in~\eqref{eq:implicit-product-function}; build its verified inverse with parameter $\tau$.
\Statex \textbf{Delete$(p)$:}
\State Set the pending point to $p$ and its rank cursor to $0$.
\For{each retained rank $k$ of $p$}
 \State Save its old value and preserve it in existing snapshot logs.
 \State Advance the pending-rank cursor to $k$, making this virtual value null.
 \State Notify the inverse of this value deletion.
\EndFor
\State Mark $p$ inactive in this generation and clear the cursor.
\Statex \textbf{Query$(z,\mathcal A)$:}
\State Enumerate full query labels in all families; search their exact inverse fibers using the thick/thin rule.
\State Return the first current identity satisfying the original predicate $\mathcal A$, or \textsc{None} after exhaustion.
\end{algorithmic}
\end{algorithm}

Queries do not interleave with a point deletion. The cursor makes a
reconstruction triggered inside that operation see the correct
intermediate function; existing snapshot logs preserve their immutable
functions. For background construction and replay, the exact order
is formalized in Section~\ref{sec:density-inverse-deamortization}.

\paragraph{Choosing the density.}
The query bound is the larger of $b_*+3v$ and $f_*+2v$.
The first term pays for query labels; the second pays for rejected
occurrences. Choose the density so that these exponents agree.
For $u\ge\sigma$, set
\[
\kappa=\frac{1+\sigma}{1+u},\qquad
 a=\frac{u(1+\sigma)}{1+u},\qquad
 v=\frac{u-\sigma}{1+u}.
\]
Then the two query exponents are equal:
\begin{equation}\label{eq:scaled-density-balanced-objective}
 b_*+3v=f_*+2v
       =(1+\sigma)\frac{q_h(u)+3u}{1+u}-3\sigma.
\end{equation}
The raw universe now has exponent $\kappa(1+u)=1+\sigma$;
the occurrence domain still contains every exact label code.

\paragraph{Optimizing the remaining parameter.}
After balancing the two query costs, only $u$ remains. The following
calculation gives the improved branch of the tradeoff and explains
why it meets the explicit-filter curve tangentially.

\begin{lemma}[Optimal density parameters]
\label{lem:optimal-density-parameters}
Let
\[
 \chi_h=2-\sqrt{1+3h^2},\qquad
 u_h=\frac{h^2(1-h^2)}{(1+h^2+\sqrt{1+3h^2})^2}.
\]
For $0\le\sigma\le\rho_h$, the balanced objective
in~\eqref{eq:scaled-density-balanced-objective}, over
$\sigma\le u\le\rho_h$, is minimized at $u=\max\{\sigma,u_h\}$.
Its value is
\[
\Lambda_h(\sigma)=
 \begin{cases}
  \chi_h-(3-\chi_h)\sigma,&0\le\sigma\le u_h,\\
  q_h(\sigma),&u_h\le\sigma\le\rho_h.
 \end{cases}
\]
\end{lemma}
\begin{proof}
The quotient
$(q_h(u)+3u)/(1+u)$ is the Rayleigh quotient on $(1,\sqrt u)$ of
\[
\begin{pmatrix}
  1-h^2&-h\sqrt{1-h^2}\\
  -h\sqrt{1-h^2}&3+h^2
 \end{pmatrix}.
\]
Its trace is $4$, determinant is $3(1-h^2)$, and its negative
off-diagonal entries give a positive least eigenvector. Thus its
minimum is $\chi_h=2-\sqrt{1+3h^2}$, attained at
\[
u_h=\frac{h^2(1-h^2)}{(1+h^2+\sqrt{1+3h^2})^2},\qquad
 x_h=\frac{3h}{1+\sqrt{1+3h^2}}.
\]
One has $h<x_h<1$ and $0<u_h<\rho_h$. To verify that this is the
only minimum on the nonnegative axis, put $r=\sqrt u$ and
$b=h\sqrt{1-h^2}$. The derivative of the quotient has the sign of
$b r^2+2(1+h^2)r-b$, which is strictly increasing for $r\ge0$
and has one positive zero. Hence the quotient increases after $u_h$,
and the constrained choice is $u=\max\{\sigma,u_h\}$.
The pieces meet tangentially: differentiating the Rayleigh quotient
at its minimizer gives $q_h'(u_h)=\chi_h-3$.
\end{proof}

The query optimization alone does not establish the update bound:
construction and deletions must also be affordable.
Section~\ref{sec:paper-parameters} verifies that construction and
deletion are no more expensive, checks the native inverse restriction,
and chooses rational parameters without increasing the space exponent.

\paragraph{What remains to obtain a fully dynamic structure.}
The present section constructs a deletion-only generation with a fixed
birth set. New points cannot simply be inserted into an already exposed
random filter family. The next section organizes separate generations
for insertions and replaces the geometric centers before their
certificates expire. For fixed $c,\sigma$ and any fixed exponent allowance $\omega>0$,
the preceding interface has $A(n)=n^{\Lambda_c(\sigma)+\omega}\mathcal P$. Its $Z_z$
variables are exactly the birth envelopes required by
Lemma~\ref{lem:adaptive-certified-generation-compiler}. That lemma
gives the zero-error adaptive theorem; the capped phase schedule in
Section~\ref{sec:density-inverse-deamortization} gives its worst-case
Monte Carlo counterpart. No future inverse-phase event is conditioned
into the birth moment calculation. Appendix~\ref{sec:uniform-subpolynomial-density}
specifies a single parameter schedule giving
$A(n)=n^{\Lambda_c(\sigma)+O_{c,\sigma}((\log n)^{-1/100})}\mathcal P$
with a fixed polynomial degree in dimension and input length.
Thus the exponent allowance can be made $o(1)$ without hiding a
dimension-dependent power in the implementation.

\section{From birth-pool search to fully dynamic diameter}
\label{sec:paper-dynamization}

The geometric and search arguments have different requirements under
updates. A center remains valid for a bounded number of changes to
its input, whereas a random search index is valid only for points
fixed before its random choices. We therefore use two clocks.
Insertion lifetimes determine when to construct a fresh search
index inside a fixed-center node. Changes to the node's input
determine when to replace its center and residual chain.

This section combines these two schedules. We first convert the
search interface from deletion-only to fully dynamic membership.
We then explain how replacement chains are built before their old
centers expire. Finally, we combine the schedules with the local
matching invariant and account for all failure events. Detailed
replacement and inverse-replay schedules appear in
Appendix~\ref{app:dynamic-schedules}.

\paragraph{The birth-pool interface.}
Write $A(n)=n^\theta\mathcal P$ for the target time bound, where
$\mathcal P$ denotes a polynomial factor in dimension, input
precision, and the relevant logarithmic parameters. Its value may
increase between bounds. An index built on $m$ points uses
$m^{1+\sigma}\mathcal P$ space, takes at most
$mA(n)\mathcal P$ construction work, and supports a deletion or
query in $A(n)\mathcal P$ work, with $\theta\ge\sigma$.
For the Monte Carlo version, these are deterministic caps, and one
success event guarantees correctness for all later subsets and all
query directions. The certified version is always correct and
supplies the conditional construction bounds and query-work variables
of Proposition~\ref{prop:scaled-density-product-index}; its expected
adaptive work is handled separately below.

\subsection{Frozen insertion intervals}

Each insertion into an unmatched search set starts a new lifetime,
even when the point was previously present and has just become
unmatched again. Group completed lifetimes into aligned intervals
of lengths $1,2,4,8,\ldots$. Once an interval is complete, no future
insertion belongs to it. Its surviving points therefore form a
valid birth pool for a fresh deletion-only index. For example, the
pool for lifetimes $1,\ldots,4$ is fixed after insertion $4$;
lifetime $5$ belongs to a different pool, while deletion of lifetime
$3$ may be applied to the old index.

\begin{lemma}[From frozen pools to dynamic search]
\label{lem:paper-frozen-dynamization}
The birth-pool interface gives a fully dynamic search index with
the same time and space exponents, up to logarithmic factors.
Capped primitives and scheduled construction give worst-case bounds.
For certified primitives, the expected-amortized guarantee follows
from Lemma~\ref{lem:adaptive-certified-generation-compiler}.
\end{lemma}
\begin{proof}
We retain small available indices while constructing larger ones.
The proof specifies their birth sets, shows that few indices coexist,
and checks correctness and space.

\paragraph{Birth snapshots.}
Keep active lifetime numbers in a persistent balanced tree. When
an aligned interval of $w=2^k$ consecutive lifetime numbers is
completed, retain its active snapshot and start a construction job.
Its size is $m\le\min\{w,n\}$. Draw the job's fresh random choices
only after fixing this snapshot. Later insertions lie outside the
interval, so the snapshot can only lose members.

\paragraph{Construction and arriving deletions.}
Give each job $O(A(n)\mathcal P)$ work per primitive operation.
Its total construction and replay work is at most
$O(mA(n)\mathcal P)$, so a sufficiently large constant budget
completes it within $w$ operations. For explicit lists, scan birth
records in the background. Index a record only if it is still
active; a deletion removes an already indexed record immediately.
Per-job flags distinguish these cases. This builds the raw
birth-pool index, without recursively invoking the present wrapper.
Singleton intervals can be served immediately.

For an implicit index, construction uses the immutable birth
function. Queue arriving deletions and replay them before activating
the result. Each birth identity is deleted at most once, so the
initial build and all replay work satisfy the same total budget.
Section~\ref{sec:density-inverse-deamortization} specifies the
snapshot oracle and the order in which a point's virtual ranks
are deleted. In particular, advancing the rank cursor before
notifying the inverse ensures that a reconstruction sees the
correct intermediate function.

\paragraph{Available indices.}
Serve queries from available intervals that partition the insertion
prefix. When a completed interval becomes available, replace its
serving descendants unless a serving ancestor already covers it.
Cancel pending descendants only at this replacement. At each length
$w$, at most one job is pending: such jobs arrive every $w$
insertions and finish within $w$ primitive operations.

There are also at most three serving intervals of any given length.
Indeed, let $h$ be the current insertion count and let a serving
interval of length $w$ end at $r$. Its parent completes by insertion
$r+w$ and finishes construction within $2w$ further operations.
If $h\ge r+3w$, that parent, or an available ancestor, must already
have replaced the interval. Thus every serving interval of length
$w$ ends in $(h-3w,h]$. Since aligned endpoints are spaced $w$
apart, there are at most three. Both the serving and pending
collections consequently have logarithmic size.

\paragraph{Updates and search correctness.}
A predecessor search among serving intervals locates the index
responsible for a deleted lifetime. Notify that index and every
pending job whose snapshot contains the lifetime. Service existing
jobs, activate completed intervals from largest to smallest, and
then start jobs triggered by the new insertion. A query checks all
serving indices. Every current point belongs to exactly one of
them, so a current target is present in one queried pool. That pool's
success event remains valid because it quantifies over every later
subset and direction, including choices made after its random
choices are exposed.

\paragraph{Space and lifetime counters.}
At one interval length, only constantly many snapshots coexist, and
each has at most $n$ birth points. Hence their retained cardinalities
sum to $O(n)$. The inequality $m^{1+\sigma}\le n^\sigma m$
gives space $n^{1+\sigma}\mathcal P$ per level, including construction
workspace. In one center lifetime there are at most
$n\mathcal P$ primitive operations, so the number of interval levels
is logarithmic in this quantity, rather than in the total age of
the stream. Restart the local lifetime counter when its center is
replaced. Reusable handles are reclaimed only after all retained
snapshots, certificates, and queued retirement records release them. The incremental reclamation described in
Appendix~\ref{app:dynamic-schedules} includes retired records in the
space bound and avoids a large deallocation during one update.
\end{proof}

\subsection{Replacing the geometric chain}

The preceding wrapper supports arbitrary membership changes while a
node's center remains fixed. A second schedule replaces the center
before too many witnesses are deleted. A node born with $m$ points
is valid for $\Theta(\alpha m)$ input changes. We start a replacement
sufficiently early, continue serving from the old node, and spread
the replacement work across the remaining changes.

A replacement consists of a whole residual chain. Constructing that
chain from top to bottom immediately would be unsafe: a small residual
has a short deletion tolerance and could become invalid before the
larger job is ready. The schedule therefore opens each small residual
late, when the common activation deadline is near. Its construction
then finishes while the old chain is still valid. More precisely,
a residual of current size $m$ is opened only when at most
$\alpha m/4096$ root-pool ticks remain; a sufficiently large work
budget finishes the level in $O(\alpha^2m)$ ticks. The verified core
removes an $\Omega(\alpha)$ fraction of the input, leaving enough
time to open its smaller child later.

Lemma~\ref{lem:paper-right-alignment} proves these deadline and age
claims. Only serving nodes start replacement jobs: a pending chain
does not recursively create more center-replacement jobs. When a
chain is activated, its nodes retain their actual birth ages.
The two scans in Section~\ref{sec:paper-center-schedule} distinguish
the immutable birth records from the current input, so insertions and
deletions arriving during construction are included exactly once.
The resulting number of serving and pending nodes, and all their
work budgets, are polynomial in $\alpha^{-1}$, $d$, and the stated
logarithmic parameters.

\subsection{Completing the guarantees}

\paragraph{Approximation and update work.}
Use Lemma~\ref{lem:paper-frozen-dynamization} for each unmatched-set
index in a local matching. A point update causes only constantly
many matching repairs per configuration. The number of configurations
and geometric levels is polynomial in $d$ and the logarithmic
parameters, so multiplying the schedules by these factors preserves
the claimed power of $n$. Lemma~\ref{lem:local-matching-diameter}
approximates every core--input distance. Assigning a diametral pair
to its first core proves the global guarantee; no approximation
factor is multiplied along the chain.

For explicit filters, use their uniform capped success event.
For implicit filters, the certified inverse phases and their queued
deletion replay give the worst-case Monte Carlo implementation
in Section~\ref{sec:density-inverse-deamortization}. For zero error,
use certified center blocks and synchronous suffix rebuilding.
Lemma~\ref{lem:adaptive-certified-generation-compiler} bounds expected
total work despite exposed random choices: it amplifies over the
whole query-direction cover and charges rare exact scans to the
finite lifetime of the generation. Its accounting holds for every
fixed finite prefix and needs no prescribed execution horizon.

\paragraph{One failure budget for the whole execution.}
For a horizon $t$, put
$S=n+t+d+B+\lceil\log(2/\delta)\rceil+2$, omitting $B$ in the
arithmetic model. The capped schedules give a predetermined
polynomial bound in $S$ on the number of center constructions,
filter constructions, verification attempts, and witness samples.
Assign conditional failure probability $\delta/S^{C_1}$ to each
relevant event, for a sufficiently large fixed $C_1$. Running
$O(\log(S/\delta))$ independent capped attempts gives this bound
when a constructor has constant success probability. At each birth,
condition on the entire previous history before drawing fresh
random choices. A conditional union bound then gives probability
at least $1-\delta$ that all required events succeed.

The work and space budgets also hold on failure outcomes. Every
reported positive answer is checked against current identities;
a failed or capped negative answer still triggers only the fixed
number of matching repairs. A center whose core is too small is
replaced by the deterministic enclosing-ball fallback from
Section~\ref{sec:paper-center-schedule}, preserving contraction and
resource bounds even though correctness is asserted only on the
good event.

Uniform rank sampling is capped as well. From a pool of size $m$,
draw $\lceil\log_2m\rceil$ bits and reject out-of-range integers.
After $O(\log(S/\delta))$ trials, use rank one if all trials fail
and charge this outcome to the failure budget. Each trial succeeds
with probability at least $1/2$. The zero-error version uses uncapped
rejection sampling, with constant conditional expected trial count.

\paragraph{The exact endpoint.}
At $c=\sqrt2$ omit the angular indices and local matchings entirely.
Maintain each node's radius maximum and a baseline witness, together
with the exact pair at each leaf. Lemma~\ref{lem:paper-sqrt-two-cache}
then gives the approximation directly. The same center-replacement
schedule gives worst-case Monte Carlo updates; certified disjoint
blocks and synchronous suffix rebuilding give zero error with
expected amortized updates. Lemma~\ref{lem:explicit-endpoint-cost}
below supplies the explicit polynomial in
Theorem~\ref{thm:paper-endpoint}.

\paragraph{An explicit polynomial at the endpoint.}
The exact $\sqrt2$ construction uses no angular filters. The following
bound makes its dependence on dimension and logarithmic parameters
explicit. We choose a deliberately generous leaf cutoff so that one
simple estimate covers center computation, sampling, and all replacement
work; the polynomial degrees are not optimized.

\begin{lemma}[Explicit endpoint cost]
\label{lem:explicit-endpoint-cost}
Put $D=d+3$ and $H=\lceil\log_2(2Dnt/\delta)\rceil$.
The arithmetic-RAM implementation at $c=\sqrt2$ has worst-case update
time $O(D^{19}H^{18})$ and space $O(nD^{19}H^{18})$.
\end{lemma}
\begin{proof}
Use the Har-Peled--Jones center constructor in the lifted dimension
$d+1$, with depth $\alpha=1/(4D^2)$
\cite{harpeledjones2019}. Its arithmetic cost with unit-cost uniform
sampling is
$O(D^7\log^3(2D)\log^3(2/\varphi))$ for failure probability
$\varphi$. Assign each construction and witness-sampling event a
failure probability $\delta/(2DntH)^{C_0}$ for a sufficiently large
absolute constant $C_0$. All relevant logarithms are then $O(H)$.
Uniform sampling from a persistent membership tree uses $O(H)$
rank-search work and a deterministic $O(H)$ cap on rejection trials.
Each trial draws $O(H)$ fair bits, so the total sampling overhead is
$O(H^2)$. Charging it to every sampling operation gives center cost
$C=O(D^7H^8)$.

A baseline uses $K_0=O(\alpha^{-1}H)=O(D^2H)$ fresh samples.
Its complete capped sampling and distance-computation work is also
at most $O(D^7H^8)$. Choose the cutoff in
\eqref{eq:paper-leaf-cutoff} as
$L=C_1D^7H^8$, with a sufficiently large absolute constant $C_1$.
This satisfies that equation, including $L\ge\alpha^{-2}$ and
$L\ge H$. A serving or pending chain has at most
$D_0=O(\alpha^{-1}\log(2n))=O(D^2H)$ nodes.
At most $D_0$ serving nodes own replacement jobs, and every such job
has at most $D_0$ pending levels. Hence $O(D_0^2)$ node contributions
suffice to account for one external update.

Each exact leaf has $O(L)$ points; recomputing all its pair distances
costs $O(dL^2)$. Radius-tree updates, fresh baseline samples, and an
immediate center computation fit the same bound. At the endpoint,
constructing a pending level scans and classifies its points and builds
radius dictionaries; it has no matching or filter indices. The
late-opening schedule allocates
$O(\alpha^{-2}(d+\log n))$ work per affected job tick to these scans.
Since $\alpha^{-2}\le L$ and $\log n\le L$, this is again
$O(dL^2)$ for $d\ge1$. Activating completed jobs and reclaiming
retired records are charged to the same node budgets. Thus the total
update work is
\[
 O(dD_0^2L^2)=O(D^{19}H^{18}).
\]

For space, there are $O(D_0^2)$ live node structures. Each retains
only a constant number of population snapshots, radius dictionaries,
construction scans, and center workspaces. Population records and
persistent-tree paths use $O(nH)$ words per such structure; center
workspace is at most $C\le L$. Even if all pair distances of a
leaf are stored, their number is at most $O(nL)$, since the leaf
contains at most $\min\{n,O(L)\}$ points. Including coordinates,
these quantities are bounded by $O(ndL^2)$ per node. Incremental
reclamation adds only a constant factor to the simultaneously live
allocation budgets. The total is therefore at most
$O(ndD_0^2L^2)=O(nD^{19}H^{18})$.
The zero-dimensional case is immediate.
\end{proof}

\section{Limits on update time and streaming space}
\label{sec:paper-lower-bounds}

Our upper bounds raise two different questions. Can the update time
be made polylogarithmic for every approximation arbitrarily close to
one? Can the maintained state use a sublinear number of bits? We first
address computation: a simple encoding and a reduction that pays for
initialization rule out a uniform near-exact scheme under the randomized
Orthogonal Vectors Conjecture. We then address memory by an explicit
INDEX reduction. The computational statements concern ordinary
randomized algorithms on inputs of bounded encoding length. The memory
statement counts the entire maintained state, including the input
points.

\subsection{Encoding inner products as an ordinary diameter}

Orthogonal Vectors and Maximum Inner Product ask for a pair with one
point from each of two sets. Diameter has no such restriction. The
following encoding ensures that a pair between the two encoded sets
attains the diameter, so no colored-diameter oracle is needed. Let
$D\ge1$, let $A,B\subseteq\{0,1\}^D$ be nonempty, and put
$M=\min_{a\in A,b\in B}\langle a,b\rangle$.

\begin{lemma}[An uncolored diameter encoding]
\label{lem:paper-uncolored-encoding}
In $O((|A|+|B|)D)$ time one can construct a strict set
$P\subseteq\{0,1\}^{5D}$ of size $|A|+|B|$ with
$\diam(P)^2=4D-2M$. A factor-$(1+\varepsilon)$ diameter estimate
therefore gives an estimate of $M$ in $[M,M+4\varepsilon D]$.
\end{lemma}
\begin{proof}
We encode the two sets separately, compare cross distances with
within-set distances, and then translate the approximation error.
Use the injective maps
\[
 F(a)=(a,\mathbf1-a,\mathbf0_D,\mathbf0_{2D}),\qquad
 G(b)=(b,\mathbf0_D,\mathbf1-b,\mathbf1_{2D}).
\]
The final $2D$ coordinates make their images disjoint. For a cross
pair, coordinatewise expansion gives
\[
 \begin{aligned}
 \|F(a)-G(b)\|^2
 &=\sum_{i=1}^D\bigl((a_i-b_i)^2+(1-a_i)^2+(1-b_i)^2+2\bigr)\\
 &=4D-2\langle a,b\rangle\ge2D.
 \end{aligned}
\]
Two points in either one image have squared distance at most $2D$.
Thus a cross pair attains the diameter of $F(A)\cup G(B)$, proving
the identity. For a one-coordinate illustration, the cross contribution
is four unless both input bits are one, when it is two.

Now let $L$ be a valid approximate diameter, so that
$\diam(P)/(1+\varepsilon)\le L\le\diam(P)$. Define
$\widehat M=(4D-L^2)/2$. The identity just proved yields
\[
 0\le\widehat M-M
   =\frac{\diam(P)^2-L^2}{2}
   \le2D\bigl(1-(1+\varepsilon)^{-2}\bigr)
   \le4\varepsilon D.
\]
This proves the approximation statement.
\end{proof}

An orthogonal cross pair exists exactly when $M=0$. The two possible
squared-diameter ranges are therefore $4D$ and at most $4D-2$.
Setting $\varepsilon=1/(8D)$ preserves this distinction, since
$4D/(1+1/(8D))^2>4D-2$.
For maximum inner product, first replace $a$ by $(a,\mathbf1-a)$
and $b$ by $(\mathbf1-b,\mathbf1)$. Their inner product becomes
$D-\langle a,b\rangle$. Applying the lemma in dimension $2D$
then yields a diameter instance in dimension $10D$ and an additive
$8\varepsilon D$ approximation to the original maximum.

\subsection{Paying for initialization only once}

A dynamic lower bound must include the time to initialize the data
structure. In particular, inserting all points into one very large
structure is not useful if that initialization already takes quadratic
time. We instead choose a smaller capacity and reuse the same
initialized structure for many block pairs. It is emptied by legal
deletions between pairs; it is never initialized again.

Let $r$ be the size of each static input set and
let $m\ge2$ be the capacity of the reusable dynamic structure. Assume empty
initialization takes $m^{K+o(1)}$ time, an update takes
$m^{\alpha+o(1)}$ amortized time, and reading an answer takes
$m^{\beta+o(1)}$ time. Empty initialization has no access to future
coordinates. Polynomial factors in logarithmic dimension,
inverse-polylogarithmic accuracy, and logarithmic execution length
are included in the $o(1)$ terms. Assume the exponents $K,\alpha,\beta$ are nonnegative and write
$s=\max\{1+\alpha,\beta\}\ge1$.

\begin{lemma}[Reusing one initialized structure]
\label{lem:paper-blocking}
If $s<2$, the corresponding static encoded problem on two
$r$-point sets can be solved in time $r^{\Psi_K(s)+o(1)}$, where
\[
 \Psi_K(s)=
 \begin{cases}
 s,&K\le s,\\
 2K/(K+2-s),&K>s.
 \end{cases}
\]
The same implication holds for expected amortized work and bounded
error on each read, with constant error in the static algorithm.
\end{lemma}
\begin{proof}
We first count operations for an arbitrary capacity $m$, then choose
$m$ to balance initialization with the work over all block pairs.
Partition each input set into blocks of size at most
$\lfloor m/2\rfloor$. For every nonempty block pair $(A_i,B_j)$,
insert its encoded points, read the approximate diameter, and delete
all those points. For example, two blocks on each side give the four
runs $(A_1,B_1),(A_1,B_2),(A_2,B_1),(A_2,B_2)$, all using the same
initialized state. The sequence of logical insertions and deletions is
fixed before the algorithm's random choices; deletion calls use the
handles returned by the corresponding insertions. The capacity is
respected in every run.

There are $O(r^2/m^2)$ runs, using $O(r^2/m)$ updates and
$O(r^2/m^2)$ reads in total. Taking the appropriate extremum of the
block answers gives the answer for the full static instance. Including
input reading, the total work is
\[
 r^{1+o(1)}+m^{K+o(1)}+r^2m^{s-2+o(1)}.
\]
If $K\le s$, choose $m=\Theta(r)$. Otherwise choose
$m=\Theta(r^{2/(K+2-s)})$, which balances the last two terms. These
choices give the stated exponent. Since $K$ and $s$ are fixed, $r$
is a fixed power of $m$, and the complete sequence has length
polynomial in $m$.

For bounded-error reads, use $O(\log r)$ independent copies and take
the median estimate for each block pair. The probability that any
block estimate fails can be made at most $1/100$, without changing
an exponent. For expected amortized work, apply the total-work bound
to this one legal sequence, including all its returns to the empty
set. Stopping after a sufficiently large constant multiple of the
expected work adds only a constant error probability by Markov's
inequality. The resulting static algorithm has bounded error and the
claimed worst-case running time.
\end{proof}

\subsection{Consequences for near-exact approximation}

We use the following randomized form of the Orthogonal Vectors
Conjecture (OVC): for every $\eta>0$, there is a constant $C$ for
which no bounded-error randomized $O(r^{2-\eta})$-time algorithm
decides whether two $r$-point sets in
$\{0,1\}^{\lceil C\log r\rceil}$ contain a pair with inner product
zero~\cite{chenwilliams}. Here bounded error means success probability
at least $2/3$ on every input.

The encoding uses accuracy proportional to $1/\log r$. Consequently,
a polynomial dependence on the inverse accuracy cannot offset a fixed
saving in the update exponent. The reuse lemma makes this observation
valid even when the initialization degree is large.

\begin{corollary}[Uniform accuracy]
\label{cor:paper-uniform-accuracy}
Assume randomized OVC. Fix $K\ge0$, $0<\gamma<1$, and a polynomial
$p$. No single fully dynamic $(1+\varepsilon)$-diameter scheme, valid
for all $0<\varepsilon<1$, can have initialization time
$n^Kp(d,\varepsilon^{-1},\log(2nt))$, expected amortized update time
$n^{1-\gamma}p(d,\varepsilon^{-1},\log(2nt))$, constant-time answer
retrieval, and success probability at least $2/3$ on each read.
This holds already for binary strict sets, $d=O(\log n)$,
$\varepsilon=\Theta(1/\log n)$, and a polynomial update horizon $t$.
\end{corollary}
\begin{proof}
The saving in the static running time is determined before the OVC
dimension constant is chosen. Put $s=2-\gamma$. For every fixed $K$,
Lemma~\ref{lem:paper-blocking} gives $\Psi_K(s)<2$. Choose
$0<\eta<2-\Psi_K(s)$ and take the dimension constant $C$ supplied by
OVC for that $\eta$.

Given the corresponding Orthogonal Vectors instance, apply
Lemma~\ref{lem:paper-uncolored-encoding} with
$D=\lceil C\log r\rceil$ and $\varepsilon=1/(8D)$. The approximate
diameter determines whether an orthogonal pair exists. The reuse
lemma chooses a capacity $m=r^{\Theta(1)}$, so the dimension,
inverse accuracy, execution horizon, and amplification factors
contribute only polynomial logarithmic overhead. It gives a
bounded-error algorithm in time $r^{\Psi_K(s)+o(1)}$, contradicting
the chosen OVC instance.
\end{proof}

The expected-work assumption concerns every fixed legal prefix
starting empty, so the corollary also applies to Las Vegas structures.
Its quantifier order matters: $K$, the polynomial $p$, and the saving
$\gamma$ are fixed before $\varepsilon$ is chosen. The statement
therefore permits an update saving that tends to zero as
$\varepsilon$ tends to zero, as in our upper bounds.

\paragraph{A fixed accuracy after fixing the initialization degree.}
A second consequence uses a constant-gap reduction. For an integer $C\ge1$ and
$0<\eta<1/2$, let $\mathsf{IP}(C,\eta)$ be the hypothesis that no
bounded-error $O(r^{2-\eta})$-time algorithm decides whether two
$r$-point binary sets in dimension $\lceil C\log r\rceil$ have a
cross pair of a specified exact inner product.
Abboud--Ron-Zewi~\cite[Lemma~4.1]{abboudronzewi2024} imply that this
hypothesis rules out additive Maximum Inner Product approximation
with error $\widetilde\Theta((\eta/C)^2)$ times the dimension in
$O(r^{2-2\eta-\zeta})$ time, for any fixed $\zeta>0$. The target
dimension is $O_{C,\eta}(\log r)$. The strict saving $\zeta$ absorbs
the logarithmic amplification needed for randomized algorithms.

Combining their reduction with our maximum-inner-product encoding
and the reuse lemma gives a constant
$\varepsilon_{C,\eta}>0$ for which a dynamic
$(1+\varepsilon_{C,\eta})$-approximation with $s<2$ must satisfy
$\Psi_K(s)\ge2-2\eta$. More explicitly, choose
$\varepsilon_{C,\eta}$ at most one eighth of the additive-error
constant in their Max-IP lower bound, since our encoding has error
at most $8\varepsilon D$. For constant-time answer retrieval this is
\[
 \alpha\ge
 1-\max\left\{2\eta,\frac{K\eta}{1-\eta}\right\}.
\]
Indeed, on the branch $K\le s=1+\alpha$ the inequality is
$s\ge2-2\eta$; on the other branch it rearranges to
$s\ge2-K\eta/(1-\eta)$.

Randomized OVC implies that for every $\eta$ there is some
$C=C(\eta)$ satisfying $\mathsf{IP}(C,\eta)$, because exact inner
product includes orthogonality. After fixing $K$, choose $\eta$
small enough that the displayed lower bound is positive. This gives
a fixed error $\varepsilon_K>0$ for which polynomial-logarithmic
updates and reads are impossible with initialization degree $K$.
Thus the error depends on the initialization degree; this argument
does not give a matching update lower bound for each prescribed
$c<\sqrt2$.

\subsection{Linear memory below \texorpdfstring{$\sqrt2$}{sqrt(2)}}

For memory, a one-way communication reduction already applies to
insertion-only streams. We adapt the INDEX construction of
Agarwal--Sharathkumar~\cite{agarwal}, using nearly orthogonal sign
vectors to obtain a logarithmic dimension bound at every fixed
$c<\sqrt2$. In INDEX, Alice receives a bit string
$x\in\{0,1\}^m$, Bob receives an index $j$, and Alice sends one
message from which Bob must recover $x_j$. Any randomized protocol
with success probability at least $2/3$ requires $\Omega(m)$ bits,
even with public randomness.

\begin{proposition}[An adapted INDEX lower bound]
\label{prop:paper-index-space}
For every fixed $1\le c<\sqrt2$, returning an estimate
$L\in[\diam(P)/c,\diam(P)]$ with probability at least $2/3$ requires
$\Omega_c(n)$ bits of state, even for insertion-only strict sets
of sign vectors in dimension $O_c(\log n)$.
\end{proposition}
\begin{proof}
Choose a fixed $0<\xi<\min\{1,2/c^2-1\}$. There exist $m$ sign
vectors $v_1,\ldots,v_m\in\{-1,1\}^d$, where
$d=O(\xi^{-2}\log(m+1))$, such that
$|\langle v_i,v_j\rangle|\le\xi d$ for $i\ne j$.
To see existence, draw the sign vectors independently. A standard
Chernoff bound gives probability at most $2e^{-\xi^2d/2}$ that one
pair violates the inequality, so a union bound over all pairs is
less than one for a sufficiently large constant in $d$. Fix such
a family and make it known to both parties.

Alice inserts $v_i$ exactly when $x_i=1$ and sends the algorithm's
entire state to Bob. Bob continues the computation by inserting
$-v_j$. The stream is a strict set: the correlation bound excludes
both duplicate and antipodal vectors among distinct $v_i$'s.
Its capacity is at most $m+1$.

If $x_j=1$, the final set contains $v_j,-v_j$, so its diameter is
$2\sqrt d$. If $x_j=0$, every pair has squared distance at most
$2(1+\xi)d$. This holds for pairs $v_i,v_k$ because their squared
distance is $2d-2\langle v_i,v_k\rangle$, and for pairs $v_i,-v_j$
because it is $2d+2\langle v_i,v_j\rangle$. Since
\[
          \frac{2\sqrt d}{c}>\sqrt{2(1+\xi)d},
\]
a valid factor-$c$ diameter estimate distinguishes the two cases.
An $s$-bit state would therefore give an $s$-bit INDEX message.
The communication lower bound yields $s=\Omega(m)$; taking
$n=m+1$ proves the proposition.
\end{proof}

With $\Theta(\log n)$-bit words, this is
$\Omega_c(n/\log n)$ words. Hence the exponent of $n$ in our
near-linear space bounds is best possible on these bounded-coordinate
inputs, up to polynomial logarithmic factors. The bound charges all
memory; it does not grant free external access to the active set.
It supplies no corresponding update-time lower bound.

For general constant approximation factors,
Padaki--Singal--Waingarten~\cite{psw2026} obtain an
$n^{\widetilde\Omega(1/c^2)}$ space lower bound in strict-turnstile
streams, where frequencies remain nonnegative after every update.
Their setting permits arbitrary stream length and multiplicities,
bounds the final support by $n$, and uses a public-randomness budget
independent of stream length. These assumptions differ from short
strict-set streams. Polynomial-length sketch simulations
\cite{jly2026} have their own model conditions, so they do not by
themselves transfer that theorem to our bounded-multiplicity setting.
The proposition above provides the memory comparison needed for our
results below $\sqrt2$ without such a transfer.

\section{Parameter choices and final resource bounds}
\label{sec:paper-parameters}

The previous sections establish the geometric guarantee, the search
operations, and their dynamic composition. We now choose their
parameters to obtain the exponents in the introduction. There are
two issues to check. In the explicit construction, rounding must not
increase the space exponent, and the geometric margins must cost
only a constant factor in time. In the implicit construction, the
cost of building and updating the inverse must stay below its query
bound. We give these verifications separately.

Throughout, fix $1<c<\sqrt2$ and write $g=c^2-1$,
$\rho=2/c^2-1$, and $L=\lceil\log_2(2n)\rceil$.
All constants in this section may depend on $c$, but not on the
coordinates or their spread.

\subsection{Explicit filters with exact exponents}

Recall that $e$ is the radial and angular rounding margin,
$\gamma$ is the strong-edge distance factor, $\kappa=\gamma+2e$
pays for radial rounding, and $\lambda$ is the ratio between
distance scales. Taking $e$ of order $1/\log n$ makes all rounding
losses in the exponent of order $1/\log n$, which contributes only
a constant factor to the running time.

\begin{lemma}[Geometric and filter parameter choices]
\label{lem:paper-explicit-parameters}
For every $0\le\sigma\le\rho$, the geometric construction has
$O_c(L^3)$ configurations per node, with a common positive
correlation $h=g-O_c(1/L)$. The explicit index can be chosen with
point-incidence exponent $u\le\sigma$ and query exponent
$q\le Q_c(\sigma)+O_c(1/L)$.
\end{lemma}
\begin{proof}
\emph{Geometric margins.}
Choose rational $e$ within a factor of two below
$\min\{(c-1)/32,1/(1024L)\}$ and rational
$\gamma\in[c-4e,c-3e]$. Set $\kappa=\gamma+2e$,
$\lambda=1+e/(4\kappa)$. Choose a fixed integer
$K\ge2/(c-1)$ by using a certified rational lower bound
$1<c_-\le c$ and taking $K=\lceil2/(c_--1)\rceil$.
Then $\gamma>1$, $\kappa\lambda\le c-3e/4<c$, and
$1+1/K<c$. These are precisely the requirements of
Section~\ref{sec:paper-geometry}. The radial mesh and the distance
scales each have $O_c(L)$ choices, so there are $O_c(L^3)$
configurations.

Use the padding margin $e$ from Lemma~\ref{lem:common-gap-lift} and set
\[
h=\gamma^2-1-11e.
\]
Since $e\le(c-1)/32$ and $\gamma\ge1+7(c-1)/8$,
$e\le(\gamma^2-1)/16$ and $h\ge45(c-1)/32>0$.
Also $h<g$ and
$g-h\le(8\sqrt2+11)e=O_c(1/L)$.
Thus the common-gap lemma applies, including its margin for query
rounding and negative correlation of rejected candidates.

\emph{Rounding the filter parameter downward.}
Put $J=128L$. Obtain a certified rational lower approximation
$\sigma_-$ with $\sigma-J^{-2}\le\sigma_-\le\sigma$, and set
\[
 \sigma_n=\max\{0,\lfloor J^2\sigma_-\rfloor/J^2\}.
\]
Then $0\le\sigma_n\le\sigma$ and
$\sigma-\sigma_n\le2J^{-2}$. This uses a floor only on a rational
number, avoiding an undecidable equality test at an arbitrary real
parameter. The ideal parameter is
$x_*=h+\sqrt{\sigma_n}\sqrt{1-h^2}$.
Choose the largest $x\in h+J^{-1}\mathbb Z_{\ge0}$ with
$x\le x_*$. Squaring the comparison decides this choice with
rational arithmetic. Since
$\sigma_n\le\rho\le(1-h)/(1+h)$, we have $h\le x\le1$.
Consequently
\[
u=\frac{(x-h)^2}{1-h^2}\le\sigma_n\le\sigma,\qquad
 q=\frac{(1-hx)^2}{1-h^2}
       \le Q_c(\sigma)+O_c(1/L).
\]
For completeness, the second inequality follows from
$\sqrt\sigma-\sqrt{\sigma_n}\le\sqrt2/J$ and $x_*-x<1/J$.
The derivative of $q$ with respect to $x$ has magnitude at most
$2$ on $[h,1]$. The function
$(\sqrt{1-s^2}-sz)_+^2$ has derivatives of magnitude at most
$4$ in $s$ and $1$ in $z$ in the interior of $[0,1]^2$.
The resulting Lipschitz bounds extend to the boundary by continuity.
Applying these bounds to $h=g-O_c(1/L)$ proves the estimate uniformly over
$0\le\sigma\le\rho$. When $\sigma=0$, choose $x=h$ exactly,
so that $u=0$ with no rounding loss.
\end{proof}

\paragraph{Construction cost and the absence of an exponent loss.}
Choose fixed tensor depth $\ell\ge16/\rho$. For the explicit
Gaussian construction, the probability estimates are uniform over
the compact range of $h$ above and $h\le x\le1$.
The exponent of each base family is at most
$I_h(x)/\ell=(1+u)/\ell\le\rho/8$.
For a birth pool of size $m$, the point work, query work, and space
are therefore bounded, respectively, by
\[
 (m^u+m^{\rho/8})\mathcal P,\qquad
 (m^q+m^{\rho/8})\mathcal P,\qquad
 (m^{1+u}+m^{\rho/8})\mathcal P,
\]
where $\mathcal P$ is polynomial in dimension and the logarithmic
parameters. Since $m\le n$,
$m^q\le m^{Q_c(\sigma)}\exp(O_c(\log m/L))
=O_c(m^{Q_c(\sigma)})$.
Moreover $Q_c(\sigma)\ge\rho\ge\sigma$, so both maintenance
and construction fit the same update budget. Only nonempty full
filter buckets are stored. The implicit direction net contributes
$O(d\log(d/e))$ to amplification, rather than its cardinality to
space, because it is never enumerated.

The dynamic composition thus gives update time
$n^{Q_c(\sigma)}\poly_{c,\sigma}(d,\log(2nt/\delta))$ and space
$n^{1+\sigma}\poly_{c,\sigma}(d,\log(2nt/\delta))$.
Putting $\sigma=\rho$ proves Theorem~\ref{thm:paper-balanced};
the same calculation proves the full explicit curve stated after
that theorem.

\subsection{The implicit tradeoff and maintenance costs}

The implicit construction may assign more labels to a point than
the final space bound allows us to store. The inverse compresses
these incidences, but recovering a bucket takes additional time.
The density-sensitive search bound makes it possible to balance
the two kinds of recovery cost. The optimization in the search
section determines the query exponent. Here we check that deletions,
construction, and workspace obey the same tradeoff.

Choose a small fixed geometric margin, so $h$ is close to $g$ and
rejected correlations remain negative. Let $\eta>0$ be the exponent
allowance used by the certified finite families. For $u\ge\sigma$,
define
\[
 \kappa=\frac{1+\sigma}{1+u},\qquad
 a=\kappa u,\qquad b=\kappa q_h(u),\qquad
 f=1+\kappa(q_h(u)-1),\qquad v=a-\sigma,
\]
where $q_h(u)=(\sqrt{1-h^2}-h\sqrt u)^2$.
Here $a$ controls point labels, $b$ controls query labels, and
$f$ controls rejected incidences in the whole birth pool.
The inverse parameter is $\tau=\lceil m^{v+2\eta}\rceil$.
The extra $2\eta$ absorbs description and route-count factors in
space; it does not introduce an extra power of $m$ in the final
space bound.

The birth-pool interface gives the following leading exponents.
Time exponents have an additional $O(\eta)$ allowance, whereas the
space exponent shown is exact.
\begin{center}
\begin{tabular}{@{}lc@{}}
\toprule
Resource & Exponent of birth-pool size $m$\\
\midrule
Space & $1+\sigma$\\
Construction per birth point & $a+3v$\\
Point deletion & $a+4v$\\
Query & $\max\{b+3v,f+2v\}$\\
\bottomrule
\end{tabular}
\end{center}
The inverse construction additionally requires $3v+6\eta<1+a$.
We next verify that the optimizer satisfies this restriction and
that maintenance never exceeds the query exponent.

Put $S_h=\sqrt{1+3h^2}$, $\chi_h=2-S_h$, and
$u_h=h^2(1-h^2)/(1+h^2+S_h)^2$.
Use the optimizer $u=\max\{\sigma,u_h\}$ from the search section.

\begin{lemma}[The optimizer pays for maintenance]
\label{lem:paper-maintenance-exponents}
With this choice, both query terms equal $\Lambda_h(\sigma)$,
while $a+3v\le a+4v\le\Lambda_h(\sigma)$.
Moreover $v<1/16$, so the inverse restriction holds with a constant
margin for sufficiently small $\eta$.
\end{lemma}
\begin{proof}
Since $\kappa(1+u)=1+\sigma$, we have $v=1-\kappa$.
It follows that $b+3v=f+2v$, and their common value is
\[
 (1+\sigma)\frac{q_h(u)+3u}{1+u}-3\sigma.
\]
The Rayleigh-quotient minimization in the preceding search section
gives $\Lambda_h(\sigma)$ at $u=\max\{\sigma,u_h\}$.

For $\sigma\le u_h$, write $a_h=u_h/(1+u_h)$.
Direct substitution gives
$a_h/\chi_h=(S_h-1)/(6S_h)<1/12$, and hence
\[
 \Lambda_h(\sigma)-(a+4v)
   =(1+\sigma)(\chi_h-5a_h)+\sigma>0.
\]
For $\sigma\ge u_h$, we have $u=\sigma$, $\kappa=1$,
$a=\sigma$, and $v=0$. On the permitted interval,
$q_h(\sigma)\ge\sigma$, proving the maintenance bound.
Construction is cheaper than deletion because $v\ge0$.
Finally, $u_h<1/16$ because its numerator
$h^2(1-h^2)$ is at most $1/4$ and its denominator is greater than
$4$. On the first branch $v\le u_h$, and on the second branch $v=0$.
Thus $3v<3/16<1+a$, leaving a fixed margin for $6\eta$.
\end{proof}

\paragraph{From fixed margins to one algorithm.}
Given any fixed $\omega>0$, choose the geometric, rational-rounding,
and construction margins so that their combined increase in the
time exponent is below $\omega$. On the explicit branch, round
the filter parameter downward to keep the implemented space
parameter at most $\sigma$. The two branches are continuous and
tangent where they meet. The resulting bounds are
$n^{\Lambda_c(\sigma)+\omega}\mathcal P$ time and
$n^{1+\sigma}\mathcal P$ space.
The adaptive compiler gives the always-correct expected-prefix
bound; capped construction and the phase schedule give worst-case
Monte Carlo updates.

This family of algorithms, one for each fixed $\omega$, does not
yet establish an $o(1)$ term for a single algorithm.
Appendix~\ref{sec:uniform-subpolynomial-density} specifies all
margins as functions of $n$, bounds the finite-cover construction
uniformly, and keeps the polynomial degrees fixed. It yields
the explicit allowance $O_{c,\sigma}((\log(2n))^{-1/100})$
in Corollary~\ref{cor:paper-uniform-overhead}, and therefore
Theorem~\ref{thm:paper-certified}.

\paragraph{Finite representations.}
The rational implementation computes centers by linear programs on
sampled original lifted coordinates. Shell membership, scale
eligibility, and the final weak-edge test use exact signed squared
comparisons. Padding coefficients are rational, and normalized
directions are represented by their unnormalized inputs. Gaussian
approximations are sampled at polynomial precision for the Monte
Carlo probability bounds; in the zero-error version, exact finite
coverage tests establish correctness independently of this coupling.
Label counts, ranks, and codes have polynomial bit length.
Consequently the arithmetic bounds acquire only a polynomial factor
in the coordinate bit length $B$. Each generation uses original
input coordinates, so repeated rebuilding does not accumulate
precision requirements.

\FloatBarrier
\section*{AI Use Statement}
This work resulted from extensive interaction between the authors and OpenAI Astra. Throughout an iterative process, the authors provided suggestions, critically evaluated Astra's responses, and refined the proposed ideas and arguments. OpenAI Astra was also used to assist with writing and revising the manuscript based on detailed prompts provided by the authors, with Gemini Argon contributing to selected parts of the process. The authors directed the entire process and critically assessed all proposed ideas and arguments. Lean files formalizing the main theorem are available at https://github.com/hajiaghayi/Diameter-Alg1.

\begingroup
\small
\bibliographystyle{alpha}
\bibliography{fully_dynamic_diameter_checked}

@inproceedings{indyk,
 author={Piotr Indyk},
 title={Better Algorithms for High-Dimensional Proximity Problems via Asymmetric Embeddings},
 booktitle={Proceedings of the Fourteenth Annual ACM-SIAM Symposium on Discrete Algorithms (SODA)},
 pages={539--545}, year={2003},
 url={https://homepage.divms.uiowa.edu/~kvaradar/sp2003/as.pdf}
}

@article{chanpathak,
 author={Timothy M. Chan and Vinayak Pathak},
 title={Streaming and Dynamic Algorithms for Minimum Enclosing Balls in High Dimensions},
 journal={Computational Geometry}, volume={47}, number={2}, pages={240--247}, year={2014},
 doi={10.1016/j.comgeo.2013.05.007},
 note={Preliminary version in WADS 2011},
 url={https://tmc.web.engr.illinois.edu/blur_wads.pdf}
}

@article{agarwal,
 author={Pankaj K. Agarwal and R. Sharathkumar},
 title={Streaming Algorithms for Extent Problems in High Dimensions},
 journal={Algorithmica}, volume={72}, number={1}, pages={83--98}, year={2015},
 doi={10.1007/s00453-013-9846-4},
 note={Preliminary version in SODA 2010},
 url={https://users.cs.duke.edu/~pankaj/publications/papers/highdim-stream.pdf}
}

@inproceedings{dynamic,
 author={Kiarash Banihashem and Jeff Giliberti and Samira Goudarzi and MohammadTaghi Hajiaghayi and Peyman Jabbarzade and Morteza Monemizadeh},
 title={Dynamic Diameter in High-Dimensions against Adaptive Adversary and Beyond},
 booktitle={Advances in Neural Information Processing Systems (NeurIPS)},
 volume={38}, year={2025}, doi={10.52202/085713-3134},
 note={Full version: arXiv:2511.01065},
 url={https://arxiv.org/abs/2511.01065}
}

@inproceedings{adaptiveafn,
 author={Kiarash Banihashem and Jeff Giliberti and Prashant Gokhale and Samira Goudarzi and Mohammadtaghi Hajiaghayi and Yuhao Liu and Morteza Monemizadeh and Sandeep Silwal},
 title={Adversarially Robust Approximate Furthest Neighbor},
 booktitle={Proceedings of the 43rd International Conference on Machine Learning},
 series={Proceedings of Machine Learning Research}, volume={306},
 pages={6281--6297}, year={2026},
 url={https://proceedings.mlr.press/v306/banihashem26b.html}
}

@inproceedings{chenwilliams,
 author={Lijie Chen and Ryan Williams},
 title={An Equivalence Class for Orthogonal Vectors},
 booktitle={Proceedings of the Thirtieth Annual ACM-SIAM Symposium on Discrete Algorithms (SODA)},
 year={2019}, note={Full version: arXiv:1811.12017},
 url={https://arxiv.org/abs/1811.12017}
}

@inproceedings{acw16,
 author={Josh Alman and Timothy M. Chan and Ryan Williams},
 title={Polynomial Representations of Threshold Functions and Algorithmic Applications},
 booktitle={57th IEEE Annual Symposium on Foundations of Computer Science (FOCS)},
 year={2016}, url={https://tmc.web.engr.illinois.edu/thr8_16.pdf}
}

@inproceedings{acw20,
 author={Josh Alman and Timothy M. Chan and Ryan Williams},
 title={Faster Deterministic and {Las Vegas} Algorithms for Offline Approximate Nearest Neighbors in High Dimensions},
 booktitle={Proceedings of the 2020 ACM-SIAM Symposium on Discrete Algorithms (SODA)},
 pages={637--649}, year={2020}, doi={10.1137/1.9781611975994.39}
}

@inproceedings{christiani,
 author={Tobias Christiani},
 title={A Framework for Similarity Search with Space-Time Tradeoffs Using Locality-Sensitive Filtering},
 booktitle={Proceedings of the Twenty-Eighth Annual ACM-SIAM Symposium on Discrete Algorithms (SODA)},
 pages={31--46}, year={2017}, doi={10.1137/1.9781611974782.3},
 url={https://arxiv.org/abs/1605.02687}
}

@inproceedings{harpeledjones2019,
 author={Sariel Har-Peled and Mitchell Jones},
 title={Journey to the Center of the Point Set},
 booktitle={35th International Symposium on Computational Geometry (SoCG)},
 series={Leibniz International Proceedings in Informatics}, volume={129},
 pages={41:1--41:14}, year={2019}, doi={10.4230/LIPIcs.SoCG.2019.41},
 url={https://drops.dagstuhl.de/entities/document/10.4230/LIPIcs.SoCG.2019.41}
}

@inproceedings{goel2001,
 author={Ashish Goel and Piotr Indyk and Kasturi Varadarajan},
 title={Reductions Among High Dimensional Proximity Problems},
 booktitle={Proceedings of the Twelfth Annual ACM-SIAM Symposium on Discrete Algorithms (SODA)},
 pages={769--778}, year={2001},
 url={https://ti.inf.ethz.ch/ew/lehre/ApproxGeom06/paper/goel_01.pdf}
}

@inproceedings{ajz2026,
 author={Alexandr Andoni and Shunhua Jiang and Stepan Zharkov},
 title={Approximate Orthogonal Vectors and Diameter via Regularity Lemma},
 booktitle={Proceedings of the 58th Annual ACM Symposium on Theory of Computing (STOC)},
 pages={1477--1488}, year={2026}, doi={10.1145/3798129.3800856},
 url={https://doi.org/10.1145/3798129.3800856}
}

@article{hausslerwelzl,
 author={David Haussler and Emo Welzl},
 title={Epsilon-Nets and Simplex Range Queries},
 journal={Discrete \& Computational Geometry}, volume={2},
 pages={127--151}, year={1987}, doi={10.1007/BF02187876}
}

@article{vc,
 author={Vladimir N. Vapnik and Alexey Ya. Chervonenkis},
 title={On the Uniform Convergence of Relative Frequencies of Events to Their Probabilities},
 journal={Theory of Probability \& Its Applications}, volume={16}, number={2},
 pages={264--280}, year={1971}, doi={10.1137/1116025}
}

@book{gls,
 author={Martin Gr{\"o}tschel and L{\'a}szl{\'o} Lov{\'a}sz and Alexander Schrijver},
 title={Geometric Algorithms and Combinatorial Optimization},
 publisher={Springer}, series={Algorithms and Combinatorics}, volume={2},
 year={1988}
}

@inproceedings{alrw2017,
 author={Alexandr Andoni and Thijs Laarhoven and Ilya Razenshteyn and Erik Waingarten},
 title={Optimal Hashing-Based Time-Space Trade-offs for Approximate Near Neighbors},
 booktitle={Proceedings of the Twenty-Eighth Annual ACM-SIAM Symposium on Discrete Algorithms (SODA)},
 pages={47--66}, year={2017}, doi={10.1137/1.9781611974782.4},
 url={https://arxiv.org/abs/1608.03580}
}

@inproceedings{chan2016kernels,
 author={Timothy M. Chan},
 title={Dynamic Streaming Algorithms for Epsilon-Kernels},
 booktitle={32nd International Symposium on Computational Geometry (SoCG)},
 series={Leibniz International Proceedings in Informatics}, volume={51},
 pages={27:1--27:11}, year={2016}, doi={10.4230/LIPIcs.SoCG.2016.27},
 url={https://drops.dagstuhl.de/entities/document/10.4230/LIPIcs.SoCG.2016.27}
}

@misc{psw2026,
 author={Ashwin Padaki and Krish Singal and Erik Waingarten},
 title={A Near-Optimal Space Lower Bound for {Euclidean} Diameter Estimation in Dynamic Streams},
 year={2026}, note={arXiv:2609.19089},
 url={https://arxiv.org/abs/2609.19089}
}

@inproceedings{kpsw2025,
 author={Sanjeev Khanna and Ashwin Padaki and Krish Singal and Erik Waingarten},
 title={A Polynomial Space Lower Bound for Diameter Estimation in Dynamic Streams},
 booktitle={2025 IEEE 66th Annual Symposium on Foundations of Computer Science (FOCS)},
 year={2025},
 note={Full version: arXiv:2510.04918},
 url={https://arxiv.org/abs/2510.04918}
}

@inproceedings{hmv2025,
 author={Magn{\'{u}}s M. Halld{\'{o}}rsson and Nicolaos Matsakis and Pavel Vesel{\'{y}}},
 title={Streaming Diameter of High-Dimensional Points},
 booktitle={33rd Annual European Symposium on Algorithms (ESA)},
 series={Leibniz International Proceedings in Informatics}, volume={351},
 pages={58:1--58:10}, year={2025}, doi={10.4230/LIPIcs.ESA.2025.58},
 url={https://drops.dagstuhl.de/entities/document/10.4230/LIPIcs.ESA.2025.58}
}

@inproceedings{expiration2026,
 author={Lotte Blank and Sergio Cabello and Mohammad Taghi Hajiaghayi and Robert Krauthgamer and Sepideh Mahabadi and Andr{\'e} Nusser and Jeff M. Phillips and Jonas Sauer},
 title={The Expiration Streaming Model: Diameter, {$k$}-Center, Counting, Sampling, and Friends},
 booktitle={53rd International Colloquium on Automata, Languages, and Programming (ICALP)},
 series={Leibniz International Proceedings in Informatics}, volume={374},
 pages={37:1--37:24}, year={2026}, doi={10.4230/LIPIcs.ICALP.2026.37},
 url={https://arxiv.org/abs/2509.07587}
}

@inproceedings{abboudronzewi2024,
 author={Elie Abboud and Noga Ron-Zewi},
 title={Finer-Grained Reductions in Fine-Grained Hardness of Approximation},
 booktitle={51st International Colloquium on Automata, Languages, and Programming (ICALP)},
 series={Leibniz International Proceedings in Informatics}, volume={297},
 pages={7:1--7:17}, year={2024}, doi={10.4230/LIPIcs.ICALP.2024.7},
 url={https://drops.dagstuhl.de/entities/document/10.4230/LIPIcs.ICALP.2024.7}
}

@inproceedings{harpeledzhou2021,
 author={Sariel Har-Peled and Timothy Zhou},
 title={Improved Approximation Algorithms for {Tverberg} Partitions},
 booktitle={29th Annual European Symposium on Algorithms (ESA)},
 series={Leibniz International Proceedings in Informatics}, volume={204},
 pages={51:1--51:15}, year={2021}, doi={10.4230/LIPIcs.ESA.2021.51},
 url={https://arxiv.org/abs/2007.08717}
}

@article{bathielagarde2025,
 author={Gabriel Bathie and Guillaume Lagarde},
 title={A $(1+\varepsilon)$-Approximation for Ultrametric Embedding in Subquadratic Time},
 journal={Proceedings of the AAAI Conference on Artificial Intelligence},
 volume={39}, number={15}, pages={15516--15523}, year={2025},
 doi={10.1609/aaai.v39i15.33703}, url={https://arxiv.org/abs/2503.13409}
}

@inproceedings{wei2019lasvegas,
 author={Alexander Wei},
 title={Optimal {Las Vegas} Approximate Near Neighbors in {$\ell_p$}},
 booktitle={Proceedings of the Thirtieth Annual ACM-SIAM Symposium on Discrete Algorithms (SODA)},
 pages={1794--1813}, year={2019},
 doi={10.1137/1.9781611975482.108},
 url={https://arxiv.org/abs/1807.07527}
}

@misc{jly2026,
 author={Cheng Jiang and Yinchen Liu and Huacheng Yu},
 title={Turnstile Streaming Algorithms Might (Still) as Well Be Linear Sketches, for Polynomial-Length Streams},
 year={2026}, eprint={2604.22052}, archivePrefix={arXiv},
 note={Version 2, October 30, 2026; accepted to FOCS 2026},
 url={https://arxiv.org/abs/2604.22052}
}

@inproceedings{mccauley2024,
 author={Samuel McCauley},
 title={Improved Space-Efficient Approximate Nearest Neighbor Search Using Function Inversion},
 booktitle={32nd Annual European Symposium on Algorithms (ESA)},
 series={Leibniz International Proceedings in Informatics},
 volume={308}, pages={88:1--88:19}, year={2024},
 doi={10.4230/LIPIcs.ESA.2024.88},
 url={https://drops.dagstuhl.de/entities/document/10.4230/LIPIcs.ESA.2024.88},
 eprint={2407.02468}, archivePrefix={arXiv}
}

@misc{kuszmaul2026,
 author={John Kuszmaul and William Kuszmaul},
 title={Dynamic and Optimal Function Inversion in the Small-Time Regime},
 year={2026}, eprint={2609.20240}, archivePrefix={arXiv},
 doi={10.48550/arXiv.2609.20240},
 url={https://arxiv.org/abs/2609.20240}
}

@inproceedings{bbm2026,
 author={Jackson Bibbens and Levi Borevitz and Samuel McCauley},
 title={Space-Efficient Text Indexing with Mismatches using Function Inversion},
 booktitle={Proceedings of the 58th Annual ACM Symposium on Theory of Computing (STOC)},
 pages={1061--1072}, year={2026},
 doi={10.1145/3798129.3800818},
 eprint={2604.01307}, archivePrefix={arXiv},
 url={https://arxiv.org/abs/2604.01307}
}

@inproceedings{aronov2024implicit,
 author={Boris Aronov and Jean Cardinal and Justin Dallant and John Iacono},
 title={A General Technique for Searching in Implicit Sets via Function Inversion},
 booktitle={2024 Symposium on Simplicity in Algorithms (SOSA)},
 pages={215--223}, year={2024},
 doi={10.1137/1.9781611977936.20},
 eprint={2311.12471}, archivePrefix={arXiv},
 note={Full version revised April 22, 2026; journal version in Algorithmica},
 url={https://arxiv.org/abs/2311.12471}
}

@misc{vannimmen2026,
 author={Mathieu Vannimmen},
 title={From Function Inversion to Preimage Reporting},
 year={2026}, eprint={2611.09656}, archivePrefix={arXiv},
 note={Version 1, November 6, 2026},
 url={https://arxiv.org/abs/2611.09656}
}
\endgroup

\clearpage
\appendix

\section{Exact construction and certification of lifted centers}
\label{sec:paper-center-constructions}

The geometric argument in Section~\ref{sec:paper-geometry} needs a
point of positive depth in the lifted space.  It also needs a way to
recover an actual data point whose distance realizes the variance
bound.  This appendix supplies both computational ingredients.
We first construct a rational centerpoint from a small sample.
We then turn a candidate center into a collection of disjoint,
explicitly checked convex combinations.  These combinations certify
the distance inequalities even if the random centerpoint computation
was unsuccessful.  The finite-search certificates in
Appendix~\ref{sec:paper-search-certificates} will serve the same purpose
for the angular and inverse-search structures.

Recall that $z\in\R^D$ has depth at least $\alpha|P|$ in $P$ if
every closed halfspace containing $z$ contains at least
$\alpha|P|$ input identities.  An $\alpha$-net for halfspaces is a
sample that meets every halfspace containing at least an
$\alpha$ fraction of $P$.  We count identities, so repeated
coordinates cause no difficulty.  A convex-hull certificate consists
of nonnegative weights summing to one whose weighted sum is $z$.
Disjoint groups with a common convex combination are often called
Tverberg partitions; related constructions are studied
in~\cite{harpeledzhou2021}.

For arbitrary real input in the arithmetic-RAM model, the first
constructor may be replaced by Har-Peled--Jones~\cite[Theorem~18]{harpeledjones2019}.
In dimension $D$, setting their constant accuracy parameter to $3/4$
gives depth $1/(4(D+2)^2)$ in
$O(D^7\log^3(D+2)\log^3(2/\varphi))$ arithmetic operations,
with failure probability at most $\varphi$ and uniform sampling access.
After lifting, $D=d+1$.  The rational constructions below additionally
bound the number of bits in all computed coordinates and permit exact
verification.  They use polynomial-time rational linear programming;
no strongly polynomial algorithm for general linear programming is
needed.

\paragraph{Why intersect sampled convex hulls?}
A sufficiently large random sample contains a fixed centerpoint in its
convex hull.  Intersecting several such hulls therefore gives a
feasible linear program with high probability.  Conversely, any point
in their intersection is contained in a halfspace only if that
halfspace meets every sampled group.  Uniform convergence turns this
sample count into a depth guarantee for the entire input.

\begin{lemma}[A polynomial-size exact centerpoint program]
\label{lem:rational-centerpoint}
Given $D\ge1$, uniform sampling access to a nonempty identified population
$P\subseteq\Q^D$ with coordinate encoding length at most $B$, and
$0<\varphi<1/2$, one can compute an exact rational
$\alpha$-centerpoint with probability at least $1-\varphi$.
Here $\alpha^{-1}=\poly(D,\log(1/\varphi))$; the number of
uniform-sampling calls, the additional bit work, and the output
encoding length are polynomial in $D,B,\log(1/\varphi)$.
The cost of implementing sampling access is charged separately.
\end{lemma}

\begin{proof}
The proof has three parts: construct the intersection program, show
that it is feasible, and show that every feasible solution has the
required depth.

\paragraph{Sampling and the linear program.}
Put $L=\log(2(D+2)/\varphi)$ and choose a sufficiently large absolute
constant $C$.  Draw $rs$ independent uniform sample indices, where
\[
 s=\lceil C(D+1)^2L\rceil,\qquad
 r=\lceil C(D+1)sL\rceil,
\]
and divide them into $r$ blocks of $s$ indices.  Multiplicities are
retained.  Find a common point of their convex hulls by the program
\begin{equation}\label{eq:paper-center-lp}
 z=\sum_{j=1}^s\lambda_{ij}p_{ij},\qquad
 \sum_{j=1}^s\lambda_{ij}=1,\qquad
 \lambda_{ij}\ge0\quad(1\le i\le r).
\end{equation}
Report failure if it is infeasible.

\paragraph{Feasibility.}
The centerpoint theorem guarantees a point $z^*$ of depth
$|P|/(D+1)$ in $P$.  A halfspace
$1/(D+1)$-net of $P$ has $z^*$ in its convex hull, by strict
separation.  The halfspace-net theorem~\cite{hausslerwelzl} implies
that one block is such a net with failure at most $\varphi/(2r)$;
the stated $s$ suffices because $\log(r/\varphi)=O(L)$.  Thus the
program is feasible with probability at least $1-\varphi/2$.
\paragraph{Depth of every feasible solution.}
Let $\mu(H)=|P\cap H|/|P|$ be the fraction of input identities in a
halfspace $H$, and let $\widehat\mu(H)$ be the corresponding fraction
among all $rs$ sampled occurrences.  The VC uniform-convergence theorem
for halfspaces gives, with probability at least $1-\varphi/2$,
\[
 \sup_H|\widehat\mu(H)-\mu(H)|\le1/(2s),
\]
provided $rs\ge C_0s^2((D+1)\log(s+1)+\log(2/\varphi))$;
our choices meet this inequality~\cite{vc}.  Every closed halfspace
containing a feasible $z$ meets each block, so its empirical mass is
at least $r/(rs)=1/s$.  Its original mass is therefore at least
$1/(2s)$.  Set $\alpha=1/(2s)$.  These two events hold simultaneously
with probability at least $1-\varphi$ by a union bound.  The uniform
statement applies to the solution selected by the program, even though
that solution depends on the entire sample.

\paragraph{Exact arithmetic.}
The program has polynomially many variables and only original input
coordinates as nonconstant coefficients.  Exact rational linear
programming gives both the stated running time and polynomial
output precision~\cite{gls}.  In particular, no long sequence of
geometric eliminations can inflate the coordinates of the center.
For a failed or infeasible attempt the dynamic structure uses its
deterministic contraction fallback; no depth claim is made then.
\end{proof}

\paragraph{From a likely center to a verifiable certificate.}
A sampled center can be wrong, but a displayed convex combination can
be checked exactly.  We therefore partition the original birth pool
into disjoint groups and retain only groups whose lifted convex hulls
contain the candidate center.  Accepting sufficiently many groups
proves the needed depth deterministically.  Disjointness is crucial:
one deletion can invalidate at most one retained group.

\begin{lemma}[Checked disjoint block certificates]
\label{lem:certified-lifted-center}
There is $s_0=\poly(d)$ such that, on a birth pool $P_0$ of
$m\ge2s_0$ rational points, a Las Vegas algorithm returns
$z=(o,s)$ and at least $m/(4s_0)$ disjoint blocks of exactly $s_0$
birth identities whose lifted convex hulls contain $z$.
It returns exact rational convex coefficients for every block.
Including fair random-bit generation, the expected construction
time and peak space are respectively
$m\poly(d,B,\log(2m))$ bit operations and
$m\poly(d,B,\log(2m))$ bits. Coordinate and coefficient encoding
lengths are polynomial in $d,B$.  Every returned certificate is
correct, and the algorithm terminates almost surely.
\end{lemma}

\begin{proof}
We generate a candidate, verify it against disjoint groups, and repeat
only when too few groups pass.  The verification gives correctness;
the probability analysis bounds the expected number of attempts.

\paragraph{Candidate and group size.}
In dimension $D=d+1$, first use
Lemma~\ref{lem:rational-centerpoint} with failure probability $1/10$.
With probability at least $9/10$ its candidate has depth
$\alpha_0=1/\poly(D)$ in the lifted pool.  Choose
\[
 s_0=\left\lceil C\alpha_0^{-1}
               ((D+1)\log(2/\alpha_0)+1)\right\rceil.
\]
A uniform $s_0$-element subset is an $\alpha_0$-net for halfspaces
with probability at least $9/10$.  This follows also from the usual
sampling-with-replacement theorem: discard repetitions and fill the
sample with uniformly chosen unused identities.  The resulting
set is uniform, and adding points preserves the net property.

\paragraph{Exact verification.}
Generate a fresh uniform random permutation of the birth identities
by Fisher--Yates shuffling, using exact rejection sampling for each
uniform integer. The shuffle uses $O(m\log(2m))$ expected fair
random bits. Partition its first $rs_0$ positions into
$r=\lfloor m/s_0\rfloor$ blocks.
For each block solve~\eqref{eq:paper-center-lp} with $z$ now fixed,
using the lifted coordinates.  Accept the attempt only if at least
$\lceil r/2\rceil$ blocks pass.  Otherwise restart the entire attempt.
\paragraph{Expected number of attempts.}
Conditioned on a genuinely deep candidate, each block fails with
probability at most $1/10$.  Markov's inequality, without any
independence between blocks, shows that at least half pass with
probability at least $4/5$.  An attempt therefore succeeds with
probability greater than $1/2$.  Each attempt uses $m\poly(d,B,\log(2m))$ expected bit work,
including the shuffle, with the stated peak space. Fresh retries
give the expected bound and almost-sure termination. Workspace is
reused after each rejected attempt; no sequence of failed attempts
can accumulate retained memory.

\paragraph{Validity of an accepted answer.}
Every accepted block is checked by an exact LP, whether or not the
candidate was deep.  Since $m\ge2s_0$, the number retained is at
least $m/(4s_0)$.  The LPs use original lifted coordinates and the
one candidate, so their rational output lengths remain polynomial.
\end{proof}

\paragraph{Example and behavior under deletions.}
For four points $(1,0),(-1,0),(0,1),(0,-1)$ in the plane, the two
opposite pairs are disjoint groups.  Their lifts both have
$z=(0,0,1)$ as their midpoint.  Thus the two pairs certify the same
center $o=(0,0)$ and variance $V=1$.  Deleting one input point
invalidates only its pair; the other pair remains an exact certificate.
The construction above gives many such disjoint
certificates in higher dimensions, without requiring symmetry.

These certificates establish depth $\beta m$, where
$\beta=1/(4s_0)$, without a probability qualification: every closed
halfspace containing $z$ meets every retained block.  They also prove
$V=s-\|o\|^2\ge0$ and~\eqref{eq:paper-certified-average} directly
from their coefficients.  During at most $\beta m/16$ input changes,
at least $15\beta m/16$ blocks remain intact.  Each supplies a core
point and an exact witness for every query.  Inserted identities
are never added to an old certificate, and a reinsertion receives
a new lifetime, so it cannot revive an invalidated block.
If desired, Carath\'eodory reduction compresses each retained block
to at most $d+2$ identities. It does not create additional blocks:
the certified depth and deletion budget remain governed by
$\beta=1/(4s_0)$ for the original polynomial block size $s_0$.

\paragraph{Failure probabilities and rebuilds.}
For the finite-horizon structure, fix the per-construction probability
before defining the depth.  A choice
$\varphi=\delta/(n+t+d+B+1)^C$ with sufficiently large constant $C$
dominates the polynomial number of centers and witness samples in
the scheduled construction; the final composition includes all its
additional parameters in this denominator.  This avoids a circular
choice involving the random hierarchy depth.  For the zero-error
structure, the constructor above has constant success probability
independently of the current history, and requires no horizon.
Conditional expected rebuild costs can therefore be charged to the
deterministic local lifetimes.  Randomness affects only construction
time, not any accepted geometric certificate or cached pair.

\section{Exact certificates and finite representations}
\label{sec:paper-search-certificates}

The zero-error structure must justify two statements by finite tests:
every target angular pair shares a filter label, and an inverse query
recovers every input with its requested label.  A high-probability
claim for a fixed query would not suffice, because future queries
may depend on the accepted random choices.

We first check angular coverage on a sufficiently fine finite net.
A margin in each inequality extends the checked statement to every
point on the sphere.  We then explain how to represent the rotations
and comparisons exactly on rational input.  Finally, we certify
inverse recovery by comparing the query routine's output with the
exact fibers in small batches.  These are the implementation details
used in Proposition~\ref{prop:finite-word-lv-angular} and
Lemma~\ref{lem:lv-dynamic-inversion}.  Gaussian probability estimates
bound expected work; correctness follows from the accepted certificates
on every outcome.

\subsection{A bounded finite Gaussian sampler}
\label{sec:bounded-gaussian-sampling}

We make precise the finite sampling used in the explicit Monte Carlo
index and in the probability analysis of the certified construction.
It requires no exact inverse-normal operation or comparison with a
transcendental rounding boundary.

For one amplified explicit birth index, let $N_G$ be a deterministic
upper bound on its number of scalar Gaussian draws and $N_C$ a bound
on all cap comparisons against its birth vectors and finite query
net. These counts are fixed before drawing the index's bits. Only
$\log N_C$ enters the precision bound; the query net is not enumerated.
Let $\varepsilon_0$ be the assigned coupling budget and put $D=d+1$.
Choose a rational clipping threshold $T\ge1$ and rational
$\Delta>0$ satisfying
\[
 2N_G e^{-T^2/2}\le\varepsilon_0/4,
 \qquad N_C\Delta\le\varepsilon_0/4.
\]
Conservative integer choices have $T^2=O(\log(2N_G/\varepsilon_0))$.
Set $e_G=\Delta/(4D)$ and
$\lambda=2^{-2(\lceil T\rceil+1)^2-3}$.
The standard normal density is at least $\lambda$ on $[-T,T]$,
so the inverse CDF clipped to this interval, denoted by $Q_T$,
is $\lambda^{-1}$-Lipschitz on $[0,1]$.

Choose an integer $b$ with $2^{-b}/\lambda\le e_G$, draw a uniform
$K\in\{0,\ldots,2^b-1\}$ using exactly $b$ fair bits, and set
$u_K=(K+1/2)2^{-b}$. Compute a rational $g_K$ within $e_G$ of
$Q_T(u_K)$ by interval bisection on $[-T,T]$. At a midpoint,
approximate the normal CDF with certified error at most
$\lambda e_G/8$. A separated comparison selects the corresponding
half interval. If the comparison intervals overlap, the density
lower bound already certifies that the midpoint is within $e_G$
of the clipped quantile, and it can be returned. Otherwise stop
once the remaining coordinate interval is short enough. There are
$O(\log(T/e_G))$ iterations. Rational Taylor approximations to the
Gaussian density, integrated on the bounded interval, evaluate its
CDF to the required error in time polynomial in
$T$ and $\log(1/(\lambda e_G))$. The normalization constant can
likewise be approximated by certified rational intervals. Thus every
bit seed gives a rational answer in bounded finite work.

For the analysis, couple $K$ to a uniform real $U\in(0,1)$ by
$K=\lfloor2^bU\rfloor$ and set $G=\Phi^{-1}(U)$. This floor is
only a map in the probability argument. Outside $|G|>T$,
\[
 |g_K-G|\le e_G+\lambda^{-1}2^{-b}\le2e_G.
\]
Independent coordinatewise couplings give vector error at most
$2De_G\le\Delta/2$. For any fixed unit vector $u$ and threshold
$a$, the ideal projection is standard normal, and
\[
 \Pr[|\langle G,u\rangle-a|\le\Delta]
 \le 2\Delta/\sqrt{2\pi}<\Delta.
\]
A union bound over all $N_C$ comparisons and $N_G$ coordinate tails
therefore preserves all birth-vector and query-net cap outcomes
with probability at least $1-\varepsilon_0/2$. Rational threshold
approximations, if needed, use the remaining $\Delta/2$ margin.
The bit budgets are polynomial in
$d,\log(2m/\varepsilon_0),\log(2/e)$ for the explicit index.
For certified filters the same sampler is used at their stated
fixed-pair precision; deterministic finite coverage tests, rather
than a continuum coupling event, establish correctness.

\subsection{The finite splitter and the small-sphere certificate}

A direct net of the whole sphere would be too large.  We first divide
a moderate-dimensional block into small blocks, using a finite family
of linear maps that includes a suitable map for every relevant pair.
We can then afford to check a net in each small block.  Recall the
angular parameters: a target pair has inner product at least $h>0$,
the data and query thresholds are $t$ and $xt$, and $h\le x\le1$.
Here $t$ denotes the Gaussian threshold, independently of the execution
horizon elsewhere in the paper.

\paragraph{Finite norm splitters.}
For moderate dimension $B_0$ and small dimension $b$, both powers of
four, let $s=B_0/b$. The splitting family in
Wei~\cite[Proposition~11]{wei2019lasvegas} can be represented exactly.
At a halving step, pair coordinates by a two-wise independent
permutation and take signed sums and differences divided by
$\sqrt2$, using four-wise independent signs. The variance of a
rescaled squared norm is $O(1/k)$ when the smaller dimension is $k$.
Chebyshev's inequality and a union bound over three vectors give a
seed preserving both halves with error $O(k^{-1/2})$. Recurse on
both halves. Along a path the errors sum to $O(b^{-1/2})$, and the
number of finite seed choices is
$\exp(O(s\log B_0))$. Affine permutations and degree-three sign
hashes give explicit seed families. Every final matrix is an integer
matrix times $s^{-1/2}$, which is rational because $s$ is a power
of four. The gate tests use squared norms exactly.

\paragraph{A finite test for coverage of a small sphere.}
Fix one small block. Write $\epsilon=1000/\sqrt b$ and
$t_0=t/\sqrt s$; the cutoff ensures $\epsilon<h/10$ and $t_0\ge1$.
Use the finite net
\[
 \mathcal Z=\{a/\|a\|:a\in\mathbb Z^b\setminus\{0\},
                                  |a_i|\le b^3\}.
\]
Rounding $b^3z$ coordinatewise and renormalizing gives net error
at most $\nu=b^{-2}$. Thus
$\log|\mathcal Z|=O(b\log b)$. Choose rational
$G=O(\sqrt{\log m})$ so that an ideal $b$-dimensional Gaussian
has norm greater than $G/2$ with probability at most $m^{-100}$.
A sampled family is accepted only if every coefficient vector has
norm at most $G$ and every ordered net pair of correlation at least
$h-3\epsilon$ has one vector $g$ satisfying
\[
\langle g,v_0\rangle\ge t_0+2G\nu,
 \qquad \langle g,z_0\rangle\ge x t_0+2G\nu.
\]
The stronger thresholds reserve room for replacing an arbitrary
sphere point by its nearest net point.  All these tests are exact
algebraic comparisons.  A pair of correlation
at least $h-2\epsilon$ rounds to a checked pair because
$2\nu\le\epsilon$. The norm cap changes each inner product by at
most $G\nu$, proving coverage at the original thresholds. This
argument is deterministic.

\paragraph{Why a random family passes with constant probability.}
We next bound the number of sampled cap vectors needed by this
finite test.  This bound controls construction time and does not
enter the deterministic coverage argument.  Put
\[
 a=h-3\epsilon,\qquad
 c_h=2^{-\lceil100/(1-h)\rceil},\qquad
 p_0=c_h t_0^{-2}\exp(-I_a(x)t_0^2/2).
\]
For correlation $a$, integrate the Gaussian density over
$[t_0+1/t_0,t_0+2/t_0]\times[x t_0+1/t_0,x t_0+2/t_0]$. At $(t_0,x t_0)$ both
density-exponent derivatives lie in $[0,t_0]$, since $a\le x\le1$.
Across the rectangle the linear and quadratic increases are at most
$4$ and $4/(1-h)$. Its area is $t_0^{-2}$ and the density prefactor
is at least $1/(2\pi)$. Thus $p_0$ is a conservative lower bound.
Joint-cap probability increases with correlation. The cutoff also
ensures $3G\nu\le1/t_0$, so this lower bound covers the stronger
thresholds with $3G\nu$ in place of $2G\nu$.

Compute rational $p_{\rm rat}\in[p_0/2,p_0]$ by certified
exponential evaluation, set $L_0=\lceil\log_2(2|\mathcal Z|)\rceil$,
and sample
\[
                      M_0=\lceil32L_0/p_{\rm rat}\rceil
\]
vectors. A union bound over net pairs gives ideal acceptance
probability at least $9/10$. Use finite Gaussian coefficients whose
coordinate error is at most $m^{-40}$ except with probability
$m^{-100}$ per coordinate. Their errors are smaller than the reserved
threshold margin. Since $M_0=m^{o(1)}$, the norm and coupling events
hold jointly with probability $1-o(1)$, and the exact certificate
accepts with probability at least $1/2$. Verification takes
$M_0|\mathcal Z|^2\poly(b,\log m)=m^{o(1)}$ time and space.
All failed attempts have this same space bound and are discarded
before the next attempt.

\paragraph{Expected label counts after acceptance.}
For a fixed full label, conditioning its $s$ independent base
families on these certificates increases probabilities by at most
$2^s$. Apply this bound to the finite unconditioned law first, then
couple the selected coefficients to independent ideal Gaussians.
The finite law is sampled by a clipped rational inverse-normal
transform on a uniform integer grid; monotonicity and certified
function evaluation supply the stated coupling and deterministic
coefficient-length bounds. For any fixed pair, coefficient error
only shifts the final positive thresholds by $m^{-O(1)}$. Gaussian
tail estimates therefore change by a subpolynomial factor. There
are at most $m^3$ raw labels per moderate index above the cutoff;
the coupling exception contributes negligibly even after summing
all label occurrences. These are fixed-pair moment estimates, not
an assertion that one continuous coupling covers every direction.
Continuum coverage already follows from the exact finite certificate.

\subsection{Exact rotations and bounded-degree comparisons}

The preceding certificate assumes exact inner products and exact
orthogonal decompositions.  Rounding the entries of an orthogonal
matrix independently would destroy orthogonality, so it would also
destroy the deterministic target-coverage argument.  We instead round
vectors that define reflections and apply the reflection formula
exactly.  We first establish exact orthogonality, then compare the
resulting distribution with an ideal random rotation to bound work,
and finally bound the cost of all algebraic comparisons.

\paragraph{Exact orthogonality from rational reflections.}
A Haar rotation is an orthogonal matrix drawn from the rotationally
invariant probability distribution.  Generate such a rotation
recursively from Gaussian directions and Householder completions. For a unit direction $u$, choose a stable
sign $s\in\{-1,1\}$ agreeing with the sign of its first coordinate,
form $w_*=u+se_1$, and use $H(w_*)D_s$, where
\[
                         D_s=\operatorname{diag}(-s,1,\ldots,1).
\]
Indeed, $H(w_*)e_1=-su$, so the diagonal factor makes the first
column equal to $u$. Approximate each Householder vector by a
rational vector $w$, keep $D_s$ exactly, and apply
\[
                         H(w)=I-2ww^{\mathsf T}/(w^{\mathsf T}w)
\]
\emph{exactly}. If a rounded vector is zero, use a fixed nonzero
vector. Each actual factor, and hence the full product, is exactly
orthogonal. Coverage of a positive pair therefore cannot fail.

\paragraph{Comparison with the ideal random rotation.}
The comparison is needed only for the expected-work estimates.
For the distributional comparison, the stable vector's norm is at
least one. The sign may be chosen from a sufficiently accurate
approximation to $u_1$; if a near-zero first coordinate changes its
sign, the norm is still bounded below by an absolute constant.
Choose the same sign in the ideal completion, including its diagonal
factor; conditional on its first column, any such
orthogonal completion followed by an independent Haar rotation
of the complement has the Haar law. On the event that all Gaussian
norms exceed $m^{-C}D^{-C}$ and all coordinates have polynomially
bounded magnitude, these operations are polynomially Lipschitz.
Gaussian small-ball and tail bounds make the exceptional probability
at most $m^{-A}$ for any prescribed constant $A$. Polynomially many
bits thus couple the rational product to the ideal one with operator
error at most $\delta/(100\sqrt D)$. Moderate block norms are
$\Omega(\sqrt{B_0/D})$ on the projection event, so normalization
changes correlations by less than the reserved negative margin.

\paragraph{Bit length of the represented rotation.}
Each factor is rounded on a fixed precision grid before its exact
Householder formula is applied. It has a polynomial-length common
denominator. A product of $D$ such factors has denominator length
at most the sum of their lengths, and numerator growth adds only
$O(D\log D)$ bits. This prevents denominator growth from being
hidden in a repeated normalization operation. Projection
concentration uses $B_0=\Theta(\delta^{-2}\log m)$; there is no
exponential dependence on $1/\delta$ in this implementation.

\paragraph{Exact comparisons without repeated normalization.}
All vector tests can be made without storing normalized coordinates.
For example, the common padding maps a rational direction record
$v\ne0$ to $(\alpha v/\|v\|,\beta)$ with rational
$\alpha,\beta$. A cap test is
\[
       \alpha\langle g,v\rangle
                       \ge (t-\beta g_{d+1})\|v\|,
\]
decidable by signs and a squared rational comparison. After multiplying
by $\sqrt r$, where $r=\|v\|^2$, rational rotations and splitter
maps leave coordinates of the form $A+B\sqrt r$. Gate denominators
cancel in squared norm ratios. In a final normalized cap test, move
the one positive norm to the other side and square only after its
sign is known. Thus each vector remains in one fixed-degree field;
no compositum is formed across the dimension. Net-pair and original
weak-edge comparisons involve only a constant number of such radicals.
A conservative degree bound of $32$ suffices for the complete
implementation, and coefficient lengths are polynomial in input
length, dimension, and logarithmic parameters. Exact root isolation
or successive sign-and-square comparisons give polynomial time.
Every generation starts from original rational records and its
rational center; a normalized output is never reused as a new
input record. Coordinate precision therefore does not accumulate
along the dynamic hierarchy.

\subsection{Certifying actual inverse recovery in small workspace}

An inverse structure must recover a complete fiber, not merely a
point that appeared somewhere in a sampled chain.  We verify the
actual recovery routine on every output name, while using only the
workspace allowed by the theorem.  This gives a direct repair
implementation of Lemma~\ref{lem:lv-dynamic-inversion}.

Recall that a fiber of a function $F$ at $y$ is
$F^{-1}(y)=\{i:F(i)=y\}$.  The derived function $F'_0$ has domain
of size $O(M)$, each nonnull fiber has at most $\tau$ members, and
one evaluation costs $E$.  A null value, denoted by $\bot$, records
an inactive virtual input.

\paragraph{Recover exact fibers in batches.}
Freeze the derived bounded-fiber function $F'_0$ and retain the prescribed raw chains
from~\cite[Section~4.3.1]{bbm2026}, initially without a repair list.
Process derived output names in batches of $\Theta(M/\tau^2)$.
A full virtual-domain scan stores all fibers for one batch within
$O(M/\tau)$ space, since each fiber has at most $\tau$ members.
There are $O(\tau^2)$ scans, costing
$\widetilde O(M\tau^2E)$.

\paragraph{Treat null inputs separately.}
The source handles null explicitly inside each chain hash. With
$M'$ the constant-padded derived domain, take
$g:[2M']\to[M']$ and iterate
\[
 H(i)=\begin{cases}
       g(F'_0(i)),&F'_0(i)\ne\bot,\\
       g(M'+i),&F'_0(i)=\bot.
      \end{cases}
\]
Thus the null inputs have private tags before hashing; an arbitrary
null fiber does not become a high-indegree chain vertex. This is
exactly the null convention in~\cite[Section~4.3.1]{bbm2026}.

\paragraph{Check recovery and store only omissions.}
For each name $y$, run its capped raw-chain recovery, validate each
reported identity by $F'_0(i)=y$, remove duplicates, and compare
with its exact stored fiber. Add exactly the missing identities
to a repair dictionary keyed by $y$. Recovery starts at $g(y)$
under the preceding chain map; the validation is on the
original derived value $y$. This fixes the composition convention
at chain endpoints and collisions. Reject an attempt when the repair
dictionary exceeds $CM/\tau$ identities. The fixed-target bounds
of~\cite[Lemmas~24--25]{bbm2026} imply an expected $O(M/\tau)$
missing identities for these actual recoveries, so sufficiently
large $C$ leaves constant acceptance probability.

\paragraph{Correctness and resource bounds.}
Every accepted raw-query-plus-repair structure now returns each
complete fiber deterministically. All $O(M)$ recovery checks cost
$\widetilde O(M\tau^3E)$, while batches, raw chains, repairs, and
one duplicate buffer use $\widetilde O(M/\tau)$ space. No forward
array of length $M$ is used in this regime. For
$\tau<C\log^2(2M)$ one may instead store explicit inverse lists:
$M=\widetilde O(M/\tau)$, and oracle evaluations and virtual
identities contribute only logarithmic factors. Thus this small
parameter case does not increase the final space exponent.

\section{Dynamic schedules and zero-error accounting}
\label{app:dynamic-schedules}
This appendix supplies the construction schedules and adaptive running-time
argument used in Section~\ref{sec:paper-dynamization}. There are three
separate tasks. First, a new inverse phase must be constructed and brought
up to date while the old phase is still valid. Second, an entire
center chain must be replaced without exhausting the age budget of its
small residuals. Third, the expected-time analysis must remain valid
when later updates depend on exposed random choices. The next three
subsections address these tasks in that order.

\subsection{A certified inverse with worst-case maintenance}
\label{sec:density-inverse-deamortization}

The clock in this subsection counts \emph{value deletions}: one tick
sets the forward-map value of one virtual identity to null. An external
point deletion can produce up to $R$ such ticks, all completed before
the next query is exposed. The inverse parameter $\tau$ controls the
space--time tradeoff and is distinct from the number of external updates.
We first schedule a single virtual inverse, then account for these $R$
virtual changes and for replay during the outer frozen-generation build.

The implicit search lemma supplies the following phase interface.
For a virtual domain of size $M$, parameter $\tau\le M^{1/3}$, and
forward-evaluation time $E$, a verified immutable inverse can be
built in expected $\widetilde O(M\tau^3E)$ time and
$\widetilde O(M/\tau)$ space. Between reconstructions it processes a
value deletion in $D=\widetilde O(\tau^3E)$ worst-case time and answers
predicate searches with the proved $(Q,F)$ bound. It remains valid
for a phase containing a constant multiple of $M/\tau$ value deletions.
All fibers used by a completed phase are certified, so these statements
are deterministic once construction finishes.

For a finite horizon, cap one complete construction attempt at a
sufficiently large constant times its conditional expectation.
This cap includes every retry, verification, and cleanup operation.
Markov's inequality gives success probability at least $1/2$.
Run $K$ independent capped attempts, sequentially, and retain the
first success. The resulting budget is
$B_0=\widetilde O(KM\tau^3E)$, with failure at most $2^{-K}$.
The same rule is applied to the entire certified-family and
point-assignment constructor, not only to the inverse. Conditioning
on one accepted birth group inflates the required geometric moments
by a constant factor. Later phase failures are union bounded
separately; we never condition those moments on every future rebuild
succeeding.

\begin{lemma}[Pipelined inverse phases]
\label{lem:paper-inverse-schedule}
With the preceding capped constructor, one current value deletion
uses $\widetilde O(K\tau^4E)$ worst-case work and
$\widetilde O(M/\tau)$ total space. Except when a cap fails, every
phase oracle and every answer remain exact.
\end{lemma}

\begin{proof}
The construction has three stages: retain an immutable birth oracle,
build the new inverse against that oracle, and replay the intervening
deletions before switching service to the new phase. The overlap is
short enough that the serving phase remains valid throughout.

Let $H=\lfloor M/(64\tau)\rfloor$, increasing the fixed small-instance
cutoff if necessary. Start a new phase every $H$ value-deletion ticks.
Its immutable oracle is the function at its starting tick. While
construction runs, store the first old value of every changed identity
in a dictionary. An oracle call consults that dictionary before the
current function, exactly reconstructing the immutable snapshot.
Give the constructor $\lceil8B_0/H\rceil$ work per tick, so it finishes
within $H/4$ ticks.

Next replay the queued deletions, with work for four deletions per
tick. New arrivals append to the queue. The replay oracle returns null
on applied deletions, the saved old value on unapplied deletions, and
the current value elsewhere. Mark a deletion applied \emph{before}
notifying the new inverse. This order ensures that any reconstruction
triggered by that notification sees the correct intermediate function.
A backlog of at most $H/4$ drains within $H/12+1$ further ticks.
The new phase is ready before $H/2$ ticks and can replace the old one
before its age reaches $3H/2$. This lies within its validity interval.

At most two phases and their retirement work coexist. Their old-value
dictionaries, replay queues, and sparse bucket remnants have
$O(H)=O(M/\tau)$ records. Construction contributes
$O(B_0/H)=\widetilde O(K\tau^4E)$ work per tick, and replay contributes
$O(D)$. A failed cap disables that copy within the same resource
bound; it is an error event, not an unbounded computation.
\end{proof}

An original point has at most $R$ virtual ranks. Delete them in order
using one pending point and one rank cursor. Record each old value in
every applicable snapshot log, then advance the cursor before notifying
the inverse; only after the last notification mark the point inactive
in this generation. Queries do not interleave with this operation.
The point's coordinates and label metadata remain accessible to every
phase that still refers to them. Thus the worst-case original-deletion
bound is $\widetilde O(RK\tau^4E)$, without a bitmap of size $mR$.

There is also an outer replay when a frozen insertion interval is
constructed. Retain its persistent root in constant time. Its birth
oracle remains immutable throughout initialization; arriving deletions
are queued by \emph{original lifetime identity}, even before the scan
has assigned its local point number. After initialization, replay those
deletions using an arrival/applied bitmap and the single rank cursor.
An arrival does not change this internal current oracle until replay
applies it. There are at most $m$ original deletions. The initial build
and all possible replay therefore cost at most $CmA(n)\mathcal P$. Give the job
a sufficiently large multiple of $A(n)\mathcal P$ per outer tick and declare it
ready as soon as initialization is complete and its queue is empty.
Every tick while it is unready pays useful work, so it finishes within
$m\le w$ ticks. This proves the constructor deadline required by
Lemma~\ref{lem:paper-frozen-dynamization}.

\subsection{Replacing a chain without aging its small residuals}
\label{sec:paper-center-schedule}

A replacement chain may contain pools much smaller than its root.
If every center were sampled when the root job started, a small
residual could exhaust its age budget before the whole chain became
ready. We therefore open each residual level only when the time
remaining until activation is proportional to that residual's own size.

Here a clock tick of a pool is an external point insertion or deletion
in that pool, rather than a virtual value deletion. Unaffected pools
receive no tick. Let $\alpha$ be the
depth parameter, $C$ the polynomial cost of computing a fresh
center from a sampled persistent snapshot, and $K_0$ the polynomial
number of fresh samples for a baseline witness. Set $C_*=4096$ and
choose a polynomial leaf cutoff $L$ satisfying
\begin{equation}\label{eq:paper-leaf-cutoff}
 L\ge2^{20}C_*\alpha^{-2},\qquad
 L\ge C+(d+\log n)K_0+64\alpha^{-1}.
\end{equation}
Small leaves are handled exactly.

A serving node born with $M$ input points starts one replacement
job at age $\lfloor\alpha M/64\rfloor$. If its current input size is
$m>4L$, give the job a deadline
$H=\lfloor\alpha m/C_*\rfloor$ further root-pool ticks; otherwise
replace the suffix immediately by an exact leaf. The old node's age
at this deadline is less than $\alpha M/16$, so it remains valid.
Only serving nodes own jobs. A pending job consists of a completed
prefix of fixed balls and one current residual; it starts no nested
replacement jobs inside that prefix.

If $r$ ticks remain and the residual currently has $m$ points, open
its next level only when every previously opened level in that job
has completed construction and $r\le\alpha m/C_*$. Thus a job
has at most one unfinished level. For $m\le2L+1$, close
it as a leaf. Otherwise retain its birth snapshot, compute a center
immediately in the opening update,
and build its radius dictionary and local search indices in the
background. If the sampled program reports infeasibility, or the
returned center has $V<0$, first choose
any birth point as $o$ and provisional $V=0$.
Check explicitly that the birth core contains at least
$\alpha m/2$ points. If a Monte Carlo center fails that check, use
the same first-coordinate center with the maximum birth radius as
the ball; this preserves contraction and all work bounds even on a
failure outcome. Correctness is asserted on the good center event.

\paragraph{The scans and their updates.}
Each pool is a persistent AVL tree, storing subtree sizes and
heights. Every node satisfies $|h_{\rm left}-h_{\rm right}|\le1$;
the height is $O(\log(|S|+1))$. Searches include the final null
node inspection, so lookup, rank selection, insertion, and deletion
take $O(1+\log(|S|+1))$ work, including on an empty pool. A first scan counts the birth core, finds the maximum
birth radius, and indexes current radii. The first two statistics examine
every immutable birth record, even if its lifetime has since ended.
Each lifetime has separate generation-local \emph{heap-indexed} and
\emph{classified} flags. For the current radius dictionary, delete a
record only if heap-indexed, index new insertions immediately, and skip
a scanned insertion if its lifetime is dead or already heap-indexed.
Once the ball is fixed, retain a snapshot of the then-current input
and scan it to classify its exact outside-ball child and construct
the local angular records. An active record is classified only if its
classified flag is unset. Concurrent insertions are classified
immediately; deletions remove classified records. Commit both matching
copies in every applicable configuration, the child entry, and the
classified flag within one external update.
Each local insertion invokes the fully dynamic search interface, whose
own flat jobs are included in its operation bound.

Over $b$ root ticks, these scans and arriving changes take
$O((m+b)A(n)\mathcal P)$ work. Give every pending job a sufficiently large
multiple of $C_*\alpha^{-2}A(n)\mathcal P$ work per affected tick. Each large
level then finishes within
\begin{equation}\label{eq:paper-level-duration}
                         b\le\alpha^2m/(64C_*)
\end{equation}
ticks. The fixed cutoff absorbs setup costs and rounding.

\begin{lemma}[Late opening meets every deadline]
\label{lem:paper-right-alignment}
Every job completes a contracting chain by its deadline. A center
born on $m$ points is at most $\alpha m/C_*$ ticks old when activated,
and one job opens at most one large level in any tick.
\end{lemma}

\begin{proof}
At a first gate crossing, the residual changes size by at most one,
and the remaining time decreases by one. Hence
\[
             \alpha m/C_*-2<r\le\alpha m/C_*.
\]
The explicitly checked core and the age bound give child size
$m'\le(1-\alpha/4)(m+b)$ at completion. Therefore
\[
 (r-b)-\alpha m'/C_*>
 \frac{\alpha^2m}{4C_*}
       -b(1+\alpha/C_*)-2>0,
\]
by~\eqref{eq:paper-leaf-cutoff}--\eqref{eq:paper-level-duration}.
The child's gate is still closed; its later opening is another first
crossing. This proves the invariant inductively and prevents two
large openings in one tick. Each completed prefix contracts by
$1-\alpha/4$, so its depth is $O(\alpha^{-1}\log n)$. At the deadline
the gate must hold, and the first-crossing inequality rules out a
large unopened residual. The terminal leaf has size at most $3L+2$.
Every center's age is bounded by the number of root ticks remaining
when it was opened, proving the activation-age assertion.
\end{proof}

\begin{algorithm}[t]
\caption{One external update and its cached diameter answer}
\label{alg:paper-external-update}
\begin{algorithmic}[1]
\Require{Current serving chain, its replacement jobs, and a valid point update.}
\State Apply the update to the serving pools, radius dictionaries, and local matchings.
\For{each affected serving node's pending replacement job}
 \State Route the same point change through its completed prefix.
 \State Advance its root clock and perform its prescribed construction budget.
 \State Process completed levels and the next gate until quiescent; open a level only after its prefix is ready.
\EndFor
\State Install completed jobs from highest root to lowest; cancel obsolete descendant jobs.
\State Preserve each installed center's birth size and true age.
\State Start due serving-node replacement jobs; handle polynomial-size leaves exactly.
\State Refresh affected baseline pairs and the maximum cached local matching witnesses.
\State Perform the scheduled retirement work.
\State \Return the largest current cached pair, including the exact leaf's pair; use the distinct-identity convention when its length is zero.
\end{algorithmic}
\end{algorithm}

A job is quiescent after all currently enabled completion and gate
actions have been processed. This rule also closes a small terminal
leaf at its deadline. Lemma~\ref{lem:paper-right-alignment} proves
that at most one large level can open in a tick; the implementation
does not assume this conclusion by truncating the loop.

The serving chain changes each shared root membership tree only once.
A pending job then updates its own radius dictionary and lower pools,
without applying that root membership change again.
An ancestor replacement supersedes completed descendant jobs; surviving
ancestor jobs are unaffected by replacement of a descendant. The new
root takes ownership of the shared input pool. Newly installed nodes
retain their true age, counted conservatively by elapsed job-root ticks.
Refreshing all installed caches costs only a polynomial factor because
the chain has polynomial depth. Every affected surviving cache is
refreshed before an answer is exposed, so a deleted identity cannot
remain in a published pair. A leaf growing to $6L+1$ is rebuilt
immediately within the polynomial leaf budget.

Only $D_0=O(\alpha^{-1}\log n)$ serving nodes own jobs, each with
at most $D_0$ pending levels. Their retained birth cardinalities sum
to $n\poly(d,\log n)$, including local frozen generations. Since
$m^{1+\sigma}\le n^\sigma m$, total space is $n^{1+\sigma}\mathcal P$.
Persistent trees and owned index records have bounded outdegree.
Retire roots onto a queue and spend a fixed multiple of allocation
work reclaiming unreachable records. Reference counts or ownership
arenas make each retired record cost constant work plus its outgoing
references. This prevents an activation from synchronously freeing
a large index. Queued records are bounded by the simultaneously live
generations and one additional construction's allocation budget;
their storage is included in the preceding bound.

\subsection{Zero error on an unbounded adaptive stream}

For expected amortized time, replace the late-opening schedule by
synchronous reconstruction at the same constant-fraction age threshold.
Use the certified center constructor and certified angular indices.
The cores are verified, all answers are checked, and every unsuccessful
construction is retried. Rebuilding a suffix born on $M$ points costs
$MA(n)\mathcal P$ in conditional expectation. Its $\Omega(\alpha M)$ preceding
changes pay for that work; the sum of sizes in its contracting chain is
$O(M/\alpha)$. Charging whole suffixes, rather than separately charging
descendants canceled early by an ancestor, gives total birth
cardinality $t\poly(d,\log n)$ over $t$ updates. Fresh constructors
have their bounds conditionally on the entire history at birth.
More explicitly, charge a triggered suffix to the old root's
age-counter updates since its birth. One past external update can
be charged only by nodes on its then-current chain, at most
$O(\alpha^{-1}\log n)$ times. A descendant canceled before its own
trigger incurs no separate charge: its construction was included
in its ancestor's suffix build.

The remaining issue is the cost of adaptive queries to exposed seeds.
A bound for each query fixed before the random choices does not alone
control an adversary that sees those choices and then selects a slow
query. We need a single event on which every direction is fast in at
least one certified copy. On its complement, exact scans preserve
correctness and the finite lifetime of the generation bounds the total
extra work.

The following is the interface provided by the certified search lemmas.
For each direction $z$ in a fixed implicit net, there is a birth-random
variable $Z_z$ of expectation at most $A(n)\mathcal P$ that deterministically
bounds the search work for every later subset, every compatible exact
predicate, and every accepted inverse phase. The variable counts
query labels and rejected incidences against the \emph{whole birth
pool}. It is independent of the later choice of subset. The net is
fixed before the family's fresh coins; it is used only in the analysis.

\begin{lemma}[Adaptive certified generation compiler]
\label{lem:adaptive-certified-generation-compiler}
Under this birth-variable interface, certified search has always-correct
expected total work $tA(n)\poly(d,B,\log n)$ over every fixed finite
adaptive prefix of $t$ external updates. It needs no prescribed horizon.
\end{lemma}

\begin{proof}
Race $r$ independent complete certified copies for a large constant
multiple of $A(n)\mathcal P$ work each. Use the first completed answer; if none
finishes, scan the current pool exactly. Certification makes both
answers correct. For a fixed net point, Markov's inequality bounds
the chance that one copy is slow by $1/4$. Independence and a union
bound give probability at most $|\mathcal N|4^{-r}$ that any direction
has no fast copy. A padded net comparison covers every actual query.
Take $r=O(\log|\mathcal N|+\log(1/\varepsilon_*))$.

A node born on $M$ points causes at most $C_0M$ primitive queries and
generation births before replacement, where $C_0$ is polynomial in
the displayed parameters. Even if a bad generation is deliberately
exploited, it causes at most $C_0^2M^2$ exact-scan work during that
node's lifetime, after enlarging $C_0$. Choose
$\varepsilon_*=(100C_0^3(n+1)^4)^{-1}$. Summing over at most
$C_0M$ births gives expected fallback work $O(M\mathcal P)$ for that node.
At each birth, condition on the preceding history; its existence and
birth pool are already fixed, while its new complete copies are
independent. The tower property therefore justifies this sum even
when earlier bad seeds affect later births.

Charge the resulting lifetime bounds to the total birth cardinality
proved above. Conditional expected bounds for inverse reconstruction
and certified center retries are charged in the same way. No step
uses a stopping-time claim about an already exposed seed, and no
failure budget is spent over the age of the stream. All retries
terminate almost surely; all work areas obey their deterministic
space caps.
\end{proof}

\section{A single algorithm with a vanishing exponent allowance}
\label{sec:uniform-subpolynomial-density}

The fixed-allowance theorem can be made uniform in the capacity
without changing the polynomial degree in dimension. A separate
algorithm for each fixed allowance $\eta>0$ does not by itself give
one algorithm with allowance tending to zero: its constants, dimension
exponent, or validity cutoff could depend uncontrollably on $\eta$.
We therefore revisit the explicit cover construction, choose all
margins as computable functions of $n$, and bound every resulting
cost with one fixed polynomial degree. The proof spends the shrinking
allowance in running time while keeping the space exponent exactly
$1+\sigma$.

\begin{theorem}[Quantified subpolynomial overhead]
\label{thm:uniform-subpolynomial-density}
Fix $1<c<\sqrt2$ and $0\le\sigma\le2/c^2-1$.
Assume these constants are rational, or can be approximated to
$k$ bits with certified error in $\operatorname{poly}(k)$ time.
There is one finite-word zero-error dynamic diameter algorithm with
all-time space
\[
       n^{1+\sigma}\operatorname{poly}_{c,\sigma}(d,B,\log(2n))
\]
and expected amortized update time
\begin{equation}
 n^{\,\Lambda_c(\sigma)+O_{c,\sigma}((\log(2n))^{-1/100})}
       \operatorname{poly}_{c,\sigma}(d,B,\log(2n)).
 \label{eq:uniform-subpolynomial-lv}
\end{equation}
The polynomial degrees are independent of $n$ and of the shrinking
allowance. The algorithm is always correct on an unbounded,
exposed-seed adaptive stream and needs no update horizon.

The capped, scheduled version has the same power of $n$, worst-case
update time, and all-time space, with
$\operatorname{poly}_{c,\sigma}(d,B,\log(2nt/\delta))$ factors, and
simultaneous correctness probability $1-\delta$ over $t$ adaptive
updates. Thus both variants may be written with exponent
$\Lambda_c(\sigma)+o(1)$, while retaining the exact space exponent
$1+\sigma$.
\end{theorem}
\begin{proof}
We specify all parameters that were previously held fixed. Write
\[
 L=\max\{2,\log(2n)\},\qquad \zeta=1/100.
\]
Constants below may depend on $c,\sigma$, but not on $n$, the input
dimension, or coordinate precision. Bounded $n$ below a fixed
$(c,\sigma)$-dependent cutoff can be handled exactly.

\paragraph{Geometry and the birth-size cutoff.}
Choose all radial, local-scale, original-query-net, and padding
reserves as conservative rationals of order $L^{-\zeta}$.
The local matching inequalities then hold with approximation at
most $c$, and the common angular problem has
\[
 h_n=g_c-O_c(L^{-\zeta}),\qquad
                  \delta_n=\Theta_c(L^{-\zeta}),
 \qquad g_c=c^2-1.
\]
Here $h_n$ is the strong correlation and $-\delta_n$ bounds rejected
correlations. Above a fixed cutoff, $g_c/2\le h_n<g_c<1$.
The number of local configurations and the reciprocal net precision
are polynomial in $L$, with fixed degrees.

A generation with birth size
\[
                         m<\exp(\sqrt L)
\]
uses exact scanning instead of filters. Its space is linear in its
birth records, deletion is immediate, and query work is
$n^{O(L^{-1/2})}\operatorname{poly}(d,B)$.
For fixed $c<\sqrt2$, $\Lambda_c(\sigma)>0$, so this fits
\eqref{eq:uniform-subpolynomial-lv} above a fixed cutoff.
As in the finite-word construction, a generation whose original
ambient dimension is at least $m$ may also be scanned:
$O(md)\le O(d^2)$. Hence it remains to treat
\[
                  \ell=\log m\ge\sqrt L,\qquad d<m.
\]

\paragraph{The finite family with explicit shrinking margins.}
Use a moderate block dimension
\[
        B_{\rm mod}=\Theta_c(L^{2\zeta}\ell),
\]
rounded up to a power of four. Spherical concentration with distortion
at most $\delta_n/16$ then makes every rejected birth pair
nonpositive in every normalized outer block, for a fixed query,
except with probability at most $m^{-30}$.
This is the only dependence on the negative margin in the projection
dimension. Pad to a power-of-four ambient dimension
$D=O(d+B_{\rm mod})$.
For the large birth sizes under consideration, $B_{\rm mod}<m$;
thus $D=O(m)$ when $d<m$, and the same polynomial-size union bound
as before applies.

Take $b=\Theta(\ell^{2/3})$, also a power of four, and
$s=B_{\rm mod}/b$. The exact splitter family $\mathcal S_{\mathrm{split}}$ has
\[
 \log|\mathcal S_{\mathrm{split}}|=O(s\log B_{\rm mod}),\qquad
 s=O_c(L^{2\zeta}\ell^{1/3}),
 \qquad
 \epsilon_{\rm split}=O(\ell^{-1/3}).
\]
The three-vector splitting proof and its rational representation do
not change. In particular, the target correlation in a small block
is at least $h_n-2\epsilon_{\rm split}$.

Choose rational angular and space parameters within
$O_{c,\sigma}(L^{-\zeta})$ of the minimizing parameters in
Section~\ref{sec:scaled-density-las-vegas}, rounding the internal
space parameter downward. Let $u,q,\kappa$ be the resulting values.
They can be chosen so that $\kappa\in[1/2,1]$ and
\[
 \max\{b_{\rm exp}+3v,\ f_{\rm exp}+2v\}
       \le\Lambda_c(\sigma)+O_{c,\sigma}(L^{-\zeta}),
\]
where
\[
 a=\kappa u,\quad b_{\rm exp}=\kappa q,\quad
 f_{\rm exp}=1+\kappa(q-1),\quad
 v=(a-\sigma_{\rm int})_+,\quad \sigma_{\rm int}\le\sigma.
\]
The corresponding construction and deletion exponents satisfy the
same bound, by the maintenance inequalities already proved.
The rational choices have polynomial encoding length in $\log n$.

Define the local Gaussian threshold by $t^2=2\kappa\ell+O(1)$,
and its small-block threshold by $t_0=t/\sqrt s$.
Uniformly for $\ell\ge\sqrt L$,
\[
 t_0^2=\Omega_c(\ell^{2/3}L^{-2\zeta})
          =\Omega_c(L^{1/3-2\zeta})\longrightarrow\infty .
\]
All small-sphere tail estimates therefore apply.
The good-pair Gaussian constant is uniform: $h_n$ lies in a fixed
compact interval inside $(0,1)$, and $h_n\le x\le1$.
The small-block target $h_n-2\epsilon_{\mathrm{split}}$ and the checked
net threshold $h_n-3\epsilon_{\mathrm{split}}$ lie in such a compact
interval as well, so the same uniform rectangle estimate applies
to the actual certificate.
In particular, the explicit constant
$2^{-\lceil100/(1-h_n)\rceil}$ is bounded below by a positive
constant depending only on $c$; it contains no exponential
dependence on $1/\delta_n$.

Use the same small-sphere net of radius $b^{-2}$, norm cutoff
$G=O(\sqrt\ell)$, and exact cover certificate. The inequality needed
for the net margin holds uniformly, since
\[
                  G\,b^{-2}t_0=O_c(\ell^{-1/2}L^{-\zeta})=o(1).
\]
The base-family size $M_{\rm base}$ satisfies
\[
 \log M_{\rm base}
  =I_{h_n}(x)\frac{t^2}{2s}
        +O_c\!\left(\epsilon_{\rm split}\frac{t^2}{s}
                                      +\log L\right).
\]
The net-verification logarithm is $O(b\log b)$.
Consequently the logarithm of the entire implicit description,
or its expected certification work, apart from a fixed polynomial
factor in $D,B$, is at most
\[
 O_c\!\left(L^{2\zeta}\ell^{1/3}\log L
                              +\ell^{2/3}\log L\right).
\]
Dividing by $\ell\ge\sqrt L$ gives
\begin{equation}
 O_c\!\left(
       L^{\,2\zeta-1/3}\log L
                      +L^{-1/6}\log L\right)
                         =o(L^{-\zeta}).
 \label{eq:uniform-density-description-loss}
\end{equation}
This explicitly controls description size, finite-net enumeration,
and the expected resampling work.

The full raw-universe logarithm has the leading term
$\kappa I_{h_n}(x)\ell$.
Its remaining terms come from splitter count, cover prefactors,
the $2^s$ conditioning loss, and the norm-gate error.
After division by $\ell$, their sum is
\[
 O_c\!\left(\ell^{-1/3}
            +L^{2\zeta}\ell^{-2/3}\log L\right)
                               =o(L^{-\zeta}).
\]
Exactly the same bound applies to the data, query, and pair moment
exponents. The bad-pair sum still contains the actual $m$ birth
points. All finite Gaussian and rotation exceptions retain their
$m^{-30}$ or smaller budgets. Their deterministic full-enumeration
bound remains polynomial in $m,D,B$, so they add no larger term.
No error budget is changed to use $m^\kappa$.

\paragraph{Exact space and fixed polynomial degrees.}
Choose a computable rational
\[
                     \eta_n=C_{c,\sigma}L^{-\zeta}
\]
within constant factors of this value, with the constant large enough
to dominate the preceding explicit losses. Above a fixed cutoff it
is less than $1/100$. Point assignment gives
\[
 R\le m^{a+\eta_n}\mathcal P,\qquad
 U\le m^{\kappa(1+u)+\eta_n}\mathcal P,\qquad U\le mR,
\]
after setting $R\ge\lceil U/m\rceil$ as before.
The implicit rank oracle costs $m^{\eta_n}\mathcal P$.
Use a power-of-two inverse parameter within a factor of two of
$m^{v+2\eta_n}$, excluding the polynomial dimension factor.
The native condition
$3v+6\eta_n<1+a$ holds uniformly above a fixed cutoff:
on the first branch $a,v$ remain near the bounded optimizer, and
on the second branch $v=0$.
The inverse and its low-workspace verifier use
\[
       \widetilde O(mR/m^{v+2\eta_n})
                   \le m^{1+\sigma_{\rm int}-\eta_n}\mathcal P
                   \le m^{1+\sigma}\mathcal P .
\]
Original point records and the global family description also fit.
Thus the shrinking allowance is spent only in time, not in space.

Here and in these estimates $\mathcal P$ has a degree independent of $n$.
To see this directly, the outer matrices have dimension
$O(d+L^{1+2\zeta})$; their exact rational construction and application
use a fixed polynomial number of operations. Splitter enumeration
and net verification are already accounted for in
\eqref{eq:uniform-density-description-loss}, rather than hidden in
the dimension polynomial. Rank selection uses integer products and
base-cap scans. Every geometric comparison has degree at most
thirty-two and polynomial coefficient length.
Required coupling precision is polynomial in
$\log m,\log D,\log(1/\delta_n)$, hence in $d,B,\log n$ with fixed
degrees. The rational center LPs, verified inverse routines, and
balanced dictionaries likewise have fixed polynomial degrees.
There is no operation of the form $d^{1/\eta_n}$, and no
uncontrolled constant of the form $\exp(1/\delta_n)$.

All earlier large-size requirements have now been checked by explicit
powers of $L$ and $\ell$: $b\le B_{\rm mod}$,
$\epsilon_{\rm split}<h_n/10$, $t_0\ge1$, and $G b^{-2}t_0=o(1)$.
Thus the remaining cutoff is one fixed number depending on
$c,\sigma$, not an unspecified cutoff depending on the current
value of $\eta_n$.

\paragraph{Dynamic composition.}
The scaled resource proposition gives initialization $mA(n)$,
original-point maintenance $A(n)$, and a birth-envelope query bound
of expectation at most $A(n)$, where
\[
 A(n)=
 n^{\,\Lambda_c(\sigma)+O_{c,\sigma}(L^{-\zeta})}
                \operatorname{poly}_{c,\sigma}(d,B,L).
\]
Small scanned generations satisfy the same interface.
The number of local configurations, query-net amplification copies,
and lifetime factors has increased only polynomially in $L$.
Apply Lemma~\ref{lem:adaptive-certified-generation-compiler}
to obtain the zero-error theorem, including unbounded adaptivity
and all-time exact space. Applying instead the capped phase and
frozen-generation schedules of
Section~\ref{sec:density-inverse-deamortization} yields the stated
worst-case Monte Carlo version. Its extra horizon and confidence
factors are polynomial logarithmic and do not enter the power of $n$.
\end{proof}

This statement concerns fixed $c,\sigma$. It does not claim polynomial
dependence on $1/(c-1)$ uniformly as $c$ tends to one, or polynomial
dependence on the reciprocal of an arbitrarily prescribed exponent
allowance.

\section{A static farthest-neighbor consequence}
\label{sec:paper-static-fns}

The same geometric reduction gives a static search structure with
near-linear preprocessing and space. We first retain a lifted-center
chain and an all-query baseline at each node. An asymmetric choice
of the explicit angular filters then makes their data cost near-linear
while retaining the desired query exponent. Finally, amplification
over direction nets makes the guarantee simultaneous for every query,
including a query chosen after the structure's seeds have been revealed.
The Monte Carlo construction uses explicit filters; the zero-error
variant below uses certified inverses to keep its exact space exponent.

\begin{corollary}[Static farthest-neighbor search]
\label{cor:paper-static-fns}
Fix $1<c<\sqrt2$ and $0<\delta<1/2$. For each nonempty indexed family of $n$ points in
$\mathbb R^d$ fixed before preprocessing randomness is drawn, there is a Monte Carlo farthest-neighbor structure with
worst-case preprocessing time and all-outcome space
\[
 n^{1+o(1)}\poly_c(d,\log(2n/\delta)),
\]
and worst-case query time
\[
 n^{\,c^2(2-c^2)+o(1)}\poly_c(d,\log(2n/\delta)).
\]
With probability at least $1-\delta$, it returns an actual
$c$-approximate farthest point for every query, including queries
chosen after the entire structure and its seeds are exposed.
The $o(1)$ terms can be taken as
$O_c(\log\log(4n)/\sqrt{\log(2n)})$ and do not depend on $d$.
The arithmetic-RAM statement holds for arbitrary real inputs.
Rational data and queries of encoding length at most $B$ have
ordinary running-time bounds with an additional fixed polynomial
factor in $B$.
\end{corollary}

\begin{proof}
An empty set reports \textsc{None}, and a singleton returns its sole
point. For all bounded $n$ below a fixed $c$-dependent cutoff, scan
exactly; this also covers the constant adjustments ensuring $t\ge1$.
Let $L=\log(2n)$. Build the static lifted-center chain from
Section~\ref{sec:paper-geometry}, using the Monte Carlo constructor
and checking that each core has size at least $\alpha m/2$.
On an unsuccessful core check use the same deterministic contraction
fallback as in the dynamic construction. This caps the depth and
total input cardinality on every outcome. A polynomial-size exact
leaf ends the chain. The rational implementation uses the sampled
centerpoint program; no strongly polynomial real linear-programming
assumption is needed.

At each node take one fresh halfspace $\alpha$-net of its lifted
fixed input pool. Its size is
\[
 O\!\left(\alpha^{-1}
       \bigl((d+2)\log(1/\alpha)+\log(n/\delta)\bigr)\right)
\]
after increasing the constant for the polynomial number of nodes.
The halfspace-net theorem~\cite{hausslerwelzl}, conditional on the
center, shows that this one sample hits every witness halfspace
in Lemma~\ref{lem:paper-lifted-certificate} simultaneously. Thus,
for every $q$, scanning the stored sample returns an actual distance
$L_q$ with
\begin{equation}\label{eq:static-fns-baseline}
                         L_q^2\ge V+Q^2,
                         \qquad Q=\|q-o\|.
\end{equation}
Center and net failures receive conditional budgets whose sum is at
most $\delta/2$. All random sampling has a deterministic work cap;
a failed cap is included in this budget. Alternatively, on rational
inputs a checked disjoint block gives the same all-query baseline,
with expected linear preprocessing per node.

Choose the local parameters of Section~\ref{sec:paper-parameters}
with $e=\Theta_c(1/L)$. At a node with $V>0$, store its core points
in their inward radial shells. Build the angular index for each
eligible left-shell, right-shell, and weak-scale configuration from
\eqref{eq:local-eligible-configuration}; its data pool is the whole
corresponding core shell. The right shell now describes a query
radius, not a second stored point set. On query $q$, obtain $L_q$.
If $V=0$ or $Q>K\sqrt V$, the baseline covers the core within
factor $c$. Otherwise round $Q$ inward on the same grid, and query
the configurations for that right shell and scales $R_\ell>L_q$.
A zero right shell makes no angular calls; its core distances are
already covered by the baseline.
Accept exactly the original angular weak predicate, and measure each
returned point's actual distance from $q$. Take the maximum of all
such distances, all node baselines, and an exact leaf scan.

Here is the local approximation proof. Let
$D_C=\max_{p\in C}\|p-q\|$. If $D_C\le cL_q$, the baseline
suffices. Otherwise $V>0$, $Q\le K\sqrt V$, and the geometric
scale grid contains $R_\ell$ such that
\[
 L_q<R_\ell<D_C/\kappa,\qquad
                         R_\ell\ge D_C/(\kappa\lambda)>D_C/c.
\]
Rounding the target and query loses at most $2e\sqrt V<2eR_\ell$,
leaving a strong rounded pair. Its squared shell radii sum to at
most $V+Q^2\le L_q^2<R_\ell^2$, so its configuration is eligible.
Neither rounded radius is zero in this case. A successful angular
query returns a weak rounded pair. Joint convexity gives
\[
 \|\widetilde p-\widetilde q\|
 \le\max\{\|p-q\|,\|p-o\|,Q\}
 \le\max\{\|p-q\|,L_q\}.
\]
Its actual distance is therefore greater than $R_\ell>D_C/c$.
Every data point belongs to a core or the leaf, proving the global
farthest-neighbor guarantee. Zero shell directions need not be indexed;
the same baseline covers them.

It remains to obtain near-linear preprocessing at the endpoint $u=0$.
Use the common lifted correlation $h=c^2-1-O_c(1/L)$ and set $x=h$.
For every nonempty shell index use the same global tensor parameters
\[
 \ell=\lceil\sqrt L\rceil,\qquad
 L\le\ell t^2/2\le L+O(1),\qquad t\ge1.
\]
The good-cap exponent is exactly $I_h(h)=1$. Hence the explicit
construction of Lemma~\ref{lem:fixed-depth-exact-space} has
\[
 \log A=L+O_c(\sqrt L\log L),\qquad
 E_u=\exp(O_c(\sqrt L\log L)),\qquad
 E_q=n^{1-h^2}\exp(O_c(\sqrt L\log L)).
\]
Base-list evaluation also costs $\exp(O_c(\sqrt L\log L))$.
Each pool has at most $n$ points, so the whole-birth bad-occurrence
bound uses $nAp_2^\ell\le E_q$, regardless of that pool's size.
Empty indices return none immediately. Thus small birth pools
require no separate asymptotic cutoff.

Amplify the capped explicit indices over the original-direction net,
all birth targets, and the polynomial number of configurations,
with remaining failure budget $\delta/2$. The union bound is over
this fixed finite family and its direction nets, rather than over a
sequence of queries. The padded-net guarantee then gives one event
valid for every later query, so no query horizon is needed. The total
stored core memberships and node inputs are $n\poly(d,\log(n/\delta))$,
so the displayed preprocessing and space bounds follow. The query
visits only polynomially many logarithmic levels and configurations.
Finally, $1-h^2=c^2(2-c^2)+O_c(1/L)$, whose extra factor in time
is constant. Tensor depth affects the logarithmic overhead, never
a power of the ambient dimension. Finite Gaussian coupling and
exact signed comparisons are as in the explicit dynamic structure.
\end{proof}

For binary points and $c=\sqrt{2-\theta}$, the corollary gives
Hamming $(2-\theta)$-FNS query exponent $2\theta-\theta^2+o(1)$,
with $n^{1+o(1)}$ preprocessing and space up to polynomial dimension
factors. This improves the $O(\sqrt\theta)$ query exponent for
Hamming farthest-neighbor search obtained by Andoni, Jiang, and
Zharkov~\cite{ajz2026}. The corollary concerns a fixed data set;
the fully dynamic diameter results do not rely on it.

\begin{corollary}[Zero-error static search]
\label{cor:paper-static-fns-las-vegas}
For rational inputs and fixed polynomial-time-computable $1<c<\sqrt2$,
there is an always-correct static farthest-neighbor structure using
$n\poly_c(d,B,\log(2n))$ space at all times. For every fixed prefix
of $t$ queries chosen adaptively after all preceding random choices
are exposed, its
expected total running time, including preprocessing, is
\[
 n^{1+o(1)}\poly_c(d,B,\log(2n))
 +t n^{\,c^2(2-c^2)+o(1)}\poly_c(d,B,\log(2n)).
\]
Here $B$ bounds data and query encoding lengths. The $o(1)$ can be
taken as $O_c((\log(2n))^{-1/100})$ with fixed polynomial degrees.
No query horizon is supplied.
\end{corollary}
\begin{proof}
We retain the certified geometry and refresh only the angular indices.
A uniformly fast collection of indices serves every later query within
budget. If a query reveals a slower collection, exact scanning answers
that query and a fresh independent construction restarts the process.
An all-query event is essential here because the query may be chosen
after the collection's seeds are exposed.

Build the static chain with checked disjoint-block certificates.
Scanning one fixed block gives~\eqref{eq:static-fns-baseline} for
every query by~\eqref{eq:paper-certified-average}; it remains intact
forever. Keep this geometry unchanged. In each angular pool use
Proposition~\ref{prop:scaled-density-product-index} with
$x=h$, $u=0$, $\kappa=1$, and $\sigma=v=0$.
Apply the explicit uniform schedule of
Appendix~\ref{sec:uniform-subpolynomial-density}.
In this application all raw counts and inverse domains use the
actual pool birth size $m$, with the stated small-pool scan cutoff.
The resulting complete angular collection has exact linear space
up to polynomial factors, expected construction cost
$\mathcal B=n^{1+o(1)}\mathcal P$, and a fixed-query birth-envelope expectation
at most $\mathcal A=n^{1-h^2+o(1)}\mathcal P$.
The polynomial $\mathcal P$ can be enlarged to cover every pool and
configuration. As before, $h=g_c-O_c((\log n)^{-1/100})$.

Let $J$ be the number of angular pools and configurations.
Within each pool, amplify independent complete copies over its
original-direction net to failure probability at most
$1/(4\max\{1,J\})$. If $J=0$, the baselines and leaf answer every
query and no angular amplification is needed.
A union bound gives one event of probability at least $3/4$ on which
every pool answers every query within its local race budget.
The sum of those budgets is $O(\mathcal A)$ after enlarging $\mathcal P$.
Coverage itself is deterministic. Query the required pools by racing
their copies separately. Each completed search is always valid;
if any pool's race exhausts its budget, scan the data exactly,
discard all angular indices, and
rebuild them with fresh coins before finishing the query.
The checked geometry is retained. All attempts obey the same
space bound and terminate almost surely.

Let $\mathrm{Reach}_j$ denote that the $j$th collection is born.
Its occurrence is determined before that collection's fresh coins.
A uniformly fast collection never triggers a rebuild, hence
$\Pr(\mathrm{Reach}_j)\le4^{1-j}$, even when queries depend on all
previous seeds and construction times. Conditional expected
construction cost at every birth is at most $\mathcal B$.
The tower property therefore bounds all construction work by
$4\mathcal B/3$ in expectation. Exact fallback scans have the same
geometric bound and cost $n\poly(d,B)$ each, absorbed in $\mathcal B$.
The local races for the first $t$ queries cost
$O(t\mathcal A)$ deterministically.
This proves the stated expected-prefix bound; it does not assert
conditional expected time for an individual query after its
structure has been exposed.
\end{proof}

\end{document}